\documentclass[a4paper,11pt]{article}
\usepackage{amsmath,amsthm,amsfonts,caption,multirow,graphicx,subfigure,float,lscape,setspace}
\usepackage[round]{natbib}

\DeclareMathOperator{\dprod}{\mathrm{\prod}}

\begin{document}

\title{Testing for Quantile Sample Selection\thanks{%
We are grateful to Bertille Antoine, Federico Bugni, Xavier D'Hautefoeuille,
Giovanni Mellace, Toru Kitagawa, Peter C.B. Phillips, and Youngki Shin for
useful discussions and comments. Moreover, we would like to thank seminar
participants at the ISNPS meeting (Salerno, 2018), the ICEEE meeting (Lecce,
2019), the CIREQ Montreal Econometrics Conference (Montreal, 2019), the
Warwick Econometrics Workshop (Warwick, 2019), and the Southampton Workshop
in Econometrics and Statistics (Southampton, 2019), and the ES World Congress (Milan, 2020) for helpful comments.}}
\author{Valentina Corradi \thanks{%
Department of Economics, University of Surrey, School of Economics,
Guildford GU2 7XH, UK. Email: \texttt{\ V.Corradi@surrey.ac.uk}} \\
Surrey University \and Daniel Gutknecht\thanks{%
Corresponding Author: Department of Economics and Business, Goethe
University Frankfurt, Theodor-W.-Adorno Platz 4, 60629 Frankfurt, Germany.
Email: \texttt{\ Gutknecht@wiwi.uni-frankfurt.de}} \\
Goethe University Frankfurt}
\date{\today  }
\maketitle

\noindent \textbf{Abstract} This paper provides nonparametric tests for
detecting sample selection in conditional quantile functions.
The first test is an omitted predictor test with the propensity score as the omitted variable, which holds uniformly across a compact set of quantile ranks. As
with any omnibus test, in the case of rejection we cannot distinguish
between rejection due to genuine selection or to misspecification. Thus, we
suggest a second test to provide supporting evidence whether the cause for rejection at the
first stage was solely due to selection or not. Using only individuals with
propensity score close to one, this second test relies on an `identification
at infinity' argument, but accommodates cases of irregular identification. Importantly, neither of the two tests requires parametric assumptions on the selection equation nor a continuous exclusion
restriction. Data-driven bandwidth procedures for both tests are proposed, and simulation evidence suggests a good finite sample performance in particular of the first test. We apply our procedure to test for selection in log hourly wages using UK
Family Expenditure Survey data as \citet{AB2017}. Our findings indicate that some of the evidence for sample selection among males may instead reflect misspecification of the quantile functions. Finally, we also derive an extension of the first test to conditional mean functions.\newline
\newline

\newline


\newpage \doublespacing

\section{Introduction}

Empirical studies using non-experimental data are often plagued by the
presence of non-random sample selection: individuals typically self select
themselves into employment, training programs etc. on the basis of
characteristics which are believed to be non-randomly distributed and
unobservable to the researcher(s) \citep{G1974,H1974}. In fact, it is well
known that ignoring selection in conditional mean models induces a bias in
the estimation, which can be additive \citep[see e.g.][]{H1979,DNV2003} or
multiplicative \citep[][]{J15} depending on the functional form of the
model. In both cases, one can deal with the selection bias by adopting a
control function approach. On the other hand, until recently little was
known about identification and estimation of conditional quantile models in
the presence of sample selection, see the recent survey by %
\citet{ABsurvey2017}. A notable exception is the case of sample selection in
`location shift' models, where only the intercept is allowed to vary across
quantile ranks, and the control function approach still applies %
\citep[e.g.,][]{HM2015}. In all other cases, however, including linear
quantile regression models, the presence of sample selection is more
difficult to deal with and control function methods can no longer be used.
In fact, \citet{AB2017} proposed an alternative three-step estimator for
parametric linear conditional quantile models, which has recently gained
popularity in the literature. However, identification and estimation in %
\citet{AB2017} require either parametric restrictions on the joint
distribution of the unobservable error terms or functional form conditions
on the latter, which are hard to verify and still rely on a parametric choice
in practice.

In this paper, we therefore propose two different nonparametric tests for
detecting sample selection in conditional quantile functions. These tests
only impose a minimal set of functional form assumptions on both the outcome
and the selection equation(s). In fact, the only additional assumption is
that selection (if present) affects the outcome through the propensity
score, the probability to be in the selected sample, a standard assumption
in the sample selection literature \citep[e.g.,][]{DNV2003}. Importantly,
the tests do also not require a continuous exclusion restriction such as a
continuous instrumental variable in the selection equation. This is of
particular relevance for empirical work as various examples from the
literature have used the three-step estimator of \citet{AB2017} without
relying on such a variable. For example, recent studies examining the
effects of sample selection in conditional wage distributions from the
Current Population Survey (CPS) in the context of nonresponse %
\citep{BHHZ2019} and the gender wage gap \citep{MW2019} rely on discrete
instrumental variables such as `month-observed-in-sample' and `number and
presence of children', respectively.

Formally, the first test we propose is a test for omitted predictors,
where the omitted predictor is the propensity score. The test statistic is
similar to that of \citet{VBDN2013}, but holds  holds
uniformly over a compact subset $\mathcal{T}\subset (0,1)$ of quantile ranks. The statistic is constructed as a weighted average
of conditional quantile errors. Since the latter are not observed, we need
to estimate them in the first place, which in turn requires estimation of
the nonparametric conditional quantile function under the null hypothesis.
This is done using the local polynomial estimator of \citet{GS2012}.
We derive an asymptotic representation for the statistic, which features estimation error from the conditional quantile function, but
not from the propensity score \citep[cf.][]{EJCL2014}. Importantly, we can test
the null of no selection over different subsets of quantile ranks simultaneously. As the
limiting distribution is non-pivotal and depends on features of the Data
Generating Process, we establish the first order validity of wild bootstrap
critical values. Moreover,  to select the
bandwidth parameter(s) for this test, we suggest a data-driven way that relies on a procedure suggested by %
\citet{LR2008} and that allows for discrete covariates. Finally, in Appendix B we also provide an
extension of this test to nonparametric conditional mean
functions. The latter are commonly used in practice and tests for sample
selection have so far relied on the correct (semi-)parametric specification
of these mean function under the null \citep[e.g.,][]{BJLST2015}.

As with other omnibus tests, if we reject the null of the first
test, we cannot distinguish between a rejection due to genuine sample
selection or to omission of other relevant predictor variable(s), not
independent of the propensity score. This distinction is crucial when
estimation of \textit{nonparametric} conditional quantile functions is the
ultimate goal. In fact, sample selection generally leads to a loss of point
identification \citep[cf.][]{AB2017}, but consistent estimation and
inference may still be carried out on a subset of observations with
propensity score close to one. By contrast, omitting relevant predictors
impedes consistent estimation and inference altogether. To understand the
heuristics of our second test, note that dependence of the conditional
quantile error on the propensity score when the latter is (close to) one
hints at the presence of omitted relevant predictors. This is so because
individuals with a propensity score equal to one are selected into the
sample almost surely and thus sample selection bias is not present in that
case. By contrast, our conditional quantile function is likely to miss out
on relevant predictor(s) correlated with the propensity score if it depends
on the latter, regardless of whether it takes on values in the interior of
the unit interval or close to one. The null hypothesis of the second test
formalizes this heuristic argument.

The test statistic of the second test is constructed as a weighted average of `quantile errors' from individuals with
(estimated) propensity score close to one, so that selection bias does no
longer `bite'.\footnote{%
Thus, to obtain power against misspecification in this test we require that
the omitted predictor(s) are correlated with the propensity score, even when
the latter is at or close to one.} However, while the second test relies on
a so called `identification at infinity' argument and thus requires
observations with estimated propensity score `close to one' (in a
nonparametric sense), our test does not need continuous instrument(s) and
allows for a thin set of observations close to the boundary, thus
accommodating cases of so called irregular identification \citep{KT2010}. In
fact, the rate of convergence of the second test depends on the degree
of irregularity of the marginal density of the propensity score. We therefore suggest a studentized version of the test
statistic, which is rate adaptive and converges weakly even if numerator and
denominator of the statistic diverge individually at the same rate.

We conduct a Monte Carlo study to asses the finite sample properties of our
two tests: while the first test performs well in terms of size and power
throughout all designs even when the exclusion restriction consists of a
discrete instrument and the tuning parameters are chosen in a completely
automated manner, the results of the second test appear to be somewhat more
sensitive to the choice of the tuning parameters as well as to the
continuity (or discreteness) of the instrument(s). This reflects the
irregularity of the underlying problem.

We apply our testing procedure to test for selection in log hourly wages of
females and males in the UK using data from the UK Family Expenditure Survey
from 1995 to 2000. The same data was recently also used by \citet{AB2017} to
analyze gender wage inequality in the UK. We run our testing procedure on
two different sub-periods of different economic performance, namely
1995-1997 and 1998-2000. As a preview of the results, we cannot find
evidence for selection among females for the 1995-1997 period, but only for
the 1998-2000 period. By contrast, while we reject the null of the first
test for males with data from 1995 to 1997, our second test strongly
suggests that this rejection may actually be due to misspecification of the
quantile function, a feature that might have remained undetected without our
testing procedure.

The rest of the paper is organized as follows. Section \ref{Set-Up} outlines
the set-up. Section \ref{First Test} then establishes
the limiting behavior of the first test for omitted variables, and the first
order validity of inference based on wild bootstrap critical values. Section %
\ref{Second Test} on the other hand derives the same results for the second test, and Section \ref{Monte Carlo Simulation} reports the findings of our Monte Carlo
Study. Finally, Section \ref{Empirical Illustration} provides an empirical
illustration in which we apply our procedure to log hourly wages of females
and males in the UK and Section \ref{Conclusion} concludes. The definition
of various nonparametric estimators used in the construction of the test
statistics, and proofs of the main Theorems 1 and 2 are provided in
Appendix A. On the other hand, a formal outline of the first test for the conditional mean, together with a set of Monte Carlo simulations can be found in Appendix B.

The proofs of several technical Lemmas as well
as of results on the first order asymptotic validity of bootstrapped
critical values have been relegated to the supplementary material. In addition, the
supplementary material contains: (i) a formal argument for a data-driven
bandwidth choice in the second test; (ii) a two-step testing procedure with a formal decision rule; (iii) a formal proof that the classification errors of
this procedure are asymptotically controlled at pre-specified levels;  (iv) proofs of the conditional mean test.

\section{Set-Up}

\label{Set-Up}

We begin by outlining the data generating process. As it is customary in the
sample selection literature, we postulate that the continuous outcome
variable of interest, $y_{i}$, is observed if and only if $s_{i}=1$, where $%
s_{i}$ denotes a binary selection indicator. A standard application example of this set-up is for
instance to the study of wages and employment: (log) wages $y_{i}$ are only
observed for individuals who participate in the labor market and who are
employed ($s_{i}=1$), and different sub-groups (e.g., males and females) may
differ in terms of their unobservable labor market attachment. Thus,
conventional measures of wage gaps or wage inequality may be biased %
\citep{H1974,H1979}. For every individual $i$, we observe $x_{i}$ and $z_{i}$. The variable(s) in $x_{i}$ affect observed outcome $y_{i}$, while the variable(s) in $z_{i}$ predict the selection variable $s_{i}$, but not  directly observed outcome $y_{i}$ once we condition on $x_{i}$.\footnote{The assumption that $z_{i}$ is statistically independent $y_{i}$ once we condition on $x_{i}$ and selection $s_{i}=1$ is testable using for example \citep{K2010}.  }  Note that the variables in $x_{i}$ and $z_{i}$ need not be disjoint as in most applications, although our testing procedure requires
at least one of the variable(s) in $z_{i}$ to be excluded from $x_{i}$ (cf. Assumption
A.1 below). This is typically referred to as instrumental exclusion restriction in the literature and we will therefore refer to the element(s) of $z_{i}$ that are excluded from $x_{i}$ as instrument(s) or instrumental variable(s) in what follows. Importantly, however, none of the instrument(s) in $z_{i}$ needs
to be continuous, an aspect that we will explore in greater detail in the
simulations in Section \ref{Monte Carlo Simulation}.

Throughout the paper, the maintained assumptions are that (i) the
variable(s) in $z_{i}$ and (ii) non-random selection (if present) enter
the conditional quantile function only through the propensity score $ \Pr \left( s_{i}=1|z_{i}\right) =p(z_{i})\equiv p_{i}$, the probability to be in the
selected sample for a given $z_{i}$. Formally, this can be expressed as follows:

\medskip

\noindent \textbf{A.Q} For all $\tau $,%
\begin{equation}
\Pr \left( y_{i}\leq q_{\tau }\left( x_{i}\right) \Bigl\vert %
x_{i},z_{i},s_{i}=1\right) =
\Pr \left( y_{i}\leq q_{\tau }\left( x_{i}\right) \Bigl\vert %
x_{i},p_{i}\right) ,  \label{EQ1a}
\end{equation}%
holds almost surely, where $q_{\tau }\left( x_{i}\right) $ denotes the
conditional $\tau $-quantile of $y_{i}$ given $x_{i}$ and selection $s_{i}=1$. \medskip

\noindent Assumption \textbf{A.Q} is a high-level assumption and indeed the quantile equivalent of Assumption 2.1(i) in \citet[][p.35]{DNV2003}. It is for instance implied by the model of \citet{AB2017} with the threshold crossing
selection equation $s_{i}=1\{p(z_{i})>v_{i}\}$ and $x_{i}\subset z_{i}$.\footnote{Indeed, \citet{AB2017} require that at least one element in $z_{i}$ exists, which is not in $x_{i}$.} In their paper: 
\begin{equation*}
y_{i}=y_{i}^{\ast }s_{i}\quad \text{iff}\quad s_{i}=1,
\end{equation*}%
with $y_{i}^{\ast}=q(x_{i},u_{i})$ where the unobservables $(u_{i},v_{i})$ are jointly statistically independent of $z_{i}$ given $x_{i}$ and are absolutely continuous (with standard uniform marginals). In this set-up, provided $p_{i}>0$ with probability one, we obtain that: 
\begin{eqnarray*}
\Pr \left( y_{i}^{\ast }\leq q_{\tau }\left( x_{i}\right) \Bigl\vert %
x_{i},z_{i},s_{i}=1\right)  &=&\Pr \left( q(x_{i},u_{i}) \leq
q_{\tau}\left( x_{i}\right) \Bigl\vert x_{i},z_{i},v_{i}<p(z_{i})\right) 
 \\
&=&\Pr \left( y_{i}\leq q_{\tau }\left( x_{i}\right) \Bigl\vert %
x_{i},p_{i}\right).  \notag
\end{eqnarray*}%
Note that in fact $q_{\tau }(x_{i})$, the `observed' $\tau$ quantile of $y_{i}$ given 
$x_{i}$ (and $s_{i}=1$), coincides with the $\tau $ quantile of $y_{i}^{\ast }$
given $x_{i}$ and selection when selection is random.\medskip

\noindent Given the existence of a valid instrument, we test the hypothesis
that the propensity score is not an omitted predictor, against its negation.
In what follows, let $\mathcal{T}=[\underline{\tau },\overline{\tau }]$
denote the compact set of quantile ranks to be examined, where $0<\underline{%
\tau }\leq \overline{\tau }<1$. Also, we use $\mathcal{X}$ to denote a
compact set in the interior of the union of the supports of covariates $%
R_{x} $, and $\mathcal{P}$ to denote a compact subset of the support of $%
p(z_{i})$. In the first step we test $H_{0,q}^{(1)}$ versus $H_{A,q}^{(1)}$
using the subset of selected individuals for which $s_{i}=1$, i.e.%
\begin{equation}
H_{0,q}^{(1)}:\Pr \left( y_{i}\leq q_{\tau }(x_{i})|x_{i}=x,p_{i}=p\right)
=\tau \quad \text{for all $\tau \in \mathcal{T}$, $x\in \mathcal{X}$, and $%
p\in \mathcal{P}$}  \label{H01q}
\end{equation}%
versus%
\begin{equation}
H_{A,q}^{(1)}: \Pr \left( y_{i}\leq q_{\tau }(x_{i})|x_{i}=x,p_{i}=p\right)
\neq \tau \quad \text{for some $\tau \in \mathcal{T}$, $x\in \mathcal{X}$,
and $p\in \mathcal{P}$}.\footnote{%
From here onwards, we make the conditioning on values of $x_{i}$ and $p_{i}$
explicit whenever required for clarity.}  \label{HA1q}
\end{equation}%
The logic behind $H_{0,q}^{(1)}$ vs. $H_{A,q}^{(1)}$ is that, given (\ref%
{EQ1a}), 
\begin{equation*}
\Pr \left( y_{i}\leq q_{\tau }(x_{i})|x_{i},p_{i}\right) =\Pr \left(
y_{i}\leq q_{\tau }(x_{i})|x_{i},p_{i},s_{i}=1\right) =\tau
\end{equation*}%
if and only if $\Pr \left( y_{i}\leq q_{\tau
}(x_{i})|x_{i},p_{i},s_{i}=1\right) =\Pr \left( y_{i}\leq q_{\tau
}(x_{i})|x_{i},s_{i}=1\right)$. Note that the null hypothesis of no omitted
predictor in (\ref{H01q}) could have been also stated in terms of
Conditional Distribution Functions (CDFs), i.e. 
\begin{equation}
F_{y|x,p}(y|x_{i}=x,p_{i}=p)=F_{y|x}(y|x_{i}=x)  \label{EQCI}
\end{equation}%
for all $x\in \mathcal{X}$, $p\in \mathcal{P}$, and $y\in \mathcal{Y}=\{q_{\tau}(x):x\in\mathcal{X},\tau\in\mathcal{T}\}$, subset of the support of $y_{i}$. A test for the null
in (\ref{EQCI}) could for instance be based on the difference between CDFs
estimated using a larger and a smaller information set. However, while this may
circumvent the issue of extreme quantile estimation, in the context of
sample selection, interest often lies in specific quantiles or a subset of
quantile ranks. For instance, we might only be interested in testing for
sample selection in the (log) wage distribution of males and females from
lower conditional quantiles such as from the 10\% to the 25\% quantiles, or
of individuals that earn below the (conditional) median wage etc.. To carry
out this type of analysis in the conditional distribution function context
would require finding corresponding values say $y_{1}$ and $y_{2}$ to
examine all $y$ such that $y_{1}\leq y\leq y_{2}$. These values are
typically unknown and require estimating the conditional quantiles in the
first place.

\section{First Test}

\label{First Test}

We now introduce a statistic for testing $H_{0,q}^{(1)}$ vs. $H_{A,q}^{(1)}$%
, as defined in (\ref{H01q}) and (\ref{HA1q}). Moreover, for notational
simplicity, from here onwards we assume that all components of $x_{i}$ are
continuous, while we make the possibility of discrete elements in $z_{i}$
explicit by partitioning it into $z_{i}=(z_{i}^{c},z_{i}^{d})$, where `$c$'
(`$d$') denotes the sub-vector of continuous (discrete) elements. The
extension to discrete elements in $x_{i}$ is immediate at the cost of more
complicated notation and more lengthy arguments in the proofs. In fact, from %
\citet{HRL2004} and \citet{LR2008} we know that discrete covariates do not contribute to
the rate at which the variance approaches zero, and hence they do not add to
the curse of dimensionality. We discuss the extension to the case of
discrete elements in $x_{i}$ in more detail in the empirical application
section. In addition, note that partitioning the vector $z_{i}$ into
continuous and discrete elements allows for the possibility that $%
x_{i}=z_{i}^{c}$, and that only $z_{i}^{d}$ satisfies the exclusion
restriction.

To implement our test, we rely on a statistic very close to that of %
\citet{VBDN2013}. This statistic has the advantage of requiring an estimate
of the conditional quantile function only under the null hypothesis, i.e.
where the conditional quantile is a function of $x_{i}$ only. To estimate
the conditional quantile function(s) at some point $x_{i}=x$, we use an $r$%
-th order local polynomial estimator, which we denote by $\widehat{q}_{\tau
}(x)$, while its corresponding probability limit is denoted by $q_{\tau
}^{\dag }(x)$, which are formally defined in the Appendix Equations (\ref%
{LPQ}) and (\ref{LPQLimit}). Moreover, define {\normalsize $\widehat{u}%
_{\tau }(x_{i})\equiv y_{i}-\widehat{q}_{\tau }(x_{i})$, $u_{\tau
}(x_{i})\equiv y_{i}-q_{\tau }^{\dag }(x_{i})$, and let }$\underline{x}=(%
\underline{x}^{1},...,\underline{x}^{d_{x}}),$ and $\overline{x}=\left( 
\overline{x}^{1},...,\overline{x}^{d_{x}}\right) $, $\underline{x},\overline{%
x}\in \mathcal{X}$, where $d_{x}$ denotes the dimension of $x_{i}$. Finally,
for notational simplicity, we assume that $\mathcal{P}$ is a connected
interval with boundary points $(\underline{p},\overline{p})$.\footnote{%
The extension to a disconnected interval is immediate at the cost of more
complex notation.} The test statistic is given by: 
\begin{equation*}
Z_{1,n}^{q}=\sup_{\tau \in \mathcal{T},(\underline{x},\overline{x})\in 
\mathcal{X},(\underline{p},\overline{p})\in \mathcal{P}}|Z_{1,n}^{q}\left(
\tau ,\underline{x},\overline{x},\underline{p},\overline{p}\right) |,
\end{equation*}%
where 
\begin{equation*}
Z_{1,n}^{q}\left( \tau ,\underline{x},\overline{x},\underline{p},\overline{p}%
\right) =\frac{1}{\sqrt{n}}\sum_{i=1}^{n}s_{i}(1\{\widehat{u}_{\tau
}(x_{i})\leq 0\}-\tau )\Pi _{j=1}^{d_{x}}1\{\underline{x}_{j}<x_{j,i}<%
\overline{x}_{j}\}1\{\underline{p}<\widehat{p}_{i}<\overline{p}\}.
\end{equation*}%
The statistc $Z_{1,n}\left( \tau ,\underline{x},\overline{x},\underline{p},%
\overline{p}\right) $ differs from \citet{VBDN2013} in two aspects. First and most importantly, our test statistic is
constructed taking the supremum also w.r.t. $\tau $ (over $\mathcal{T}$). We
therefore test for selection across
all quantile ranks in a compact set $\mathcal{T}$ simultaneously, while the test of \citet{VBDN2013} would have only allowed for a pointwise search. Second, the omitted regressor $p_{i}$ is not observable and thus replaced by a
nonparametric estimator, $\widehat{p}_{i}$.  Heuristically, the uniformity of our test is achieved via the use of
a local polynomial quantile estimator for which \citet{GS2012} established a
Bahadur representation uniform over compact sets $\mathcal{X}$ \textit{and} $%
\mathcal{T}$. As for the nonparametric estimator of the propensity score, under regularity and bandwidth
conditions outlined below, we show  that the estimation error arising
from $\widehat{p}_{i}$ is asymptotically negligible. This is a well known
result for estimates affecting the statistic only through a weight function %
\citep[cf.][]{EJCL2014}. 

 Finally, note that one could construct an alternative
statistic in which the quantile estimator is replaced by a conditional CDF
estimator, say: 
\begin{equation*}
\frac{1}{\sqrt{n}}\sum_{i=1}^{n}s_{i}(1\{y_{i}-y\leq 0\}-\widehat{F}%
_{y|x,s_{i=1}}(y|x_{i}))\Pi _{j=1}^{d_{x}}1\{\underline{x}_{j}<x_{j,i}<%
\overline{x}_{j}\}1\{\underline{p}<\widehat{p}_{i}<\overline{p}\}.
\end{equation*}%
with $\widehat{F}_{y|x,s_{i=1}}(y|x_{i})$ denoting a nonparametric estimator of the CDF $F%
_{y|x,s_{i=1}}(y|x_{i})$. However, as discussed in the Section \ref{Set-Up}, such a CDF based formulation
will not allow to test the null of no selection for specific subsets of the
(conditional) distribution.

In the sequel, we make the following assumptions:\medskip

\noindent \textbf{A.1}\label{A1} $(y_{i},x_{i}^{\prime },z_{i}^{c\prime
},z_{i}^{d\prime },s_{i})\subset R_{y}\times R_{x}\times R_{z}^{c}\times
R_{z}^{d}\times \{0,1\}$ are identically and independently distributed. Let $%
\mathcal{X}\equiv \mathcal{X}_{1}\times \ldots \times \mathcal{X}_{d_{x}}$
denote a compact subset of the interior of $R_{x}$. $z_{i}$ contains at
least one variable which is not contained in $x_{i}$ and which is not $x_i$%
-measurable. The variables $x_{i}$ and $z_{i}^{c}$ (conditional on all
values of $z_{i}^{d}$) have probability density functions with respect to
Lebesgue measure which are strictly positive and continuously differentiable
(with bounded derivatives) over the interior of their respective support.
Also, assume that the joint density function of $y_{i}$, $x_{i}$ and $p_{i}$
is uniformly bounded everywhere, and that $\Pr (s_{i}=1|x,p)=\Pr
(s_{i}=1|p)>0$ for all $x\in\mathcal{X}$ and $p\in\mathcal{P}$.\medskip

\noindent \textbf{A.2}\label{A2} The distribution function $%
F_{y|x,s=1}(\cdot |\cdot ,\cdot )$ of $y_{i}$ given $x_{i}$ and selection $%
s_{i}=1$ has a continuous probability density function $f_{y|x,s=1}(y|x,s=1)$
w.r.t. Lebesgue measure which is strictly positive and bounded for all $y\in
R_{y}$, $x\in \mathcal{X}$. The partial derivative(s) $\nabla
_{x}F_{y|x,s=1}(y|x,s=1) $ are continuous on $R_{y}\times \mathcal{X}$.
Moreover, there exists a positive constant $C_{1}$ such that: 
\begin{equation*}
|f_{y|x,s=1}(y|x,s=1)-f_{y|x,s=1}(y^{\prime }|x^{\prime },s=1)|\leq
C_{1}\Vert (y,x)-(y^{\prime },x^{\prime })\Vert
\end{equation*}%
for all $(y,x),(y^{\prime },x^{\prime })\in R_{y}\times \mathcal{X}$. Also
assume that $q_{\tau }(x)$ is $r+1-$th times continuously differentiable on $%
\mathcal{X}$ for all $\tau \in \mathcal{T}$ with $r>\frac{1}{2}d_{x}$.
\medskip

\noindent \textbf{A.3}\label{A3} There exists an estimator $\widehat{p}%
(z_{i}^{c},z_{i}^{d})$ such that for any value in the support of $%
z_{i}^{d}=z^{d}$, it holds that $\sup_{z^{c}\in\mathcal{Z}}|\widehat{p}%
(z^{c},z^{d})-p(z^{c},z^{d})|=o_{p}(n^{-\frac{1}{4}})$ with $\mathcal{Z}$ a
compact subset of $R_{z}^{c}$, and that: 
\begin{equation*}
\Pr\left(\exists i: z_{i}^{c}\in R_{z}^{c}\setminus \mathcal{Z},
p(z_{i}^{c},z_{i}^{d})\in\mathcal{P}\right)=o(n^{-\frac{1}{4}}).
\end{equation*}%
\medskip

\noindent \textbf{A.4}\label{A4} For some positive constant $C_{2}$, it
holds that: 
\begin{equation*}
|F_{p|x,u_{\tau },s=1}(p|x,0,s=1)-F_{p|x,u_{\tau },s=1}(p^{\prime
}|x^{\prime },0,s=1)|\leq C_{2}\Vert (p,x)-(p^{\prime },x^{\prime })\Vert
\end{equation*}%
for all $\tau \in \mathcal{T}$, $(p,p^{\prime })\in \mathcal{P}$, and $%
(x,x^{\prime })\in \mathcal{X}$, where $F_{p|x,u_{\tau },s=1}(p|x,0,s=1)$
denotes the conditional distribution function of $p_{i}$ given $x_{i}=x$, $%
u_{\tau }(x)=0$, and $s_{i}=1$. \medskip

\noindent \textbf{A.5}\label{A5} The non-negative kernel function $K(\cdot )$
is a bounded, continuously differentiable function with uniformly bounded
derivative and compact support on $[-1,1]$. It satisfies $\int K(v)dv=1$ as
well as $\int vK(v)dv=0$.\medskip

\noindent Assumption \textbf{A.1} imposes the existence of at least one,
continuous or discrete, element excluded from $x_{i}$. It guarantees the
existence of selected observations for all values in $\mathcal{X}$ and $%
\mathcal{P}$. Assumptions \textbf{A.2} and \textbf{A.4} on the other hand
are rather standard smoothness assumptions, while \textbf{A.3} is a
high-level condition, which ensures that $p(z_{i}^{c},z_{i}^{d})$ can be
estimated at a specific rate uniformly over $\mathcal{Z}$ so that estimation
error in $\widehat{p}(z_{i}^{c},z_{i}^{d})$ is asymptotically negligible, while the second part of \textbf{A.3} ensures that values of $z_{i}^{c}$ outside that set are asymptotically negligible for the estimation of the propensity score. In
fact, in the case where $\widehat{p}(z_{i}^{c},z_{i}^{d})$ is a local
constant kernel estimator, the use of a second order kernel imposes
restrictions on the dimensionality of the number of continuous regressors, $%
d_{z}$, namely $d_{z}<4$.\footnote{%
Let $h_{z}=(h_{zc},h_{zd})$ and $d_{z}=(d_{zc},d_{zd})$. If we estimate $%
\widehat{p}_{i}$ using a local constant estimator and set $h_{zc}=O(n^{-%
\frac{1}{4+d_{zc}}})$ and $h_{zd}=O(n^{-\frac{2}{4+d_{zc}}}),$ then for $%
d_{zc}<4$ Assumption \textbf{A.3} is satisfied. This is because for $d_{zc}<4
$, the bias of the continuous component is of order $n^{-\frac{2}{4+d_{zc}}%
}=o\left( n^{-1/4}\right) $, and the standard deviation for the continuous
component is of order $(\sqrt{nh_{d_{zc}}})^{-1}=o\left( n^{-1/4}\right) $.
As for the discrete component, the bias is of order $n^{-\frac{1}{4+d_{zc}}%
}=o\left( n^{-1/4}\right) $ and it does not contribute to the variance, see
e.g. Theorem 2.1 in \citet{LR2008}.} Note also that a sufficient condition
for the second part of \textbf{A.3} is the existence of sufficient moments.
Letting `$\Rightarrow $' denote weak convergence, we establish the
asymptotic behavior of $Z_{1,n}^{q}$.

In the sequel, let $h_{x}$ be the bandwidth used in the estimation of the
conditional quantile.

\noindent \textbf{Theorem 1: } Let Assumptions \textbf{A.1}-\textbf{A.5} and 
\textbf{A.Q} hold. Moreover, if as $n\rightarrow \infty ,$ $%
(nh_{x}^{2d_{x}})/\log n\rightarrow \infty $, $nh_{x}^{2r}\rightarrow 0,$
then

\noindent (i) under $H_{0,q}^{(1)}$,%
\begin{equation*}
Z_{1,n}^{q}\Rightarrow Z_{1}^{q},
\end{equation*}
where $Z_{1}^{q}$ is the supremum of the absolute value of a zero mean Gaussian process whose
covariance kernel is defined in the proof of Theorem 1.

\noindent (ii) under $H^{(1)}_{A,q},$ there exists $\varepsilon >0,$ such
that%
\begin{equation*}
\lim_{n\rightarrow \infty }\Pr \left( Z_{1,n}^{q}>\varepsilon \right) =1.
\end{equation*}
\medskip

\noindent The results of Theorem 1 rely on an appropriate choice of $h_{x}$.
As common in the nonparametric testing literature, our rate conditions
require undersmoothing, and thus cross-validation is not directly applicable
in our setting.\footnote{%
In fact, the order of the bandwidth selected by cross-validation is too
large for $nh_{x}^{2r}\rightarrow 0$.} However, to still pick $h_{x}$ in a
data-driven manner ensuring minimal bias at the same time, one possibility
to select $h_{x}$ in practice could be to choose $h_{x}$ on the basis of
cross-validation for a local polynomial estimator of order smaller than the
one assumed for the test. For instance, if the assumed polynomial order for
the test was $r=3$ as an example, $h_{x}$ could be chosen by
cross-validation for a local linear estimator, i.e. $h_{x}=O\left( n^{-\frac{%
1}{4+d_{x}}}\right) $. This in turn implies that $nh_{x}^{2r}\rightarrow 0$
as well as $nh_{x}^{2d_{x}}/\log (n)\rightarrow \infty $ whenever $d_{x}<4$.
In the Monte Carlo simulations of Section \ref{Monte Carlo Simulation}, we
demonstrate that this procedure in fact seems to perform well for the chosen
designs.

Finally, note that in the proof of Theorem 1 we show that, under $%
H_{0,q}^{(1)}$, $Z_{1,n}^{q}\left( \tau ,\underline{x},\overline{x},%
\underline{p},\overline{p}\right) $ has the following asymptotic
representation: 
\begin{eqnarray}
&&\frac{1}{\sqrt{n}}\sum_{i=1}^{n}s_{i}(1\{y_{i}\leq q_{\tau }(x_{i})\}-\tau
)\Pi _{j=1}^{d_{x}}1\{\underline{x}_{j}<x_{j,i}<\overline{x}_{j}\}\left(
(1\{p_{i}<\overline{p}\}-1\{p_{i}<\underline{p}\})\right. \notag\\
&&\left. -(F_{p|x,u_{\tau },s=1}(\overline{p}|x_{i},0,s_{i}=1)-F_{p|x,u_{%
\tau },s=1}(\underline{p}|x_{i},0,s_{i}=1))\right) ,\label{Arep}
\end{eqnarray}%
which holds uniformly over $\mathcal{T}$, $\mathcal{X}$ and $\mathcal{P}$.
The term involving the difference of conditional distribution functions $%
(F_{p|x,u_{\tau },s=1}(\overline{p}|x_{i},0,s_{i}=1)-F_{p|x,u_{\tau },s=1}(%
\underline{p}|x_{i},0,s_{i}=1))$ stems from the contribution of the quantile
estimation error to the asymptotic representation. It therefore becomes
evident that estimation error from the estimated propensity score, $\widehat{%
p}_{i}$, on the other hand does not play a role in this representation, a
finding that was also corroborated by \citet{EJCL2014}.

Since the limiting distribution $Z_{1}^{q}$ depends on features of the data
generating process, we derive a bootstrap approximation for it. In
particular, we follow \citet{HZ2003}, and use the bootstrap statistic: 
\begin{eqnarray}
&&Z_{1,n}^{\ast q}\left( \tau ,\underline{x},\overline{x},\underline{p},%
\overline{p}\right)  \notag \\
&=&\frac{1}{\sqrt{n}}\sum_{i=1}^{n}s_{i}(B_{i,\tau }-\tau )\Pi
_{j=1}^{d_{x}}1\{\underline{x}_{j}<x_{j,i}<\overline{x}_{j}\}\left( \left(
1\{\widehat{p}_{i}<\overline{p}\}-1\{\widehat{p}_{i}<\underline{p}\}\right)
\right.  \label{Z*q} \\
&&\left. -\left( \widehat{F}_{p|x,u_{\tau },s=1}\left( \overline{p}%
|x_{i,}0,s_{i}=1\right) -\widehat{F}_{p|x,u_{\tau },s=1}\left( \underline{p}%
|x_{i,}0,s_{i}=1\right) \right) \right) ,  \notag
\end{eqnarray}%
where $B_{i,\tau }=1\left\{ U_{i}\leq \tau \right\} $ with $U_{i}\overset{%
i.i.d.}{\sim }U(0,1)$, and independent of the sample, and $\widehat{F}%
_{p|x,u_{\tau },s=1}\left( p|x_{i,}0,s_{i}=1\right) $ denotes a
nonparamemtric kernel estimator with corresponding bandwidth sequence $h_{F}$
satisfaying $h_{F}\rightarrow 0$ as $n\rightarrow \infty $ (see Equation (%
\ref{EQCDF}) in the Appendix for a formal definition). The bootstrap test
statistic is then given by: 
\begin{equation*}
Z_{1,n}^{\ast q}=\sup_{\tau \in \mathcal{T},(\underline{x},\overline{x})\in 
\mathcal{X},\underline{p},\overline{p}\in \mathcal{P}}|Z_{1,n}^{\ast
q}\left( \tau ,\underline{x},\overline{x},\underline{p},\overline{p}\right)
|.
\end{equation*}%
Let $c_{(1-\alpha ),n,R}^{\ast (1)}$ be the $(1-\alpha )$ percentile of the
empirical distribution of $Z_{1,n}^{\ast q,1},...,Z_{1,n}^{\ast q,R},$where $%
R$ is the number of bootstrap replications. The following Theorem
establishes the first order validity of inference based on the bootstrap
critical values, $c_{(1-\alpha ),n,R}^{\ast }$.\medskip

\noindent \textbf{Theorem 1$^{\ast }$: }Let Assumption \textbf{A.1}-\textbf{%
A.5} and \textbf{A.Q} hold. If as $n\rightarrow \infty$, {\normalsize $%
(nh_{x}^{2d_{x}})/\log n\rightarrow \infty $, $nh_{x}^{2r}\rightarrow 0,$ }$%
h_{F}\rightarrow 0,$ $nh_{F}^{d_{x}+1}\rightarrow \infty ,$ {\normalsize and 
}$R\rightarrow \infty ,$ then

\noindent (i) under $H_{0,q}^{(1)}$%
\begin{equation*}
\lim_{n,R\rightarrow \infty }\Pr \left( Z_{1,n}^{q}\geq c_{(1-\alpha
),n,R}^{\ast (1)}\right) =\alpha
\end{equation*}%
(ii) under $H_{A,q}^{(1)}$%
\begin{equation*}
\lim_{n,R\rightarrow \infty }\Pr \left( Z_{1,n}^{q}\geq c_{(1-\alpha
),n,R}^{\ast (1)}\right) =1.
\end{equation*}
\medskip

\section{Second Test}

\label{Second Test}

Under (\ref{EQ1a}) and the assumptions outlined in the previous section,
failure to reject $H_{0,q}^{(1)}$ rules out sample selection asymptotically,
with probability approaching one. By contrast, rejection in this first test
could in principle occur either due to genuine sample selection or due to an
omitted variable in the outcome equation, which happens to be correlated
with the propensity score. This is so since the omitted predictor test, as
any omnibus test, does not possess directed power against specific
alternatives. Since this distinction is crucial for the estimation of
nonparametric conditional quantile functions as outlined in the
Introduction, we design a second test which has directed power against
detecting misspecification. To this end, let $\widetilde{q}_{\tau
}(x_{i},\pi _{i})$ and $q_{\tau }(x_{i})$ denote probability limits of two
local polynomial quantile estimators for the selected subsample of $y_{i}$
on $x_{i}$ and $\pi _{i}$ as well as on $x_{i}$ only, respectively. We say
that $\pi _{i}$ is a relevant predictor if for some $\tau\in\mathcal{T}$ and
some value $\pi$ in the support of $\pi_{i}$, $\widetilde{q}_{\tau }(x,\pi
)\neq q_{\tau }(x)$ for at least all $x$ in a subset of $\mathcal{X}$ with
non-zero Lebesgue measure.

We want to disentangle selection from relevant omitted predictors correlated
with the propensity score, which may (or may not) be present simultaneously
with sample selection. In order to impose no-selection as maintained
hypothesis, we require that at least one value $z$ in the support of $z_{i}$
s.t. $p(z)=1$ exists, which is indeed one of the identification assumptions in \citet[][p.6]{AB2017}. Then, in the absence of relevant omitted predictors $%
\pi _{i}$ whenever $p\rightarrow 1$, the selection bias approaches zero, and 
$\lim_{p\rightarrow 1}\Pr \left( y_{i}\leq q_{\tau }(x_{i})|p_{i}=p\right)
=\tau $. By contrast, when $\pi _{i}$ is also a relevant omitted predictor,
correlated with $p_{i},$ when the latter is close to one, then $%
\lim_{p\rightarrow 1}\Pr \left( y_{i}\leq q_{\tau }(x_{i})|p_{i}=p\right)
\neq \tau $ with positive probability.

The requirement that $p(z)=1$ for at least one value $z$ is typically
labelled `identification at infinity' in the nonparametric identification
literature \citep[e.g.][]{C1986}. In fact, note that `identification at infinity' may
in principle hold when all elements of $x_{i}$ and $z_{i}$ are the same, see %
\citet{MR2008} for an application of `identification at infinity' to
quantile models without an exclusion restriction. However, as simulation evidence suggested a rather poor test performance in this case we continue to maintain the assumption of a discrete or continuous exclusion restriction.

Since on the event $p_{i}=1$, the individual is selected into the sample
with certainty, and so selection is not present, in a second step, we test
the null hypothesis that the propensity score is an omitted predictor when $%
p_{i}=1$.\footnote{%
Note that the test has power when the omitted predictor is relevant also
when $p$ is close to one.} That is, we test that: 
\begin{equation}
H_{0,q}^{(2)}:\Pr \left( y_{i} \leq q_{\tau }(x_{i})|p_{i}=1\right) =\tau \text{
for all }\tau \in \mathcal{T}  \label{H02q}
\end{equation}%
versus%
\begin{equation}
H_{A,q}^{(2)}:\Pr \left( y_{i} \leq q_{\tau }(x_{i})|p_{i}=1\right) \neq \tau 
\text{ for some }\tau \in \mathcal{T}.  \label{HA2q}
\end{equation}%
Thus, if selection is the sole cause for rejection of $H_{0,q}^{(1)}$, we do
not expect to reject $H_{0,q}^{(2)}$ (at least asymptotically). By contrast,
if we reject $H_{0,q}^{(2)}$, we take this as an indication that
misspecification was likely to be the or at least one driver of the
rejection at the first stage. Of course, as we discuss in the supplement in the context of the Decision Rule, we cannot rule out
selection if both misspecification and selection are present and lead to a
rejection simultaneously.

A common concern in the context of `identification at infinity' is so called
irregular identification \citep{KT2010}, where, although conditional
quantiles are point identified, they cannot be estimated at a regular
convergence rate as the marginal density of $p_{i}$ may not be bounded away
from zero at the evaluation point $p(z)=1$. That is, heuristically, even if
`identification at infinity' holds, and for some value $z\in R_{z}$, $p(z)$
can reach one, it is still possible that observations in the neighborhood of
one are very sparse in practice (`thin density set'), and so convergence
occurs at an irregular rate \citep{KT2010}. To address this issue, we only
use observations from parts of the support where the density of $p_{i}$ is
bounded away from zero. Formally, this is implemented by introducing a
trimming sequence, converging to zero at a sufficiently slow rate so that
irregular identification is no longer a concern. Thus, let $\delta =1-H$
with $H\rightarrow 0$ and $H/h_{p}\rightarrow \infty $ as $n\rightarrow
\infty $, where $H$ governs the speed of the trimming sequence $\delta $,
while $h_{p}$ defines the window width around $\delta $. Then, reaclling the definition of $\widehat{u}_{\tau}(x_{i})=y_{i}-\widehat{q}_{\tau}(x_{i})$, the second
test is based on the statistic 
\begin{equation}
Z_{2,n,\delta }^{q}\left( \tau \right) =\frac{\sum_{i=1}^{n}s_{i}(1\{%
\widehat{u}_{\tau }(x_{i})\leq 0\}-\tau )K\left( \frac{\widehat{p}%
_{i}-\delta }{h_{p}}\right) }{\left( \int K^{2}(v)dv\sum_{i=1}^{n}s_{i}(1\{%
\widehat{u}_{\tau }(x_{i})\leq 0\}-\tau )^{2}K\left( \frac{\widehat{p}%
_{i}-\delta }{h_{p}}\right) \right) ^{1/2}}.  \label{Z2}
\end{equation}%
This statistic only uses observations with (estimated) propensity score $%
\widehat{p}_{i}\in (1-h_{p}-H,1+h_{p}-H)$, and thus overcomes the issue of
possible irregular identification as long as a sufficient number of
observations are assumed to exist in this set (see below). Note here that
the convergence speed of $H$ is inherently pegged to the tail behavior of
the density of $p_{i}$ in the neighborhood of $p=1$, which is of course
unknown in practice. That is, the thinner the density tail of $p_{i}$, the
slower $H$ has to go to zero. We discuss below this issue and a potential
data-driven way to select $H$ and $h_{p}$. Finally, as pointed out above, in
order for the test to possess directed power against misspecification we
require that relevant omitted predictor(s) $\pi _{i}$, if present, are correlated with
the event $\{p_{i}=1\}$, or, more specifically $\{p_{i}\in
(1-h_{p}-H,1+h_{p}-H)\}$.

In what follows, let $G_{u_{\tau}}(\tau ,1-H)\equiv \Pr (u_{\tau
}(x_{i})\leq 0|p_{i}=1-H)$, and note that under $H_{0,q}^{(2)}$, it holds
that $\lim_{H\rightarrow 0}G_{u_{\tau}}(\tau ,1-H)=\tau $ for every $\tau
\in \mathcal{T}$, and that $\lim_{H\rightarrow 0}\Pr (s_{i}=1|p_{i}=1-H)=1$.
We make the following additional assumptions: \medskip

{\normalsize \noindent \textbf{A.6}\label{A6}} There exists at least one $%
z\in R_{z}$ such that $p(z)=1$. Moreover, there exists a strictly positive,
continuous, and integrable function $g_{u_{\tau },p}(u_{\tau },1)$ and $%
g_{p}(1)$ such that for all $\tau \in \mathcal{T}$: 
\begin{equation*}
\sup_{\tau \in \mathcal{T}}\left( \left\vert \frac{f_{u_{\tau },p}\left(
u_{\tau },1-H\right) }{g_{u_{\tau },p}(u_{\tau },1)H^{\eta }}-1\right\vert
\right) \rightarrow 0\quad \text{and}\quad \left( \left\vert \frac{%
f_{p}\left( 1-H\right) }{g_{p}(1)H^{\eta }}-1\right\vert \right) \rightarrow
0
\end{equation*}%
as $n\rightarrow \infty $ for some $0\leq \eta <\overline{\eta }<1$, where $f_{u_{\tau},p}(\cdot,\cdot)$ and $f_{p}(\cdot)$ are the joint and marginal densities of $u_{\tau}$ and $p$, respectively.

{\normalsize \noindent \textbf{A.7}\label{A7} The distribution function $%
F_{u_{\tau}|p,s=1}(\cdot |\cdot ,\cdot )$ of $u_{\tau}(x_{i})$ given $p_{i}$%
, and selection $s_{i}=1$ has a continuous probability density function $%
f_{u_{\tau}|p,s=1}(u_{\tau}|p,s=1)$ w.r.t. Lebesgue measure for all $\tau\in%
\mathcal{T}$. The functions $f_{u_{\tau}|p,s=1}(u_{\tau}(x_{i})|p,s=1)$, $%
\Pr(s_{i}=1|p) $, and $f_{p}(p)$ are continuously differentiable w.r.t. to $p
$ for all $\tau\in\mathcal{T}$ with bounded partial derivatives. Moreover,
assume that these functions are left-continuous at $p=1$.\medskip }

\noindent \textbf{A.8}\label{A8} Assume that for all $\tau \in \mathcal{T}$,
there exist positive constants $C$ and $C^{\prime}$ such that: 
\begin{equation*}
\left\vert G_{u_{\tau}}(\tau ,1-H)-G_{u_{\tau}}(\tau ,1)\right\vert \leq C
H^{1-\eta }
\end{equation*}%
as well as 
\begin{equation*}
\left\vert \Pr (s_{i}=1|1-H)-1\right\vert \leq C^{\prime} H^{1-\eta }.
\end{equation*}%
\medskip

\noindent \textbf{A.9}\label{A9} Assume that the support of $x_{i}$, $R_{x}$, is equal to $\mathcal{X}$, and that:
\[
 \sup_{(y,x)\in R_{y}\times\mathcal{X}}\left\lvert \frac{\partial f_{y|x,s=1}(y|x,s=1)}{\partial y}\right\rvert<\infty\quad\text{and}\quad  \sup_{(y,x)\in R_{y}\times\mathcal{X}}\left\lvert \frac{\partial f_{y|x,s=1}(y|x,s=1)}{\partial x}\right\rvert<\infty.
\]
\medskip

\noindent Assumption \textbf{A.6} requires identification at infinity, for
at least one value $z$ of $z_{i}$, and this can be achieved provided some
common covariate and/or the  exclusion restriction are continuous.\footnote{Note that assumption \textbf{A.6} could be relaxed to unbounded supports at the cost of more complex notation and proofs.}
Importantly, our set-up \textit{does deal} with the pratically relevant case
of irregular identification, in the sense that $f_{p}(1)$ may not
necessarily be bounded away from zero at $p=1$. More precisely, when $\eta =0
$, $\lim_{H\rightarrow 0}f_{p}(1-H)$ is bounded away from zero, while $\eta
>0$ corresponds to the case of irregular support (with a larger value of $%
\eta $ representing thinner tails). That is, if $\eta >0$, we allow for a
thin set of observations with a propensity score close to one. Similarly,
when $\eta =0$, the first part \textbf{A.8} becomes a standard Lipschitz
condition, while as $\eta $ gets closer to one and the tails of the
densities in \textbf{A.6} become thinner, we allow $G_{u_{\tau }}(\tau ,1-H)$
and $\Pr (s_{i}=1|1-H)$ to approach $G_{u_{\tau }}(\tau ,1)$ and 1,
respectively, at a slower rate.\footnote{%
Note that under $H_{0,q}^{(2)}$, we have that $G_{u_{\tau }}(\tau ,1)=\tau $
almost surely.} Finally, Assumption \textbf{A.9} is a technical condition that ensures that the uniform local Bahadur representation continues to hold also at the boundary of the support of $x_{i}$. More specifically, together with \textbf{A.1}, \textbf{A.2}, and \textbf{A.5}, it allows the application of Proposition 4.4 in \citet{FG2016} and avoids the use of trimming, which would in turn require further steps in the proofs and more complex assumptions.

The rate of convergence of the numerator in (\ref{Z2}) depends on $H^{\eta }$%
, the tail behavior of the density $f_{p}(p)$ around $p=1$, which is of
course unknown in practice. In fact, we are generally ignorant about the
rate of convergence given by $\sqrt{nh_{p}H^{\eta }}$, which may in
principle be as fast as $\sqrt{nh_{p}}$. To address this problem, we use a
studentized statistic, which allows the convergence rate to vary depending
on the sparsity of observations around $p$ close to one. That is, as we
cannot infer the appropriate scaling factor, it is crucial that, regardless
the degree of thinness of the set of observations with propensity score
close to one, $Z_{2,n,\delta}\left( \tau  \right) $ and $\sqrt{\widehat{var}%
(Z_{2,n,\delta }\left( \tau \right) )}$ diverge at the same rate, so that
the ratio still converges in distribution.

In the sequel, we study the asymptotic behavior of $\frac{Z_{2,n,\delta
}^{q}\left( \tau \right) }{\sqrt{\widehat{var}(Z_{2,n,\delta }^{q}\left(
\tau \right) )}}$ as an empirical process over $\tau \in \mathcal{T}$.
\medskip 

\noindent \textbf{Theorem 2: } Let Assumption \textbf{A.1}, \textbf{A.3}, 
\textbf{A.5}, \textbf{A.6}, \textbf{A.7}, \textbf{A.8}, \textbf{A.9}, and \textbf{A.Q}
hold. If as $n\rightarrow \infty ,$ $(nh_{x}^{2d_{x}})/\log n\rightarrow
\infty $, $nh_{x}^{2r} \rightarrow 0,$ $H\rightarrow 0,$ $H/h_{p}\rightarrow
\infty $, $nh_{p}H^{2-\eta }\rightarrow 0$, and $nh_{p}H^{\eta }\rightarrow
\infty ,$ then

\noindent (i) under $H_{0,q}^{(2)}$ , 
\begin{equation*}
\sup_{\tau \in \mathcal{T}}\left\vert \frac{Z_{2,n,\delta }^{q}\left( \tau
\right) }{\sqrt{\widehat{var}(Z_{2,n,\delta }^{q}\left( \tau \right) )}}%
\right\vert \Rightarrow Z_{2}^{q},
\end{equation*}%
where $Z_{2}^{q}$ is the supremum of the absolute value of a zero mean Gaussian process with
covariance kernel defined in the proof of Theorem 2.

\noindent (ii) under $H_{A,q}^{(2)}$ , there exists $\varepsilon >0,$ such
that%
\begin{equation*}
\lim_{n\rightarrow \infty }\Pr \left( \sup_{\tau \in \mathcal{T}}\left\vert 
\frac{Z_{2,n,\delta }^{q}\left( \tau \right) }{\sqrt{\widehat{var}%
(Z_{2,n,\delta }^{q}\left( \tau \right) )}}\right\vert >\varepsilon \right)
=1.
\end{equation*}%
\medskip 

\noindent Theorem 2 establishes the limiting distribution of the studentized
statistic. As the theoretical results crucially hinge on the tuning
parameters $H$ and $h_{p}$, whose rates depend in turn on the unknown $\eta $%
, a discussion of their choice in practice is warranted. In fact, a possible
data-driven choice of these parameters, without claiming optimality of a
specific kind, could be as follows: as shown in the supplementary material,
one may re-write $h_{p}$ and $H$, which is a function of $h_{p}$ itself, as
functions of $\eta $ only, i.e. $h_{p}\left( \eta \right) =Cn^{-\frac{1+\varepsilon}{%
1 +\varepsilon+ \eta }}\log (n)$ and $H(\eta
)=h_{p}(\eta )^{1/(1+\varepsilon )}$ for some arbitrary $\varepsilon >0$ and $%
\eta <\overline{\eta }$, and some scaling constant $C$. Here, $\overline{%
\eta }$ represents the threshold value with the slowest possible convergence
rate still satisfying the rate conditions of Theorem 2. Thus, in order to
`choose' the smallest possible $\eta $ in practice, which in turn
corresponds to the fastest possible convergence rate, one could for instance
plot $\frac{1}{\left( nh_{p}\left( \eta \right) \right) ^{1-\eta }}%
\sum_{i=1}^{n}K\left( \frac{\widehat{p}_{i}-(1-H(\eta ))}{h_{p}\left( \eta
\right) }\right) $ for a given $\varepsilon >0$ (e.g., $\varepsilon =.1$) on
a grid of different $\eta $ values with $\eta \in \lbrack 0,1)$, and select $%
\hat{\eta}$ as the smallest value for which the estimated density is bounded
away from zero, e.g. above a minimum threshold value such as $0.1$. In
fact, with this procedure, if the set of $\widehat{p}_{i}$ close to $1$ is
not `thin', we would expect to select $\hat{\eta}=0$ in large enough samples. We investigate this procedure further in Section \ref{Monte Carlo Simulation}.

As outlined in the proof of Theorem 2, quantile estimation error vanishes.
This is because under appropriate rate conditions it approaches zero at a
rate which is faster than the convergence rate of the statistic$.$ Hence,
when constructing the wild bootstrap statistic we do not have to `subtract'
an estimator of the conditional distribution of $p_{i}.$ On the other hand,
as the rate of convergence depends on the `degree' of irregular
identification at $p$ close to 1, we also need an appropriately studentized
bootstrap statistic, i.e.%
\begin{equation}
Z_{2,n}^{\ast q}=\sup_{\tau \in \mathcal{T}}\left\vert \frac{Z_{2,n,\delta
}^{\ast q}\left( \tau \right) }{\sqrt{\widehat{var^{\ast }}%
(Z_{2,n}^{q}\left( \tau ,\delta \right) )}}\right\vert ,  \label{boot2}
\end{equation}%
where%
\begin{equation*}
Z_{2,n,\delta }^{\ast q}\left( \tau \right) =\sum_{i=1}^{n}s_{i}(B_{i,\tau
}-\tau )K\left( \frac{\widehat{p}_{i}-\delta }{h_{p}}\right) 
\end{equation*}%
with $B_{i,\tau }=1\left\{ U_{i}\leq \tau \right\} $ with $U_{i}\overset{%
i.i.d.}{\sim }U(0,1)$ and independent of the sample, and 
\begin{eqnarray}
&&\widehat{var}^{\ast }(Z_{2,n,\delta }^{q}\left( \tau \right) )  \notag \\
&=&\left( \frac{1}{n}\sum_{i=1}^{n}(B_{i,\tau }-\tau )^{2}\right) \left(
\int K^{2}(v)dv\right) \sum_{i=1}^{n}s_{i}K\left( \frac{\widehat{p}%
_{i}-\delta }{h_{p}}\right) .  \label{var*}
\end{eqnarray}%
By noting that $\frac{1}{n}\sum_{i=1}^{n}(B_{i,\tau }-\tau )^{2}=\tau
(1-\tau )+o_{p}^{\ast }(1),$ given (\ref{var*}), we see that whenever
`identification at infinity' holds and the number of observations with
propensity score in the interval $(1-h_{p}-H,1+h_{p}-H)$ grows at rate $%
nh_{p},$ then both numerator and denominator in (\ref{boot2}) are bounded in
probability, otherwise they diverge at the same rate.

Let $c_{(1-\gamma ),n,R}^{\ast (2)}$ be the $(1-\gamma )$ percentile of the
empirical distribution of $Z_{2,n}^{\ast q,1},...,Z_{2,n}^{\ast q,R},$where $%
R$ is the number of bootstrap replications. The following Theorem
establishes the first order validity of inference based on the bootstrap
critical values, $c_{(1-\gamma ),n,R}^{\ast (2)}$.\medskip

\noindent \textbf{Theorem 2$^{\ast }$: } Let Assumption \textbf{A.1}, 
\textbf{A.3}, \textbf{A.5}, \textbf{A.6}, \textbf{A.7}, \textbf{A.8}, \textbf{A.9}, and 
\textbf{A.Q} hold. If as $n\rightarrow \infty ,$ $(nh_{1}^{2d_{x}})/\log
n\rightarrow \infty $, $nh_{x}^{2r}\rightarrow 0,$ $H\rightarrow 0,$ $%
H/h_{p}\rightarrow \infty $, $nh_{p}H^{2-\eta }\rightarrow 0$, $%
nh_{p}H^{\eta }\rightarrow \infty$, and $R\rightarrow \infty$, then

\noindent (i) under $H_{0,q}^{(2)}$%
\begin{equation*}
\lim_{n,R\rightarrow \infty }\Pr \left( Z_{2,n}^{q}\geq c_{(1-\gamma
),n,R}^{\ast (2)}\right) =\gamma
\end{equation*}%
(ii) under $H_{A,q}^{(2)}$%
\begin{equation*}
\lim_{n,R\rightarrow \infty }\Pr \left( Z_{2,n}^{q}\geq c_{(1-\gamma
),n,R}^{\ast (2)}\right) =1.
\end{equation*}

\noindent Theorem 2* establishes the first order validity of inference based
on wild bootstrap critical values. Under $H_{0,q}^{(2)}$, the studentized
statistic and its bootstrap counterpart have the same limiting distribution.
Under, $H_{A,q}^{(2)},$ the statistic diverges, as the numerator is of
larger probability order than the denominator, while the bootstrap statistic
remains bounded in probability.

As a final remark, note that when the ultimate goal is to estimate the conditional quantile function(s) nonparametrically, both tests may be used in a testing procedure (see the supplement for a formal outline). That is, if we fail to reject the null hypothesis of the first test, one may interpret this as evidence against sample selection and decide to rely on
nonparametric estimators of the conditional quantiles using all selected
individuals in the data. On the other hand, if we reject the first test, but
fail to reject the second one, estimation of the conditional
quantile function(s) may still be carried out using only individuals with propensity score close to
one, e.g. as in (\ref{LPQP1}) in the Appendix. By contrast, if the null
hypotheses of both tests are rejected, there is evidence for relevant
omitted predictor(s) (and possibly endogenous selection), and neither the
estimator using all selected individuals nor the one using only those with
propensity score close to one will deliver estimates consistent for the
conditional quantile function(s) of interest. Of course, as pointed out before,
the distinction between sample selection and relevant omitted predictors is only possible when the
omitted predictors are correlated with $p_{i}$ when $p_{i}$ takes on values close to
one. However, even when omitted predictors are present, but uncorrelated with the event $p_{i}$ close to one, one may make the `right' decision deciding in favor of selection, and by estimating
the quantiles using only observations with propensity score close to one.
This is so as for observations with propensity score close to one, omitted predictor bias
is not present. 

\section{Monte Carlo Simulation}

\label{Monte Carlo Simulation}

In this section we examine the finite sample properties of our tests via a
Monte Carlo study. Results for the conditional mean can be found in the
supplement. The outcome equation of our simulation design (given selection $%
s_{i}=1$) is given by: 
\begin{equation*}
y_{i}=\left(x_{i}-0.5\right)^3 + \left(x_{i}-0.5\right)^2 +
\left(x_{i}-0.5\right) +\gamma_{1}\widetilde{z}_{i}+ 0.5\varepsilon_{i},
\end{equation*}
where $x_{i}\sim U(0,1)$, the distribution of $\widetilde{z}_{i}$ varies according to
the design (see below), and the marginal distribution of $\varepsilon_{i}$
is standard normal. In the above equation, the parameter $\gamma_{1}$ determines the level of misspecification. Thus, when $\gamma_{1}$ is non-zero, $\widetilde{z}_{i}$ becomes an omitted relevant predictor as outlined in the previous section. On the other hand, when $\gamma_{1}$ is zero, the conditional
quantile function $q_{\tau}(x_{i})$ is given by the third order polynomial
function: 
\begin{equation}  \label{PolyFun}
q_{\tau}(x_{i})=\left(x_{i}-0.5\right)^3 + \left(x_{i}-0.5\right)^2 +
\left(x_{i}-0.5\right) + 0.5 q_{\tau}(\varepsilon_{i}),
\end{equation}
where $q_{\tau}(\varepsilon)$ denotes
the unconditional $\tau$ quantile of $\varepsilon_{i}$. Selection enters
into this set-up via: 
\begin{equation*}
s_{i}=1\{0.75(x_{i}-0.5)+0.75 \widetilde{z}_{i}>\sigma v_{i}\},
\end{equation*}
with: 
\begin{equation*}
\left(%
\begin{array}{c}
\varepsilon_{i} \\ 
v_{i}%
\end{array}%
\right) \sim N\left(\left(%
\begin{array}{c}
0 \\ 
0%
\end{array}%
\right),\left(%
\begin{array}{cc}
1 & \rho \\ 
\rho & 1%
\end{array}%
\right) \right).
\end{equation*}
Thus, $\rho$ controls the degree of selection and we consider three
scenarios, namely the case of `no selection' ($\rho=0$), `moderate
selection' ($\rho=0.25$), and `strong selection' ($\rho=0.5$), while we set
the scaling factor $\sigma$ of $v_{i}$ to one and $\gamma_{1}=0$ for most cases of the first test. The instrument $\widetilde{z}_{i}$ is simulated according to one of
the following four designs: 
\begin{eqnarray*}
\mathbf{\text{(i)}}: && \widetilde{z}_{i}\sim N(0,1), \\
\mathbf{\text{(ii)}}: && \widetilde{z}_{i}\sim Binom(0.5)-0.5, \\
\mathbf{\text{(iii)}}: &&\widetilde{z}_{i}\sim 1.5-Poisson(1.5), \\
\mathbf{\text{(iv)}}: && \widetilde{z}_{i}\sim Discrete Unif.(0,7).
\end{eqnarray*}
Design (i) is the benchmark scenario and will illustrate the performance of
the first test when the instrument exhibits continuous variation as for
instance in the empirical illustration of Section \ref{Empirical
Illustration}. Design (ii) on the other hand represents the opposite (extreme)
case where the excluded instrumental variable $\widetilde{z}_{i}$ only takes on two
values. Designs (iii) and (iv) are intermediate scenarios where $\widetilde{z}_{i}$ follows
a (discrete) Poisson distribution with an average of around 7 support
points, and a discrete uniform distribution with 7 support points (and
equally distributed point mass).\footnote{%
The relationship between the `signal' variance of $0.75(x_{i}-0.5)+0.75
\widetilde{z}_{i} $ and the `noise' variance of $v_{i}$ therefore ranges from around 0.2
to 0.6 across cases.} Importantly, the last two distributions are meant to
be stylized examples of set-ups with discrete instruments such as `number of
kids' or `month-observed-in-the-sample'.

We consider three quantiles $\mathcal{T}=\{0.3,0.5,07\}$ and two sample
sizes $n=\{1000 , 2000\}$, which, given a selection probability of
approximately $0.5$, imply an effective sample size for the outcome equation
of around 500 to 1000 observations, respectively. Throughout, we estimate $%
q_{\tau}(x_{i})$ using the selected sample and a third order local
polynomial estimator in line with (\ref{PolyFun}) and the conditions of
Theorem 1.\footnote{%
In order to restrict ourselves to a compact subset $\mathcal{X}$, we trim
the outer 2.5\% observations of the selected sample.} In total, we consider
six different cases: since our theoretical results suggest that estimation
error from the propensity score does not feature in the asymptotic
representation of (\ref{Arep}) under $H_{0,q}^{(1)}$, in the first four
cases, Cases I-IV, we consider designs (i) to (iv) using the oracle propensity score. For
these cases, we choose the bandwidths $h_{x}$ and $h_{F}$ ad-hoc in line
with Theorem 1 and 1* as $h_{x}=c\cdot \text{sd}(x_{i})\cdot n^{-\frac{1}{3}}$, $%
c=\{3.5,4,4.5\}$, and $h_{F}=2.2\cdot \text{sd}(x_{i},\widehat{u}%
_{\tau}(x_{i}))\cdot n^{-\frac{1}{6}}$, respectively, where $\text{sd}(\cdot)$ denotes the standard deviation. In Case V, we still
consider the oracle propensity score using design (i), but instead choose $%
h_{F}$ and $h_{x}$ via cross-validation ($\widehat{h}_{x}$). More
specifically, for $\widehat{h}_{x}$, we employ the method described after
Theorem 1 and pick the bandwidth running cross-validation for a lower local
polynomial order (i.e., $\widetilde{r}=1$) than the one assumed. In Case VI, we
replace the oracle propensity score from Case V by an estimate using a
standard local constant estimator with second order Epanechnikov kernel and
a cross-validated bandwidth as permitted by our theory when the number of continuous variables in the selection equation is less than four (cf. footnote 6).\footnote{%
To construct this estimator as well as the estimators for the conditional
mean in the supplement, we use routines from the \texttt{np package} of %
\citet{HR2008}. This package allows to construct the bandwidth according the cross-validation procedure outlined in Section 2 of \citet{LR2008}. For the local polynomial quantile estimator, we use a
routine from the \texttt{quantreg package} \citep{K2017}.} 

Finally, in
Case VII and VIII we examine the power of the first test under no selection,
but misspecification. More specifically, we set $\rho=0$ and vary $%
\gamma_{1} $, the misspecification parameter, for the design of Case I in
Case VII and of Case II in Case VIII with cross-validation, respectively.
Throughout, we use the `warp speed' procedure of \citet{GPW2013} with 999
Monte Carlo replications and fix the nominal level of the test to $%
\alpha_{1}=0.05$ and $\alpha_{2}=0.1$. The results of the first test can be
found in Table \ref{Table MC1} below.

Turning to the results, observe that under $H_{0,q}^{(1)}$ (i.e., $\rho=0$),
we have overall a good size control and rejection rates converge to the
nominal levels as the sample size increases. Interestingly, the performance
of test in Cases III and IV with a discrete multi-valued instrument is
slightly worse in terms of size than for the binary instrument of Case II.
Moreover, we can see that the Cases V and VI, where the bandwidths are
chosen in an automated manner, are in line with the rest, suggesting that
the data-driven procedures to pick the tuning parameters may be a good
alternative for the application of the test in practice. Turning to power ($%
\rho=0.25$ and $\rho=0.5$), note that power picks up rather quickly with the
sample size when sample selection is `moderate' ($\rho=0.25$) and is close
to one when $\rho=0.5$ throughout. As expected, we observe that power is
generally lower when discrete instruments are used, although the gap between
Case II (binary instrument) and Case I (continuous instrument) is never
above 10 percentage points when $n=2,000$. Finally, note that when we turn
to the misspecification in Cases VII and VIII, we observe that even at $%
\gamma_{1}=0.25$ the rejection rate of the first test is immediately very
high and very close to one suggesting that misspecification may be a concern
when potentially relevant predictors have been omitted.

For the second test, we consider the case of misspecification by
manipulating $\gamma_{1}$ while operating under the alternative of the first
test. More specifically, we set $\rho=0.25$ throughout and examine three
different values of $\gamma_{1}$ when $\widetilde{z}_{i}$ is normally distributed (CASE I%
$^{\ast}$-III$^{\ast}$) with $\widetilde{z}_{i}\sim N(0,0.5)$, and three alternative
cases with $\widetilde{z}_{i}\sim Binom(0.5)-0.5$ (Cases IV$^{\ast}$-VI$^{\ast}$),
respectively. For each distribution of $\widetilde{z}_{i}$, we start with the case under
the null of the second test ($\gamma_{1}=0$), and then consider cases under
the alternative with `moderate' misspecification ($\gamma_{1}=0.25$) and
`strong' misspecification ($\gamma_{1}=0.5$), respectively. In the case where $\widetilde{z}_{i}\sim N(0,0.5)$ we choose $h_{p}\in%
\{0.075,0.05,0.03,0.02\}$ and $\delta\in\{0.95,0.975,0.98\}$. On the other
hand, for the binomial variable, we set $h_{p}\in\{0.075,0.05\}$ and $%
\delta\in\{0.95,0.975\}$ reflecting the fact that the discrete nature of $%
\widetilde{z}_{i}$ in the Cases IV$^{\ast}$-VI$^{\ast}$ leads to less observations with
a propensity score value close to one.\footnote{%
We also set the scaling factor $\sigma$ to $0.5$ in this case.} Finally, to analyze the performance under the data-driven procedure outlined after Theorem 2, we added to each case the performance when $\widehat{h}_{x}$ and $\widehat{h}_{p}$ are chosen in a data-driven manner. More specifically, $\widehat{h}_{x}$ is chosen via cross-validation as outlined for the first test, while for $\widehat{h}_{p}$ we set $h_{p}(\eta)=\log(n)n^{-\frac{1+\varepsilon}{1+\varepsilon+\eta}}$ and $H(\eta)=h_{p}^{1/(1+\varepsilon)}$ for $\varepsilon=0.1$ and select the smallest possible $\eta$ from a grid $\{0.1,0.2,0.3,0.4,0.5,0.6,0.7,0.8,0.9\}$ such that $\frac{1}{\left( nh_{p}\left( \eta \right) \right) ^{1-\eta }}%
\sum_{i=1}^{n}K\left( \frac{p_{i}-(1-H(\eta ))}{h_{p}\left( \eta
\right) }\right) >0.1$. All results
can be found in Table \ref{Table MC2}.

Starting with the size results, observe that in Case I$^{\ast}$ with $%
z_{i}\sim N(0,0.5)$ the empirical rejection rate of the test appears to be
somewhat sensitive to the choices of $h_{p}$ and $\delta$, respectively.
More specifically, note that when $h_{p}=0.075$ and $\delta=.95$ the test
over-sizes substantially at $n=2,000$. In fact, as expected by the
theoretical results, this problem is alleviated when $\delta$ is chosen
closer to one and $h_{p}$ smaller as the sample size increases, see for
instance $\delta=0.98$ and $h_{p}=0.02$. By contrast, when $\widetilde{z}_{i}\sim
Binom(0.5)-0.5$, the size results become generally very conservative and are
close to zero. Turning to power, we can observe that power is generally very
good when $\widetilde{z}_{i}$ is normally distributed, but rather poor when $%
\gamma_{1}=0.25$ and $\widetilde{z}_{i}$ follows the binomial distribution. This is of
course to be expected and suggests that when only elements in $x_{i}$ are
continuous, larger sample sizes might be required for good power results. Finally, observe that the data-driven choice of $h_{p}$ and $h_{x}$ delivers results that are largely of a similar order of magnitude in terms of size and power as the results with a fixed bandwidth, suggesting that this choice might also be an option in practice. 

Summarizing this section, we obtain a somewhat mixed picture of our tests in
finite samples. While the first test performs well throughout the designs
even when the instrument is discrete and the tuning parameters are chosen in
a completely automated manner, the second test appears, as expected, to be somewhat more sensitive to the choice of the tuning parameters $\delta$
and $h_{p}$ as well as to the continuity (or discreteness) of $\widetilde{z}_{i}$, which
seems to reflect the irregularity of the underlying problem.

\begin{table}[H]
\footnotesize
\begin{center}
\begin{tabular}{|c|c|c|c|c|c|c|c|c|c|c|}
\hline\hline
\multicolumn{11}{|c|}{Conditional Quantile - First Test} \\ \hline\hline
CASE I &$\gamma_{1}=0$	& \multicolumn{3}{|c|}{$\rho=0$} &  \multicolumn{3}{|c|}{$\rho=0.25$} &  \multicolumn{3}{|c|}{$\rho=0.5$} \\ \hline

&&$	$c=3.5$	$&$	$c=4$	$&$	$c=4.5$	$&$	$c=3.5$	$&$	$c=4$	$&$	$c=4.5$	$&$	$c=3.5$	$&$	$c=4$	$&$	$c=4.5$	$\\	
\multirow{2}{*}{$\alpha=0.05$}&	$n=1000$	&$	0.071	$&$	0.090	$&$	0.080	$&$	0.381	$&$	0.393	$&$	0.379	$&$	0.888	$&$	0.902	$&$	0.892	$\\	
&	$n=2000$	&$	0.065	$&$	0.060	$&$	0.054	$&$	0.631	$&$	0.614	$&$	0.517	$&$	0.986	$&$	0.986	$&$	0.987	$\\	\hline	
\multirow{2}{*}{$\alpha=0.10$}&	$n=1000$	&$	0.143	$&$	0.157	$&$	0.147	$&$	0.497	$&$	0.517	$&$	0.476	$&$	0.940	$&$	0.950	$&$	0.937	$\\	
&	$n=2000$	&$	0.114	$&$	0.108	$&$	0.126	$&$	0.762	$&$	0.715	$&$	0.679	$&$	0.999	$&$	0.995	$&$	0.994	$\\	\hline
																					
CASE II	&$\gamma_{1}=0$ & \multicolumn{3}{|c|}{$\rho=0$} &  \multicolumn{3}{|c|}{$\rho=0.25$} &  \multicolumn{3}{|c|}{$\rho=0.5$} \\ \hline

&&$	$c=3.5$	$&$	$c=4$	$&$	$c=4.5$	$&$	$c=3.5$	$&$	$c=4$	$&$	$c=4.5$	$&$	$c=3.5$	$&$	$c=4$	$&$	$c=4.5$	$\\	
\multirow{2}{*}{$\alpha=0.05$}&	$n=1000$	&$	0.076	$&$	0.068	$&$	0.066	$&$	0.240	$&$	0.243	$&$	0.245	$&$	0.652	$&$	0.668	$&$	0.654	$\\	
&	$n=2000$	&$	0.041	$&$	0.049	$&$	0.058	$&$	0.388	$&$	0.362	$&$	0.341	$&$	0.905	$&$	0.898	$&$	0.920	$\\	\hline
\multirow{2}{*}{$\alpha=0.10$}&	$n=1000$	&$	0.130	$&$	0.120	$&$	0.112	$&$	0.352	$&$	0.377	$&$	0.330	$&$	0.784	$&$	0.787	$&$	0.771	$\\	
&	$n=2000$	&$	0.090	$&$	0.101	$&$	0.111	$&$	0.505	$&$	0.489	$&$	0.443	$&$	0.950	$&$	0.951	$&$	0.955	$\\	\hline
																					
CASE III&$\gamma_{1}=0$ 	& \multicolumn{3}{|c|}{$\rho=0$} &  \multicolumn{3}{|c|}{$\rho=0.25$} &  \multicolumn{3}{|c|}{$\rho=0.5$} \\ \hline

&&$	$c=3.5$	$&$	$c=4$	$&$	$c=4.5$	$&$	$c=3.5$	$&$	$c=4$	$&$	$c=4.5$	$&$	$c=3.5$	$&$	$c=4$	$&$	$c=4.5$	$\\	
\multirow{2}{*}{$\alpha=0.05$}&	$n=1000$	&$	0.119	$&$	0.099	$&$	0.094	$&$	0.338	$&$	0.375	$&$	0.338	$&$	0.858	$&$	0.902	$&$	0.870	$\\	
&	$n=2000$	&$	0.086	$&$	0.072	$&$	0.065	$&$	0.624	$&$	0.550	$&$	0.558	$&$	0.994	$&$	0.988	$&$	0.992	$\\	\hline
\multirow{2}{*}{$\alpha=0.10$}&	$n=1000$	&$	0.174	$&$	0.162	$&$	0.133	$&$	0.492	$&$	0.494	$&$	0.478	$&$	0.933	$&$	0.942	$&$	0.931	$\\	
&	$n=2000$	&$	0.141	$&$	0.133	$&$	0.149	$&$	0.743	$&$	0.696	$&$	0.692	$&$	0.997	$&$	0.995	$&$	0.997	$\\	\hline
CASE IV &$\gamma_{1}=0$	& \multicolumn{3}{|c|}{$\rho=0$} &  \multicolumn{3}{|c|}{$\rho=0.25$} &  \multicolumn{3}{|c|}{$\rho=0.5$} \\ \hline
&&$	$c=3.5$	$&$	$c=4$	$&$	$c=4.5$	$&$	$c=3.5$	$&$	$c=4$	$&$	$c=4.5$	$&$	$c=3.5$	$&$	$c=4$	$&$	$c=4.5$	$\\	
\multirow{2}{*}{$\alpha=0.05$}&	$n=1000$	&$	0.083	$&$	0.083	$&$	0.077	$&$	0.466	$&$	0.421	$&$	0.437	$&$	0.937	$&$	0.944	$&$	0.931	$\\	
&	$n=2000$	&$	0.096	$&$	0.092	$&$	0.088	$&$	0.683	$&$	0.698	$&$	0.663	$&$	0.998	$&$	0.996	$&$	0.996	$\\	\hline
\multirow{2}{*}{$\alpha=0.10$}&$n=1000$	&$	0.152	$&$	0.156	$&$	0.135	$&$	0.589	$&$	0.528	$&$	0.543	$&$	0.968	$&$	0.974	$&$	0.968	$\\	
&	$n=2000$	&$	0.150	$&$	0.154	$&$	0.181	$&$	0.794	$&$	0.798	$&$	0.778	$&$	1.000	$&$	0.999	$&$	0.999	$\\	\hline
CASE V &$\gamma_{1}=0$	& \multicolumn{3}{|c|}{$\rho=0$} &  \multicolumn{3}{|c|}{$\rho=0.25$} &  \multicolumn{3}{|c|}{$\rho=0.5$} \\ \hline
&&\multicolumn{3}{|c|}{$	\widehat{h}_{x}	$}&	\multicolumn{3}{|c|}{$	\widehat{h}_{x}	$}&				\multicolumn{3}{|c|}{$	\widehat{h}_{x}	$}\\
\multirow{2}{*}{$\alpha=0.05$}&	$n=1000$	&\multicolumn{3}{|c|}{$	0.072	$}&				\multicolumn{3}{|c|}{$	0.331	$}&				\multicolumn{3}{|c|}{$	0.876	$}\\					
&	$n=2000$	&\multicolumn{3}{|c|}{$	0.045	$}&				\multicolumn{3}{|c|}{$	0.546	$}&				\multicolumn{3}{|c|}{$	0.989	$}\\	\hline				
\multirow{2}{*}{$\alpha=0.10$}&	$n=1000$	&\multicolumn{3}{|c|}{$	0.123	$}&				\multicolumn{3}{|c|}{$	0.482	$}&			\multicolumn{3}{|c|}{$	0.930	$}\\	
&$n=2000$	&\multicolumn{3}{|c|}{$	0.111	$}&				\multicolumn{3}{|c|}{$	0.686	$}&				\multicolumn{3}{|c|}{$	0.996	$}\\	\hline		
	CASE VI& $\gamma_{1}=0$& \multicolumn{3}{|c|}{$\rho=0$} &  \multicolumn{3}{|c|}{$\rho=0.25$} &  \multicolumn{3}{|c|}{$\rho=0.5$} \\ \hline
	
&	&\multicolumn{3}{|c|}{$	\widehat{h}_{x}	$}&				\multicolumn{3}{|c|}{$\widehat{h}_{x}		$}&			\multicolumn{3}{|c|}{$\widehat{h}_{x}		$}\\
	\multirow{2}{*}{$\alpha=0.05$}&	$n=1000$	&\multicolumn{3}{|c|}{$	0.060	$}&		\multicolumn{3}{|c|}{$	0.325	$}&		\multicolumn{3}{|c|}{$	0.842	$}\\		
&	$n=2000$	&\multicolumn{3}{|c|}{$0.040	$}&				\multicolumn{3}{|c|}{$	0.565	$}				&\multicolumn{3}{|c|}{$	0.991	$}\\	\hline				
\multirow{2}{*}{$\alpha=0.10$}&	$n=1000$	&\multicolumn{3}{|c|}{$	0.130	$}&			\multicolumn{3}{|c|}{$	0.426$}&		\multicolumn{3}{|c|}{$	0.908	$}\\			
&$n=2000$	&\multicolumn{3}{|c|}{$	0.104	$}&		\multicolumn{3}{|c|}{$	0.673	$}&		\multicolumn{3}{|c|}{$	0.997	$}\\	\hline
\hline
	CASE VII& $\rho=0$& \multicolumn{3}{|c|}{} & \multicolumn{3}{|c|}{$\gamma_{1}=0.25$} &  \multicolumn{3}{|c|}{$\gamma_{1}=0.5$} \\ \hline
	&	&\multicolumn{3}{|c|}{}&				\multicolumn{3}{|c|}{$\widehat{h}_{x}		$}&			\multicolumn{3}{|c|}{$\widehat{h}_{x}		$}\\
	\multirow{2}{*}{$\alpha=0.05$}&	$n=1000$	&\multicolumn{3}{|c|}{}&		\multicolumn{3}{|c|}{$	1.000	$}&		\multicolumn{3}{|c|}{$	1.000	$}\\		
&	$n=2000$	&\multicolumn{3}{|c|}{}&				\multicolumn{3}{|c|}{$	1.000	$}				&\multicolumn{3}{|c|}{$	1.000	$}\\	\hline				
\multirow{2}{*}{$\alpha=0.10$}&	$n=1000$	&\multicolumn{3}{|c|}{}&			\multicolumn{3}{|c|}{$	1.000$}&		\multicolumn{3}{|c|}{$	1.000	$}\\			
&$n=2000$	&\multicolumn{3}{|c|}{}&		\multicolumn{3}{|c|}{$	1.000	$}&		\multicolumn{3}{|c|}{$	1.000	$}\\	\hline

	CASE VIII& $\rho=0$& \multicolumn{3}{|c|}{} & \multicolumn{3}{|c|}{$\gamma_{1}=0.25$} &  \multicolumn{3}{|c|}{$\gamma_{1}=0.5$} \\ \hline
	&	&\multicolumn{3}{|c|}{}&				\multicolumn{3}{|c|}{$\widehat{h}_{x}		$}&			\multicolumn{3}{|c|}{$\widehat{h}_{x}		$}\\
	\multirow{2}{*}{$\alpha=0.05$}&	$n=1000$	&\multicolumn{3}{|c|}{}&		\multicolumn{3}{|c|}{$	0.997	$}&		\multicolumn{3}{|c|}{$	1.000	$}\\		
&	$n=2000$	&\multicolumn{3}{|c|}{}&				\multicolumn{3}{|c|}{$	1.000	$}				&\multicolumn{3}{|c|}{$	1.000	$}\\	\hline				
\multirow{2}{*}{$\alpha=0.10$}&	$n=1000$	&\multicolumn{3}{|c|}{}&			\multicolumn{3}{|c|}{$	0.999$}&		\multicolumn{3}{|c|}{$	1.000	$}\\			
&$n=2000$	&\multicolumn{3}{|c|}{}&		\multicolumn{3}{|c|}{$	1.000	$}&		\multicolumn{3}{|c|}{$	1.000	$}\\	\hline
\hline

\end{tabular}%
\end{center}
\normalsize
\caption{First Test - \footnotesize Notes: (i) Cases I-IV: Designs (i)-(iv), respectively, with oracle propensity score $p_{i}$ and ad-hoc choices for $h_{x}$ and $h_{F}$. (ii) Case V: Design (i) with oracle propensity score $p_{i}$ and cross-validated $\widehat{h}_{x}$ and $\widehat{h}_{F}$. (iii) Case VI: Design (i) with estimated propensity score $\widehat{p}_{i}$ (cross-validated bandwidth) and cross-validated $\widehat{h}_{x}$ and $\widehat{h}_{F}$. (iv) Case VII: misspecification, Design (i), with oracle propensity score $p_{i}$ and cross-validated $\widehat{h}_{x}$ and $\widehat{h}_{F}$. (v) Case VIII: misspecification, Design (ii), with oracle propensity score $p_{i}$ and cross-validated $\widehat{h}_{x}$ and $\widehat{h}_{F}$. }
\label{Table MC1}
\end{table}

\begin{table}[H]
\scriptsize
\begin{center}
\begin{tabular}{|c|c|c|c|c|c|c|c|c|c|c|c|c|c|}
\hline\hline
\multicolumn{14}{|c|}{Conditional Quantile - Second Test} \\ \hline\hline
CASE I$^{\ast}$ & $\gamma_{1}=0$& \multicolumn{3}{|c|}{$\delta=0.95, h_{p}=0.075$} &  \multicolumn{3}{|c|}{$\delta=0.95, h_{p}=0.05$	} &  \multicolumn{3}{|c|}{$\delta=0.975, h_{p}=0.03$	}& \multicolumn{3}{|c|}{$\delta=0.98, h_{p}=0.02$} \\ \hline
&	&$	$c=3.5$	$&$	$c=4$	$&$	$c=4.5$	$&$	$c=3.5$	$&$	$c=4$	$&$	$c=4.5$	$&$	$c=3.5$	$&$	$c=4$	$&$	$c=4.5$	$&$	$c=3.5$	$&$	$c=4$	$&$	$c=4.5$	$\\	
\multirow{2}{*}{$\alpha=0.05$}& 	$n=1000$	&$	0.067	$&$	0.076	$&$	0.067	$&$	0.066	$&$	0.060	$&$	0.067	$&$	0.079	$&$	0.070	$&$	0.080	$&$	0.075	$&$	0.083	$&$	0.089	$\\	
&	$n=2000$	&$	0.088	$&$	0.129	$&$	0.131	$&$	0.121	$&$	0.096	$&$	0.105	$&$	0.102	$&$	0.080	$&$	0.072	$&$	0.068	$&$	0.080	$&$	0.078	$\\	\hline
\multirow{2}{*}{$\alpha=0.10$}& 	$n=1000$	&$	0.113	$&$	0.112	$&$	0.124	$&$	0.112	$&$	0.111	$&$	0.118	$&$	0.125	$&$	0.120	$&$	0.122	$&$	0.141	$&$	0.140	$&$	0.143	$\\	
&	$n=2000$	&$	0.215	$&$	0.210	$&$	0.227	$&$	0.174	$&$	0.167	$&$	0.169	$&$	0.124	$&$	0.127	$&$	0.129	$&$	0.138	$&$	0.144	$&$	0.141	$\\	\hline
 &  & \multicolumn{12}{|c|}{$\widehat{h}_{x}, \widehat{h}_{p}$} \\

\multirow{2}{*}{$\alpha=0.05$}& 	$n=1000$	& \multicolumn{12}{|c|}{$0.076$} \\ 
&	$n=2000$	&\multicolumn{12}{|c|}{$0.0831$} \\	\hline
\multirow{2}{*}{$\alpha=0.10$}& 	$n=1000$	&\multicolumn{12}{|c|}{$0.115$} \\	
&	$n=2000$	&\multicolumn{12}{|c|}{$0.140$} \\	\hline

CASE II$^{\ast}$ & $\gamma_{1}=0.25$& \multicolumn{3}{|c|}{$\delta=0.95, h_{p}=0.075$} &  \multicolumn{3}{|c|}{$\delta=0.95, h_{p}=0.05$	} &  \multicolumn{3}{|c|}{$\delta=0.975, h_{p}=0.03$	}& \multicolumn{3}{|c|}{$\delta=0.98, h_{p}=0.02$} \\ \hline							
&	&$	$c=3.5$	$&$	$c=4$	$&$	$c=4.5$	$&$	$c=3.5$	$&$	$c=4$	$&$	$c=4.5$	$&$	$c=3.5$	$&$	$c=4$	$&$	$c=4.5$	$&$	$c=3.5$	$&$	$c=4$	$&$	$c=4.5$	$\\	
							
\multirow{2}{*}{$\alpha=0.05$}& 	$n=1000$	&$	0.998	$&$	0.998	$&$	0.997	$&$	0.991	$&$	0.987	$&$	0.986	$&$	0.875	$&$	0.858	$&$	0.876	$&$	0.761	$&$	0.777	$&$	0.805	$\\	
&	$n=2000$	&$	1.000	$&$	1.000	$&$	1.000	$&$	1.000	$&$	1.000	$&$	1.000	$&$	0.999	$&$	0.999	$&$	0.998	$&$	0.984	$&$	0.988	$&$	0.986	$\\	\hline
\multirow{2}{*}{$\alpha=0.10$}& 	$n=1000$	&$	0.998	$&$	0.999	$&$	0.998	$&$	0.994	$&$	0.993	$&$	0.994	$&$	0.924	$&$	0.922	$&$	0.922	$&$	0.884	$&$	0.882	$&$	0.879	$\\	
&	$n=2000$	&$	1.000	$&$	1.000	$&$	1.000	$&$	1.000	$&$	1.000	$&$	1.000	$&$	1.000	$&$	1.000	$&$	1.000	$&$	0.996	$&$	0.996	$&$	0.995	$\\	\hline

 &  & \multicolumn{12}{|c|}{$\widehat{h}_{x}, \widehat{h}_{p}$} \\

\multirow{2}{*}{$\alpha=0.05$}& 	$n=1000$	& \multicolumn{12}{|c|}{$0.821$} \\ 
&	$n=2000$	&\multicolumn{12}{|c|}{$0.857$} \\	\hline
\multirow{2}{*}{$\alpha=0.10$}& 	$n=1000$	&\multicolumn{12}{|c|}{$0.888$} \\	
&	$n=2000$	&\multicolumn{12}{|c|}{$0.927$} \\	\hline

CASE III$^{\ast}$ & $\gamma_{1}=0.5$ & \multicolumn{3}{|c|}{$\delta=0.95, h_{p}=0.075$} &  \multicolumn{3}{|c|}{$\delta=0.95, h_{p}=0.05$	} &  \multicolumn{3}{|c|}{$\delta=0.975, h_{p}=0.03$	}& \multicolumn{3}{|c|}{$\delta=0.98, h_{p}=0.02$} \\ \hline
							
&	&$	$c=3.5$	$&$	$c=4$	$&$	$c=4.5$	$&$	$c=3.5$	$&$	$c=4$	$&$	$c=4.5$	$&$	$c=3.5$	$&$	$c=4$	$&$	$c=4.5$	$&$	$c=3.5$	$&$	$c=4$	$&$	$c=4.5$	$\\	

\multirow{2}{*}{$\alpha=0.05$}& 	$n=1000$	&$	1.000	$&$	1.000	$&$	1.000	$&$	1.000	$&$	1.000	$&$	1.000	$&$	0.998	$&$	0.998	$&$	0.997	$&$	0.967	$&$	0.973	$&$	0.980	$\\	
&	$n=2000$	&$	1.000	$&$	1.000	$&$	1.000	$&$	1.000	$&$	1.000	$&$	1.000	$&$	1.000	$&$	1.000	$&$	1.000	$&$	1.000	$&$	1.000	$&$	1.000	$\\	\hline
\multirow{2}{*}{$\alpha=0.10$}& 	$n=1000$	&$	1.000	$&$	1.000	$&$	1.000	$&$	1.000	$&$	1.000	$&$	1.000	$&$	1.000	$&$	1.000	$&$	1.000	$&$	0.995	$&$	0.995	$&$	0.995	$\\	
&	$n=2000$	&$	1.000	$&$	1.000	$&$	1.000	$&$	1.000	$&$	1.000	$&$	1.000	$&$	1.000	$&$	1.000	$&$	1.000	$&$	1.000	$&$	1.000	$&$	1.000	$\\	\hline

 &  & \multicolumn{12}{|c|}{$\widehat{h}_{x}, \widehat{h}_{p}$} \\

\multirow{2}{*}{$\alpha=0.05$}& 	$n=1000$	& \multicolumn{12}{|c|}{$0.993$} \\ 
&	$n=2000$	&\multicolumn{12}{|c|}{$0.981$} \\	\hline
\multirow{2}{*}{$\alpha=0.10$}& 	$n=1000$	&\multicolumn{12}{|c|}{$1.000$} \\	
&	$n=2000$	&\multicolumn{12}{|c|}{$0.994$} \\	\hline

CASE IV$^{\ast}$ & $\gamma_{1}=0$&  \multicolumn{3}{|c|}{$\delta=0.95, h_{p}=0.075$} &  \multicolumn{3}{|c|}{$\delta=0.95, h_{p}=0.05$	} &  \multicolumn{3}{|c|}{}& \multicolumn{3}{|c|}{} \\ \hline
&	&$	$c=3.5$	$&$	$c=4$	$&$	$c=4.5$	$&$	$c=3.5$	$&$	$c=4$	$&$	$c=4.5$	$&&&&&& \\	
\multirow{2}{*}{$\alpha=0.05$}&$n=1000$	&$	0.005	$&$	0.003	$&$	0.004	$&$	0.007	$&$	0.005	$&$	0.005	$&&&&&&\\
&	$n=2000$	&$	0.002	$&$	0.002	$&$	0.003	$&$	0.001	$&$	0.003	$&$	0.003	$&&&&&&\\\hline
\multirow{2}{*}{$\alpha=0.10$}&	$n=1000$	&$	0.013	$&$	0.009	$&$	0.012	$&$	0.010	$&$	0.007	$&$	0.012	$&&&&&&\\
&	$n=2000$	&$	0.011	$&$	0.014	$&$	0.015	$&$	0.011	$&$	0.009	$&$	0.006	$&&&&&&\\\hline

 &  & \multicolumn{12}{|c|}{$\widehat{h}_{x}, \widehat{h}_{p}$} \\

\multirow{2}{*}{$\alpha=0.05$}& 	$n=1000$	& \multicolumn{12}{|c|}{$0.001$} \\ 
&	$n=2000$	&\multicolumn{12}{|c|}{$0.003$} \\	\hline
\multirow{2}{*}{$\alpha=0.10$}& 	$n=1000$	&\multicolumn{12}{|c|}{$0.008$} \\	
&	$n=2000$	&\multicolumn{12}{|c|}{$0.011$} \\	\hline

CASE V$^{\ast}$ & $\gamma_{1}=0.25$&  \multicolumn{3}{|c|}{$\delta=0.95, h_{p}=0.075$} &  \multicolumn{3}{|c|}{$\delta=0.95, h_{p}=0.05$	} &  \multicolumn{3}{|c|}{}& \multicolumn{3}{|c|}{} \\ \hline
							
&	&$	$c=3.5$	$&$	$c=4$	$&$	$c=4.5$	$&$	$c=3.5$	$&$	$c=4$	$&$	$c=4.5$	$&&&&&&\\	
							
\multirow{2}{*}{$\alpha=0.05$}&	$n=1000$	&$	0.177	$&$	0.156	$&$	0.184	$&$	0.084	$&$	0.088	$&$	0.078	$&&&&&&\\
&	$n=2000$	&$	0.536	$&$	0.554	$&$	0.518	$&$	0.271	$&$	0.250	$&$	0.253	$&&&&&&\\\hline
\multirow{2}{*}{$\alpha=0.05$}&	$n=1000$	&$	0.256	$&$	0.278	$&$	0.291	$&$	0.129	$&$	0.130	$&$	0.150	$&&&&&&\\
	&$n=2000$	&$	0.721	$&$	0.716	$&$	0.720	$&$	0.425	$&$	0.421	$&$	0.424	$&&&&&&\\\hline

 &  & \multicolumn{12}{|c|}{$\widehat{h}_{x}, \widehat{h}_{p}$} \\

\multirow{2}{*}{$\alpha=0.05$}& 	$n=1000$	& \multicolumn{12}{|c|}{$0.209$} \\ 
&	$n=2000$	&\multicolumn{12}{|c|}{$0.472$} \\	\hline
\multirow{2}{*}{$\alpha=0.10$}& 	$n=1000$	&\multicolumn{12}{|c|}{$0.350$} \\	
&	$n=2000$	&\multicolumn{12}{|c|}{$0.641$} \\	\hline

CASE VI$^{\ast}$ & $\gamma_{1}=0.5$&  \multicolumn{3}{|c|}{$\delta=0.95, h_{p}=0.075$} &  \multicolumn{3}{|c|}{$\delta=0.95, h_{p}=0.05$	} &  \multicolumn{3}{|c|}{}& \multicolumn{3}{|c|}{} \\ \hline
							
&	&$	$c=3.5$	$&$	$c=4$	$&$	$c=4.5$	$&$	$c=3.5$	$&$	$c=4$	$&$	$c=4.5$	$&&&&&&\\	

\multirow{2}{*}{$\alpha=0.05$}&	$n=1000$	&$	0.779	$&$	0.739	$&$	0.760	$&$	0.457	$&$	0.472	$&$	0.438	$&&&&&&\\
&	$n=2000$	&$	1.000	$&$	0.997	$&$	0.998	$&$	0.899	$&$	0.915	$&$	0.910	$&&&&&&\\\hline
\multirow{2}{*}{$\alpha=0.05$}&	$n=1000$	&$	0.868	$&$	0.864	$&$	0.870	$&$	0.571	$&$	0.573	$&$	0.588	$&&&&&&\\
&	$n=2000$	&$	1.000	$&$	1.000	$&$	0.999	$&$	0.960	$&$	0.962	$&$	0.962	$&&&&&&\\\hline
																				
 &  & \multicolumn{12}{|c|}{$\widehat{h}_{x}, \widehat{h}_{p}$} \\

\multirow{2}{*}{$\alpha=0.05$}& 	$n=1000$	& \multicolumn{12}{|c|}{$0.612$} \\ 
&	$n=2000$	&\multicolumn{12}{|c|}{$0.996$} \\	\hline
\multirow{2}{*}{$\alpha=0.10$}& 	$n=1000$	&\multicolumn{12}{|c|}{$0.679$} \\	
&	$n=2000$	&\multicolumn{12}{|c|}{$1.000$} \\	\hline
															
\hline
\end{tabular}%
\end{center}
\normalsize
\caption{Second Test - \footnotesize Notes: (i) Cases I$^{\ast}$-III$^{\ast}$: Design (i) with different degrees of misspecification ($\gamma_{1}=0$, $\gamma_{1}=0.25$, $\gamma_{1}=0.5$, respectively) and oracle propensity score $p_{i}$. (ii) Case IV$^{\ast}$-VI$^{\ast}$: Design (ii) with different degrees of misspecification ($\gamma_{1}=0$, $\gamma_{1}=0.25$, $\gamma_{1}=0.5$, respectively) and oracle propensity score $p_{i}$.  }
\label{Table MC2}
\end{table}

\section{Empirical Illustration}

\label{Empirical Illustration}

Our illustration is based on a subsample of the UK wage data from the Family
Expenditure Survey used by \citet{AB2017}.\footnote{%
For the exact construction of the sample see their paper and references
therein.} As pointed out by these authors, due to changes in employment
rates over time, simply examining wage inequality for females and males at
work over time may provide a distorted picture of market-level wage
inequality. We will therefore run our selection testing procedure on two
different subsets of the data, namely 1995 to 1997, a period of increasing
gross domestic product (GDP) growth rates, and 1998 to 2000, a period of
high, but stable GDP growth rates. Unlike \citet{AB2017}, however, our
testing procedure for selection will not rely on a parametric specification
of the conditional log-wage quantile functions, but remain completely
nonparametric.

The covariates we include in $x_{i}$ are dummies for marital status,
education (end of schooling at 17 or 18, and end of schooling after 18),
location (eleven regional dummies), number of kids (split by six age
categories), time (year dummies), as well as age in years. This set of
covariates is identical to the one used by \citet{AB2017}, but for the fact
that the latter used cohort dummies instead of age in years. The continuous
instrumental variable is given by the measure of potential out-of-work
(welfare) income, interacted with marital status. This variable, which was
also used by \citet{AB2017}, builds on \citet{BRS2003} and is constructed
for each individual in the sample (employed and non-employed) using the
Institute of Fiscal Studies (IFS) tax and welfare-benefit simulation model.

The final sample for the years 1995-1997 comprises 21,263 individuals,
11,647 of which are females and 9,616 of which are males, respectively. The
number of working females (males) with a positive log hourly wage in that
sample is 7,761 (7,623). By contrast, for the 1998-2000 period we obtain
16,350 observations, 8,904 females and 7,446 males. The number of working
females (males) in that sample are 5,931 (6,058).

All estimates are constructed using routines from the \texttt{np package} of %
\citet{HR2008}. That is, we estimate the propensity score $%
\Pr(s_{i}=1|z_{i})$ fully nonparametrically using a standard local constant
estimator with an Epanechnikov second order kernel function for the
continuous instrument and discrete kernel functions as in Equations (1) and (2) of \citet{LR2008} for
the remaining discrete controls. The bandwidth is determined in a
data-driven manner using cross-validation according to the procedure in Section 2 of \citet{LR2008} as permitted by A.3 since $d_{zc}<4$ (cf. footnote 6). More specifically, to reduce computation complexity, we follow the method outlined in \citet{R1993} conducting
cross-validation on random subsets of the data (size $n=450$), and select the
median values over 50 replications. 

The conditional
quantile function $q_{\tau}(x_{i})$ on the other hand is estimated as in Equation (19) of %
\citet{LR2008}. Note that in this case, all covariates in $x_{i}$ are discrete, and the rate conditions
of Theorem 1 and 1$^{\ast}$ do therefore not apply directly as discrete predictors do not contribute to the asymptotic variance. As a result, we choose to select the bandwidth parameters for the  discrete kernel functions according to the same method outlined in \citet{R1993} using cross-validation for discrete covariates as automated by the \texttt{np package}. The same holds true for the conditional distribution function $F_{p|x,u_{\tau},s=1}(\cdot|\cdot,\cdot,\cdot)$, which is constructed as in Equation
(4) of the same paper. Finally, the
quantile grid is chosen to be $\mathcal{T}=\{.1,.2,.3,.4,.5,.6,.7,.8,.9\}$.

\begin{figure}[H]
\subfigure[$\tau=10\%$]{
\includegraphics[width=7.5cm,height=3.75cm]{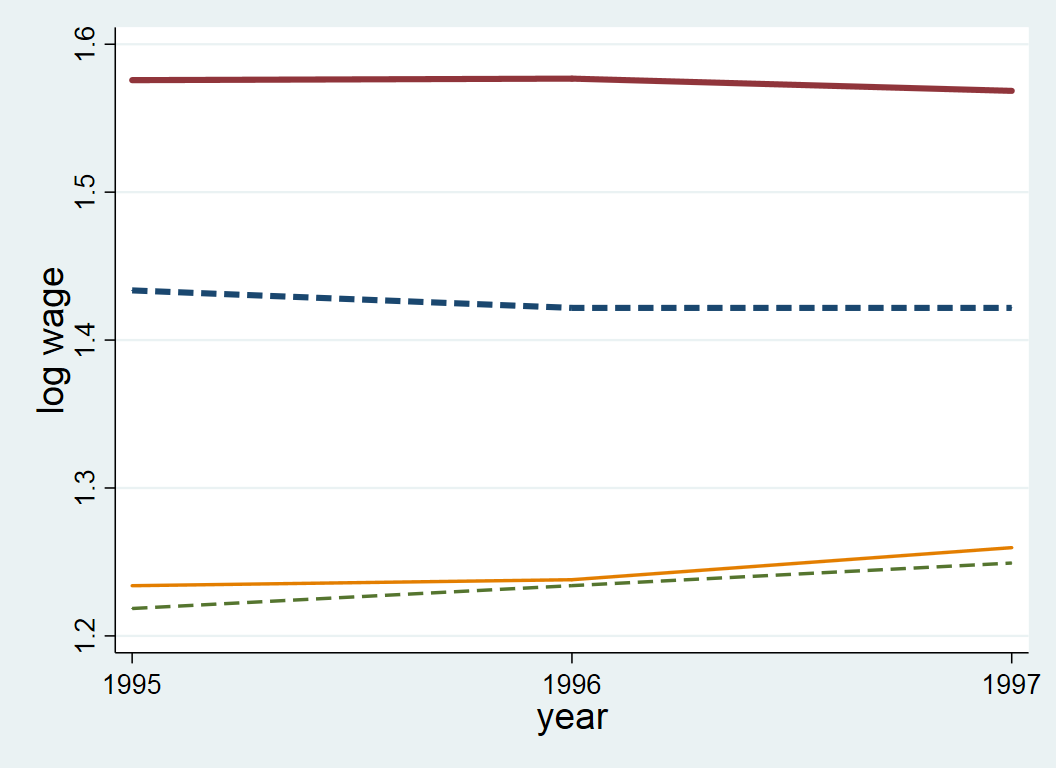}
        \label{Fig 1a}}\quad 
\subfigure[$\tau=40\%$ ]{
\includegraphics[width=7.5cm,height=3.75cm]{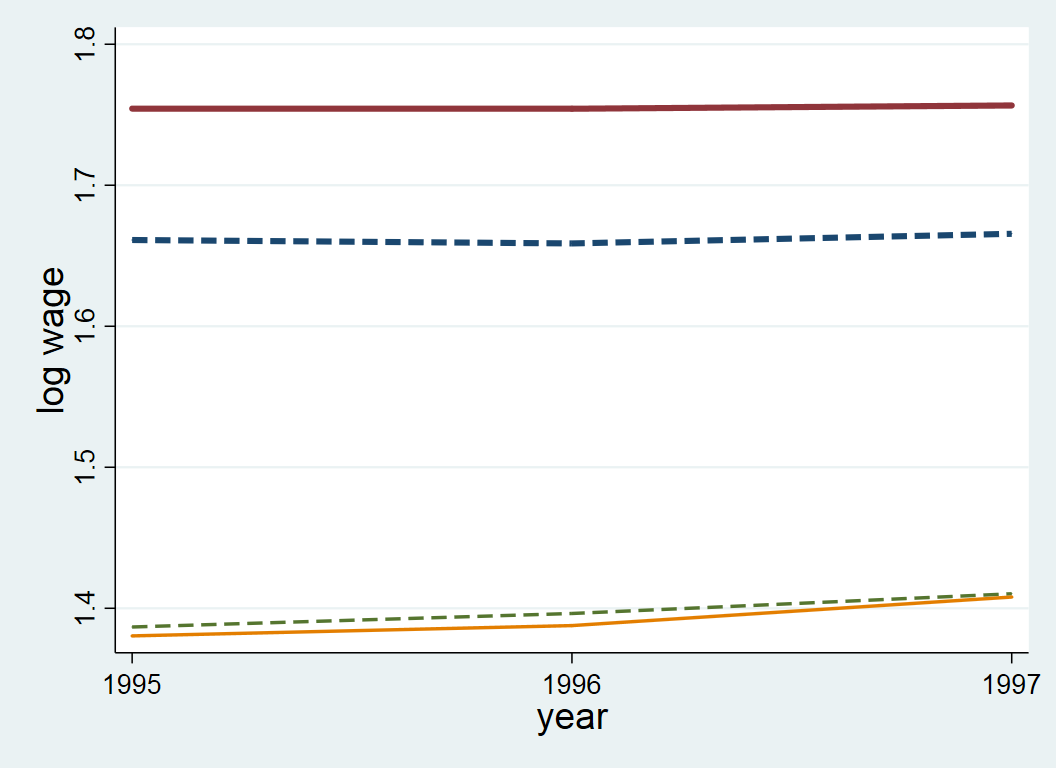}
        \label{Fig 1b}}
\par
\subfigure[$\tau=30\%$]{
\includegraphics[width=7.5cm,height=3.75cm]{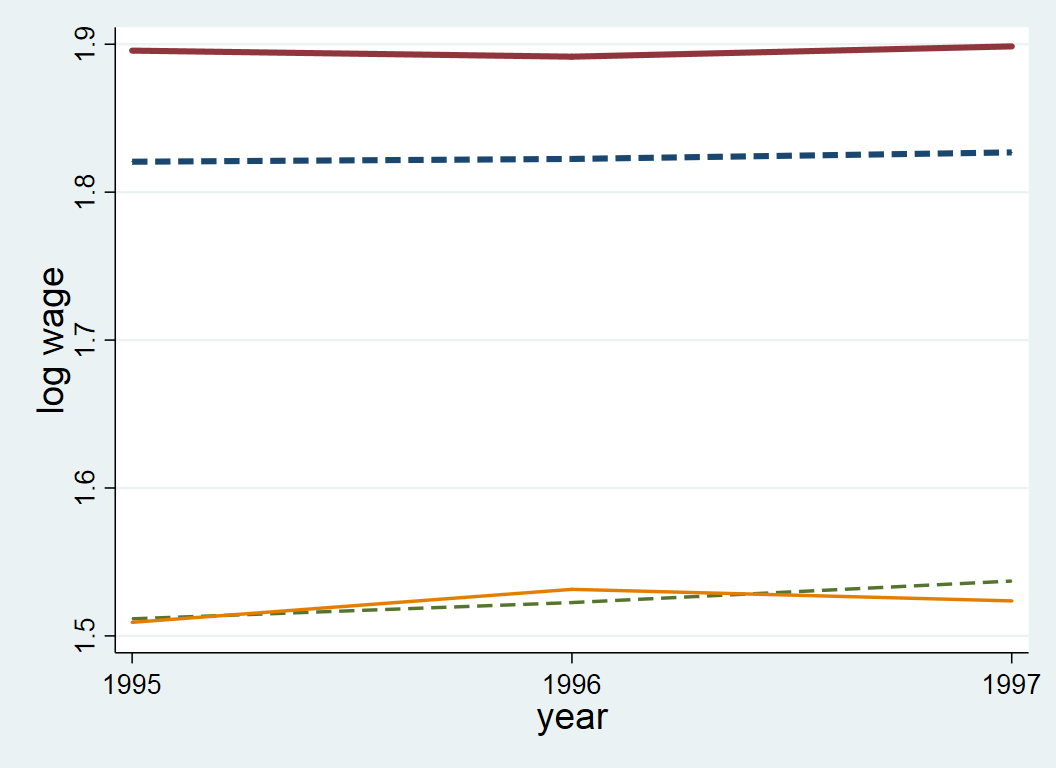}
        \label{Fig 1c}}\quad 
\subfigure[$\tau=40\%$ ]{
\includegraphics[width=7.5cm,height=3.75cm]{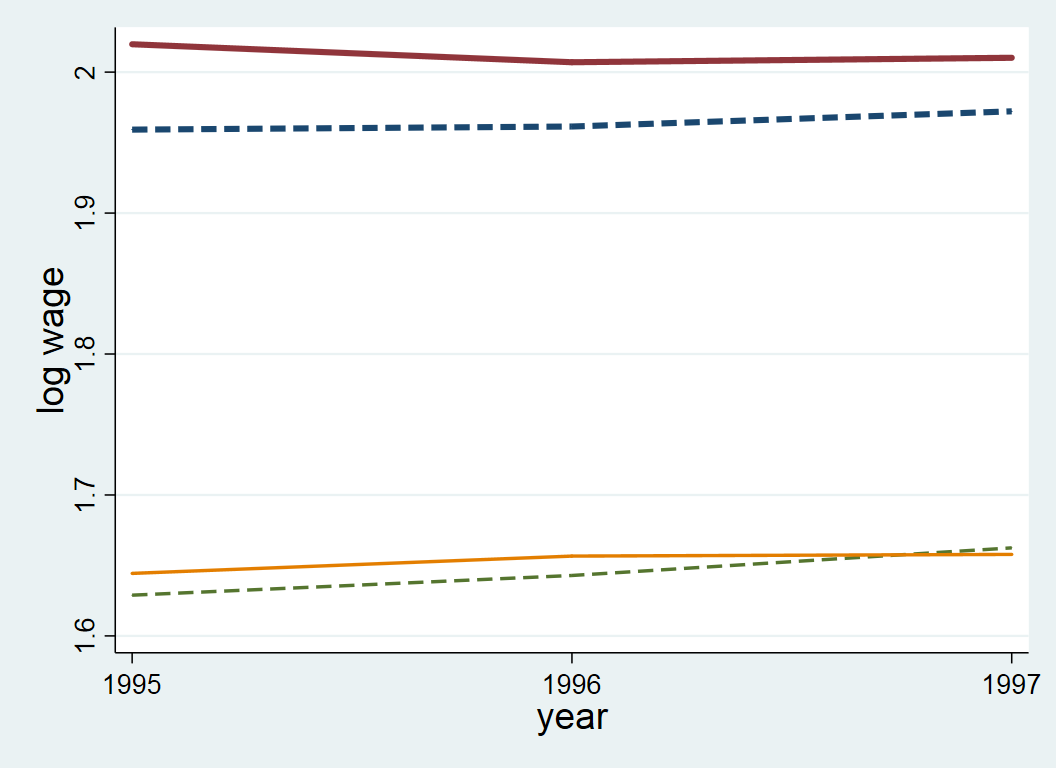}
        \label{Fig 1d}}
\par
\subfigure[$\tau=50\%$ ]{
\includegraphics[width=7.5cm,height=3.75cm]{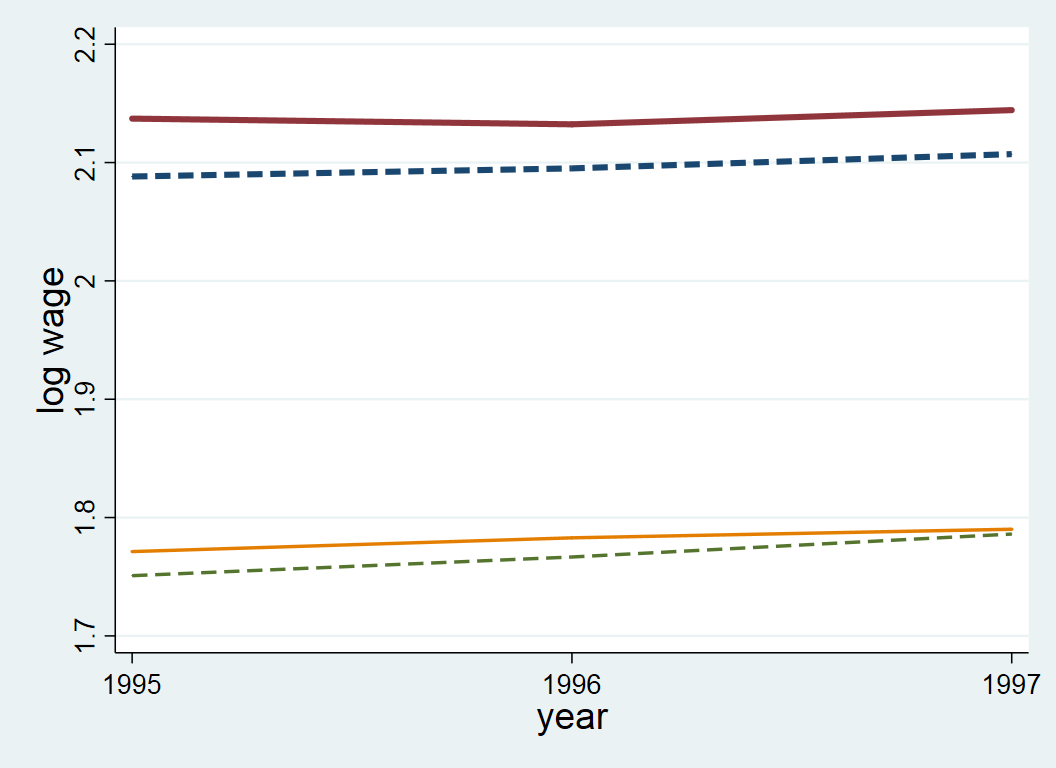}
        \label{Fig 1e}}\quad 
\subfigure[$\tau=60\%$ ]{
\includegraphics[width=7.5cm,height=3.75cm]{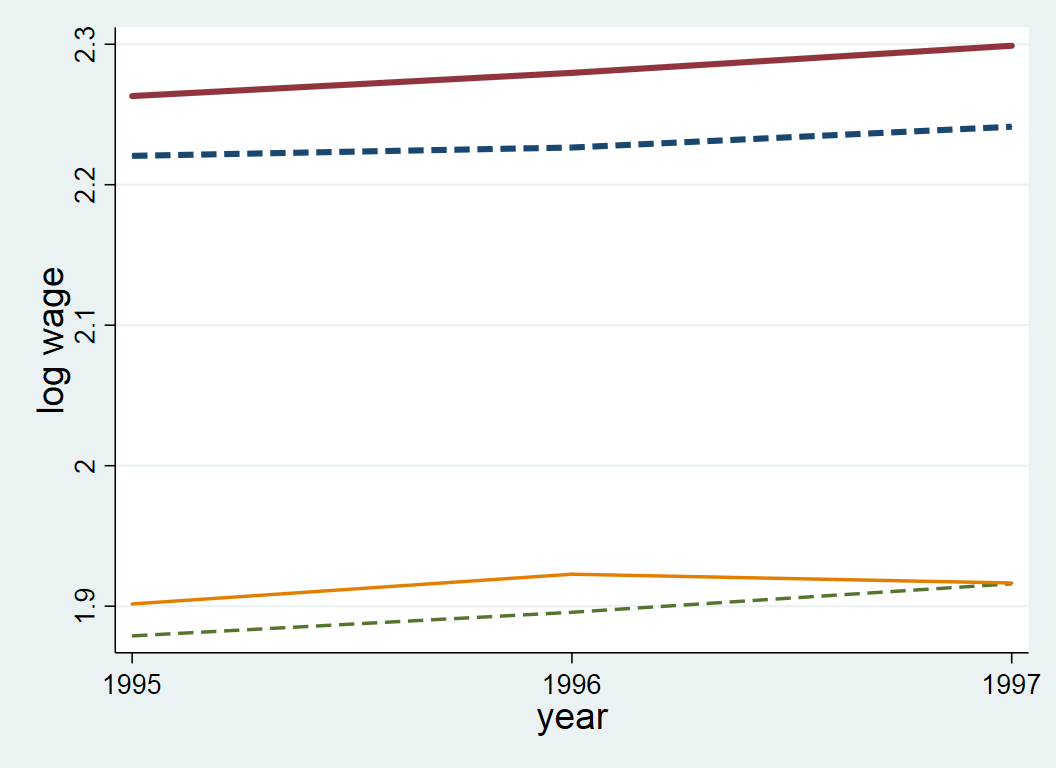}
        \label{Fig 1f}}
\par
\subfigure[$\tau=70\%$ ]{
\includegraphics[width=7.5cm,height=3.75cm]{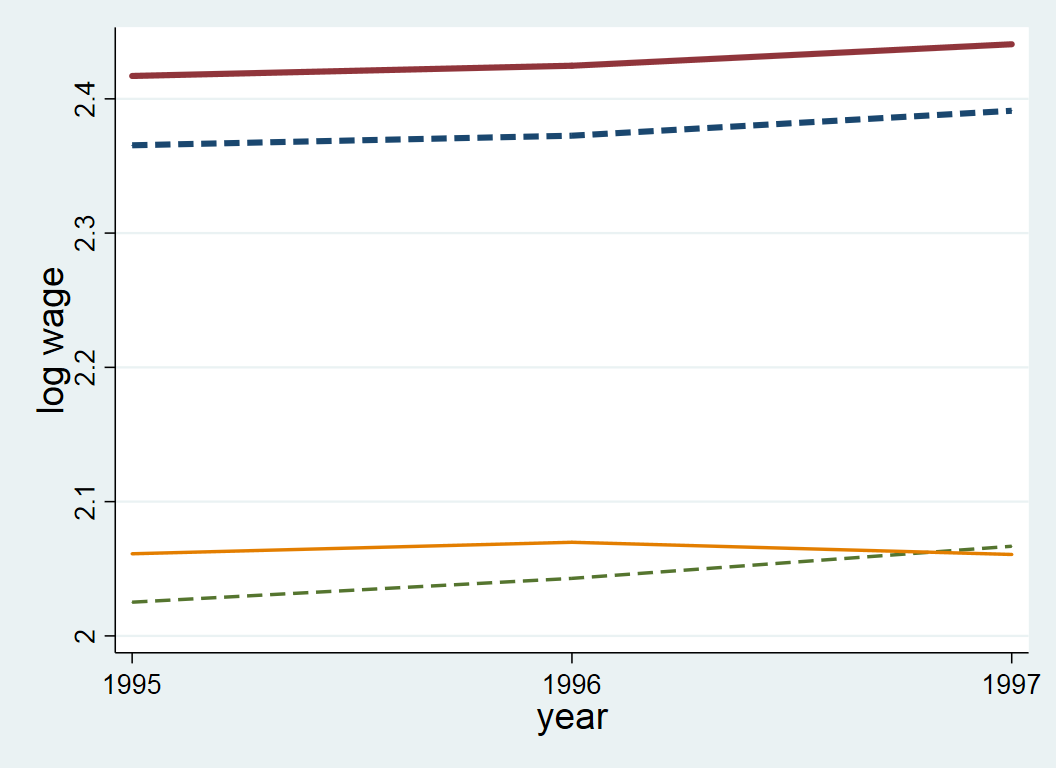}
        \label{Fig 1g}} 
\subfigure[$\tau=80\%$ ]{
\includegraphics[width=7.5cm,height=3.75cm]{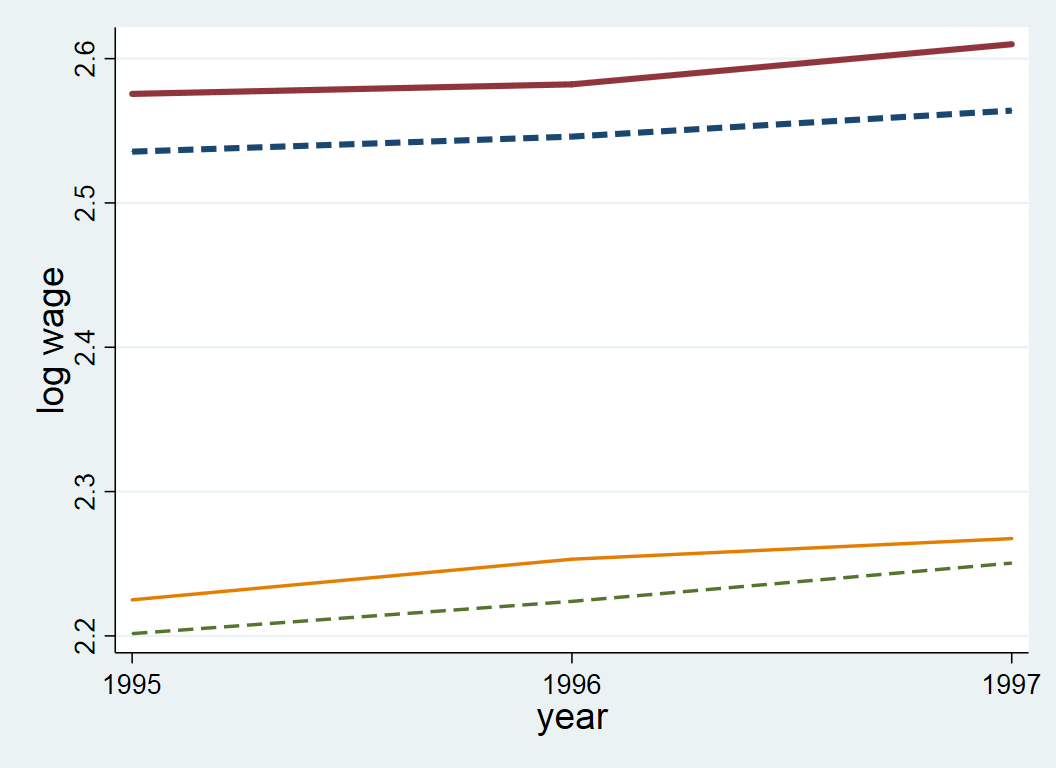}
        \label{Fig 1h}}
\par
\subfigure[$\tau=90\%$ ]{
\includegraphics[width=7.5cm,height=3.75cm]{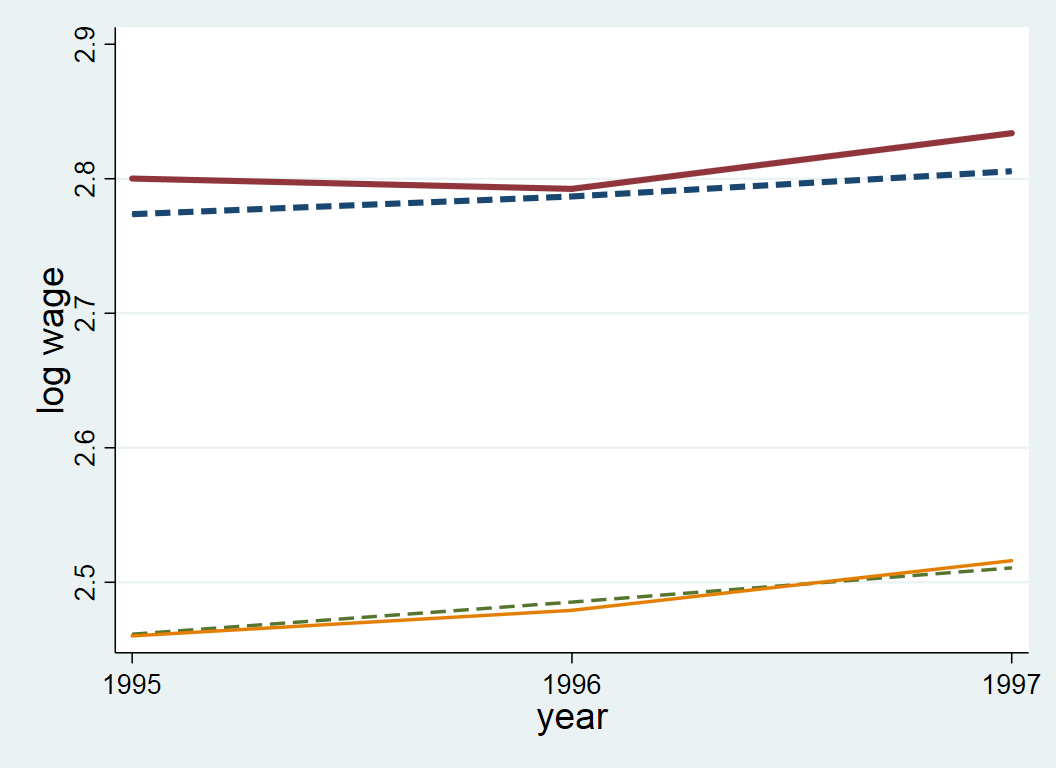}
        \label{Fig 1i}}
\caption{Corrected and Uncorrected Log Hourly Wage Quantiles by Gender
1995-1997 \citep{AB2017}. \textit{Note: male quantiles are always at the
top, female ones at the bottom (solid lines: uncorrected quantiles; dashed
lines: selection corrected quantiles)}}
\label{Fig 1}
\end{figure}

\begin{figure}[H]
\subfigure[$\tau=10\%$]{
\includegraphics[width=7.5cm,height=3.75cm]{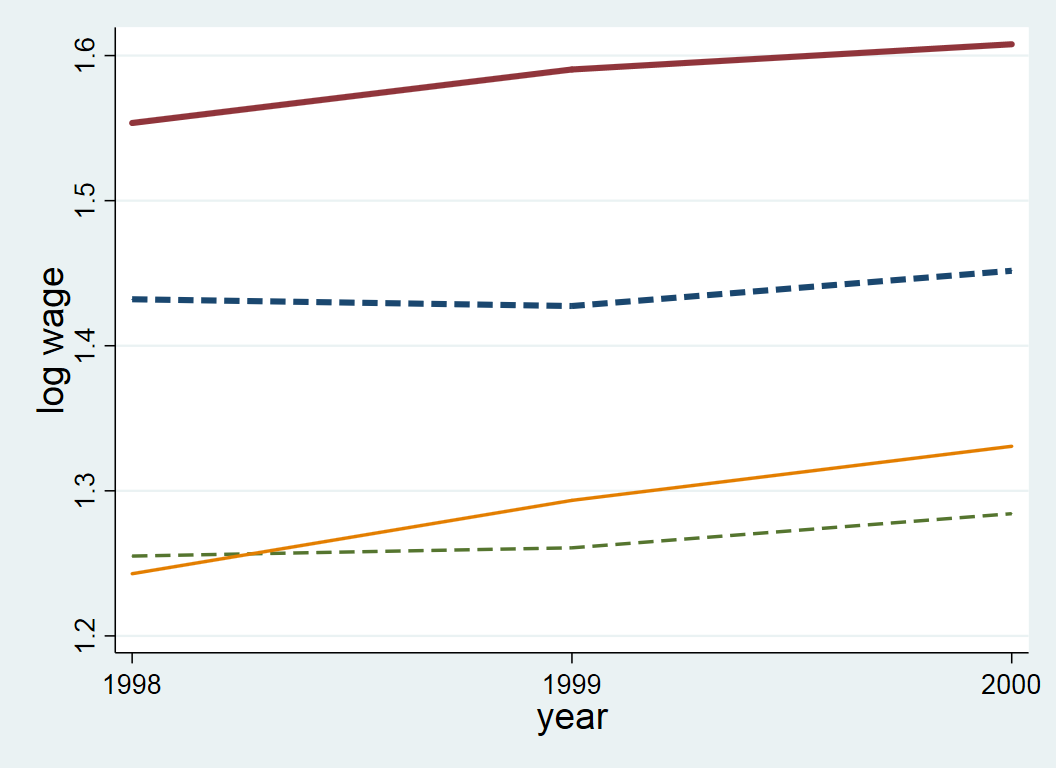}
        \label{Fig 2a}}\quad 
\subfigure[$\tau=20\%$ ]{
\includegraphics[width=7.5cm,height=3.75cm]{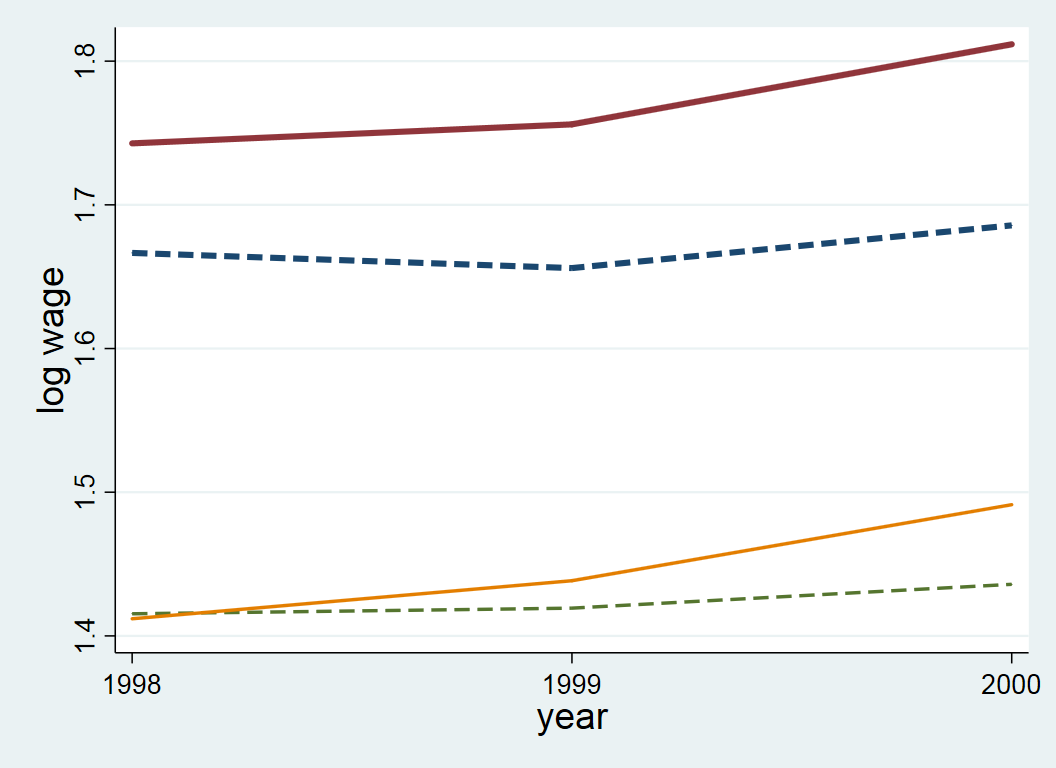}
        \label{Fig 2b}}
\par
\subfigure[$\tau=30\%$]{
\includegraphics[width=7.5cm,height=3.75cm]{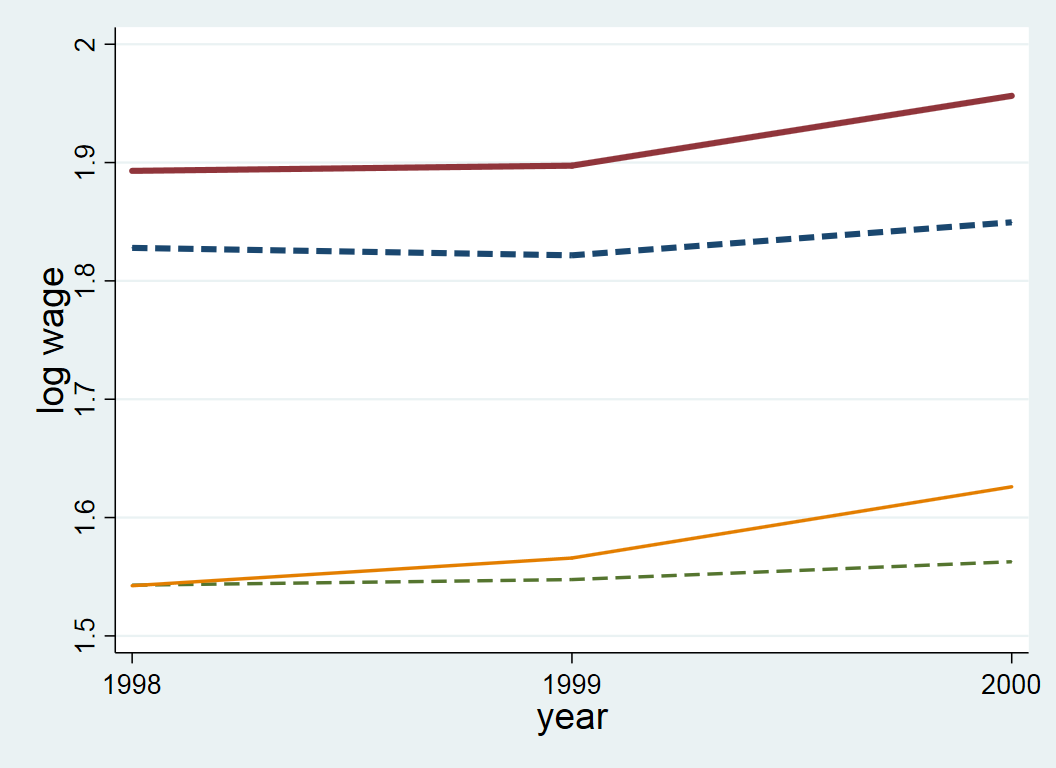}
        \label{Fig 2c}}\quad 
\subfigure[$\tau=40\%$ ]{
\includegraphics[width=7.5cm,height=3.75cm]{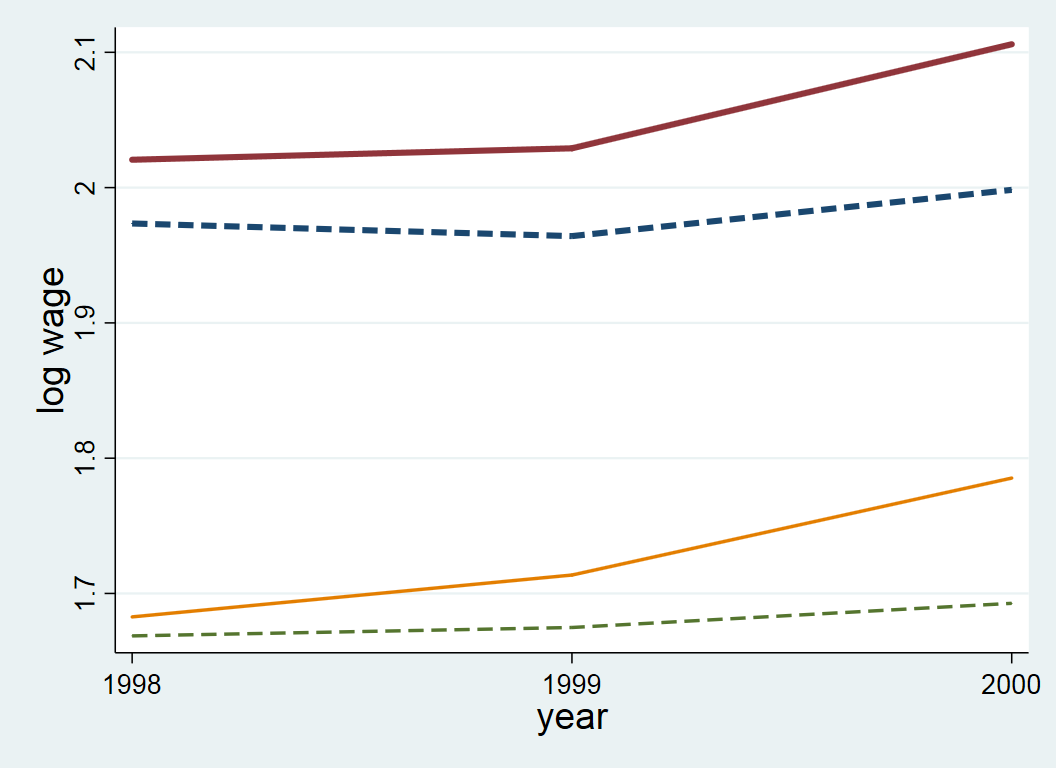}
        \label{Fig 2d}}
\par
\subfigure[$\tau=50\%$ ]{
\includegraphics[width=7.5cm,height=3.75cm]{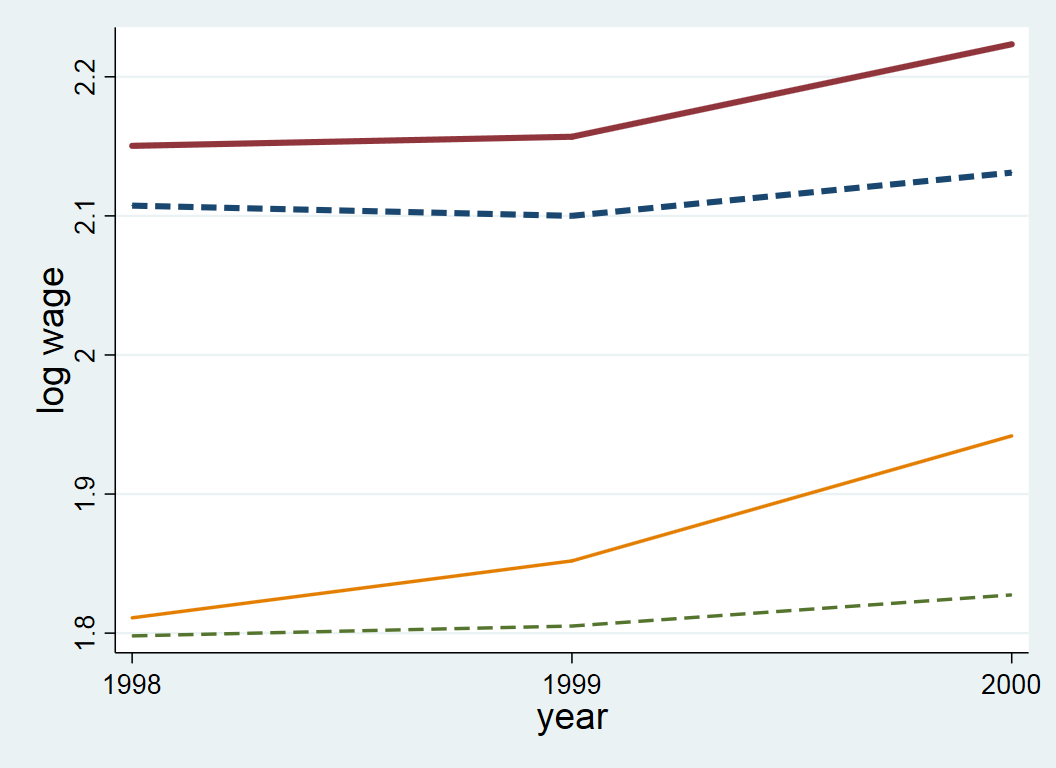}
        \label{Fig 2e}}\quad 
\subfigure[$\tau=60\%$ ]{
\includegraphics[width=7.5cm,height=3.75cm]{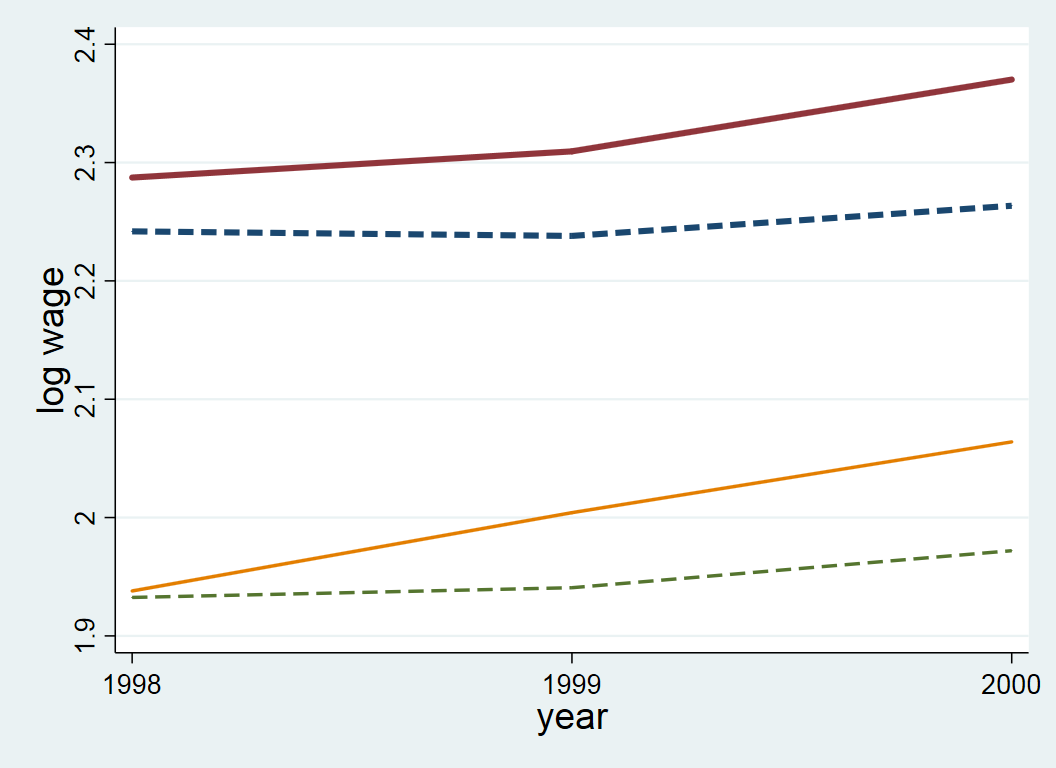}
        \label{Fig 2f}}
\par
\subfigure[$\tau=70\%$ ]{
\includegraphics[width=7.5cm,height=3.75cm]{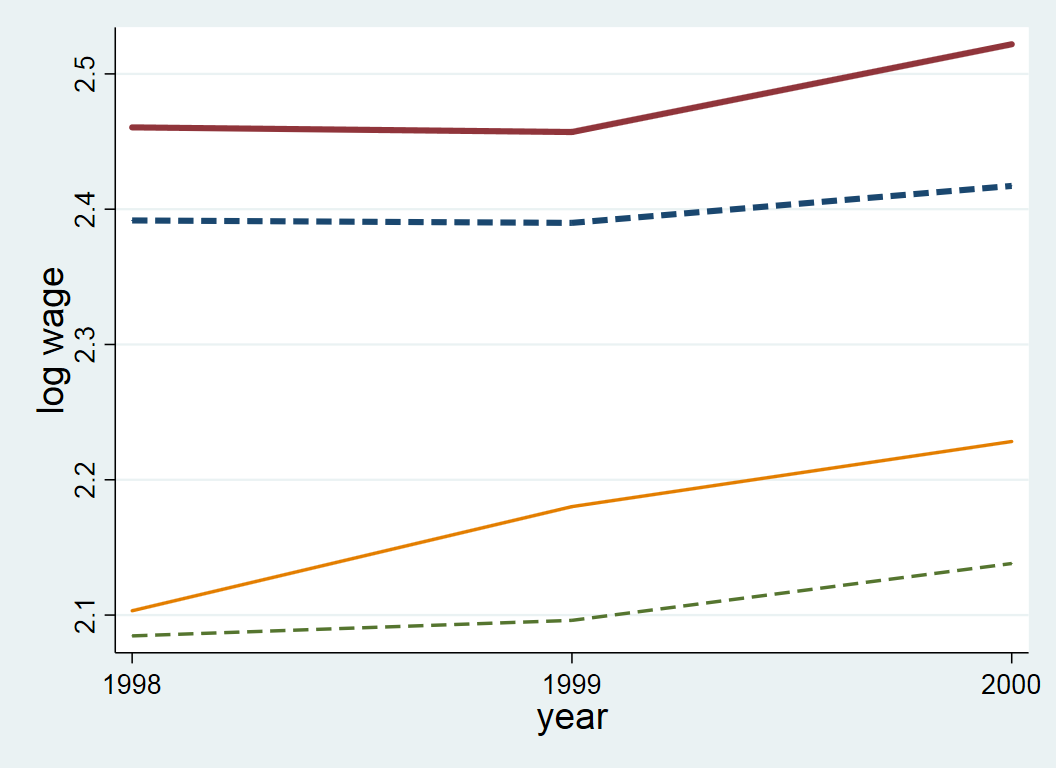}
        \label{Fig 2g}}\quad 
\subfigure[$\tau=80\%$ ]{
\includegraphics[width=7.5cm,height=3.75cm]{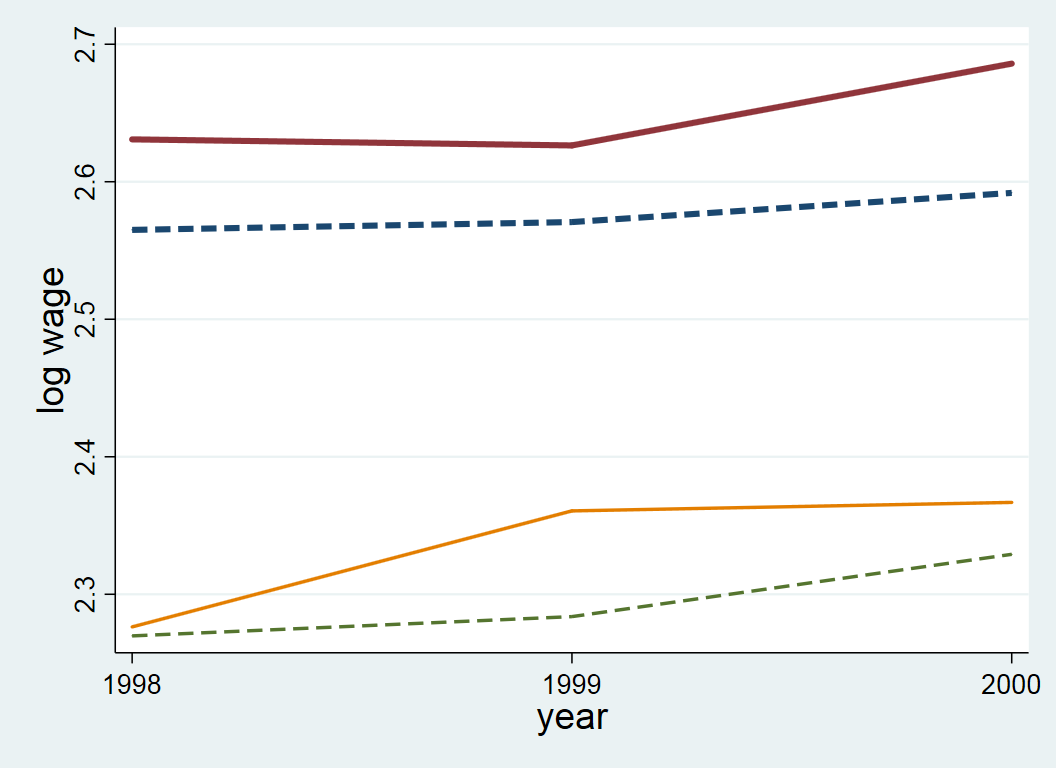}
        \label{Fig 2h}}
\par
\subfigure[$\tau=90\%$ ]{
\includegraphics[width=7.5cm,height=3.75cm]{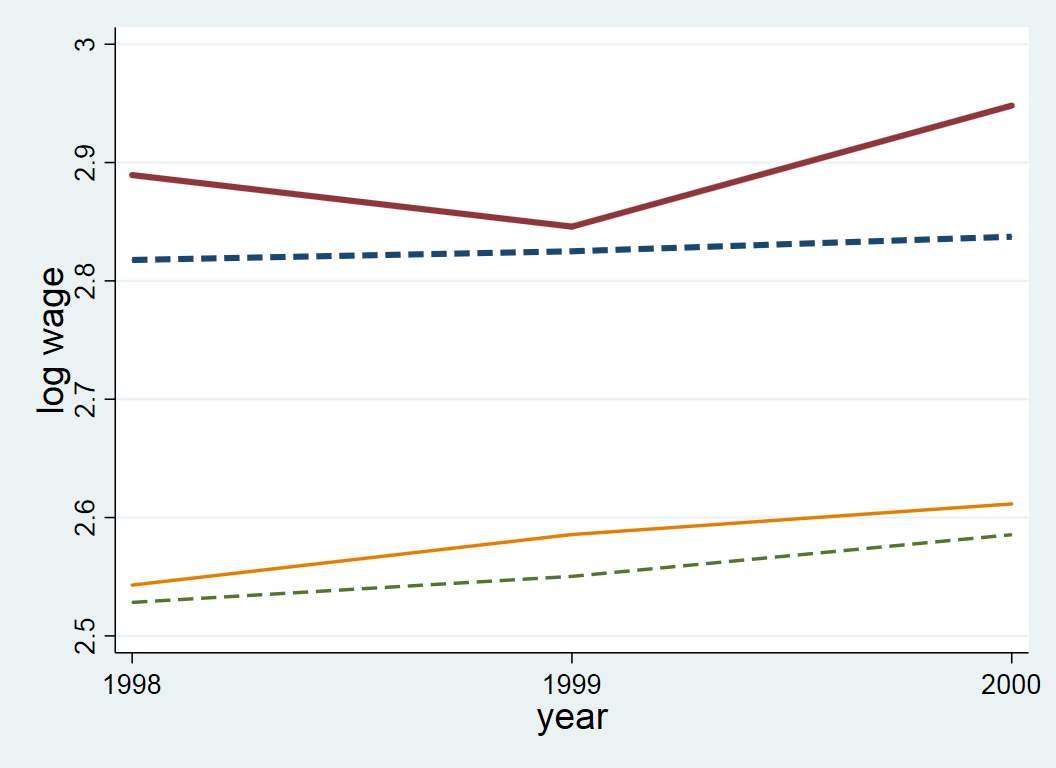}
        \label{Fig 2i}}
\caption{Corrected and Uncorrected Log Hourly Wage Quantiles by Gender
1998-2000 \citep{AB2017}. \textit{Note: male quantiles are always at the
top, female ones at the bottom (solid lines: uncorrected quantiles; dashed
lines: selection corrected quantiles)}}
\label{Fig 2}
\end{figure}

To provide the reader with a better illustration of the potential magnitudes
of selection into work, we replicate the predictions from the (estimated)
parametric conditional quantile functions from Figure 1 of %
\citet[][p.16]{AB2017} for the sub-periods 1995-1997 and 1998-2000, see
Figures \ref{Fig 1} and \ref{Fig 2}, respectively. In these pictures, solid
lines represent estimated uncorrected (for selection) conditional log-wage
quantile functions, while dashed lines are the ones corrected for sample
selection.\footnote{%
For the exact specification used, see \citet{AB2017}.} Throughout, female
quantile lines lie below the male quantile lines. The figures, which display
selection corrections based on a linear quantile regression model and
parametric selection correction, show little difference between the original
and the corrected lines for both males and females for the 1995-1997 period
(except for males at lower percentile levels), but more pronounced
differences for the subsequent 1998-2000 period. In terms of magnitude, the
effect of correction appears to be generally bigger for males than for
females in both subperiods.

Turning to the test results in Table \ref{Table 1}, we see that while we
cannot find any evidence of selection during the 1995-1997 period for
females at conventional significance levels, there is some evidence for
males at the 10\% significance level. In fact, taking a closer look at the
results in Table \ref{Table 1}, we observe that rejection for males occurs
on the basis of the 10th percentile, which is in line with the graphical
evidence in Figure \ref{Fig 1}. Switching over to the right panel of Table %
\ref{Table 1}, however, we obtain a different picture: for females, $%
H_{0,q}^{(1)}$ is rejected at any conventional level, and rejection is most
pronounced at the 20th and the 30th percentile. On the other hand, we cannot
reject $H_{0,q}^{(1)}$ for males. This failure to reject $H_{0,q}^{(1)}$ for
males is in contrast to the graphical evidence in Figure \ref{Fig 2}, and
highlights the importance of formal testing under a more flexible
specification.

To determine whether misspecification may be the cause of rejection, we perform
the second test for males in the 1995-1997 period, and for females in the
1998-2000 period (see Table \ref{Table 2}). Turning to the results, we
strongly reject the null of no misspecification of the conditional quantile
function $q_{\tau}(x_{i})$ for males, but fail to reject that null for
females at any conventional levels. Both results appear to be robust to
different choices of $\delta$ and $h_{p}$. Thus, under the assumption that
out-of-work income is a valid instrument and indeed selection enters outcome
as postulated in Equation (\ref{EQ1a}), our test results suggest that there
is evidence for selection among females for the 1998-2000, but not for
males. In fact, what appears to be selection among males during the
1995-1997 period could actually be attributed to misspecification of the
conditional quantile function.\footnote{%
In a related paper, \citet{K2010} tested for the validity of the same
instrumental variable (but for the interaction with marital status) in a
similar data set on the basis of the UK Family Expenditure Survey used by %
\citet{BGIM2007}. Although his test results are not directly informative
here as his test is run on a much coarser set of covariates $x_{i}$ not
including e.g. regional, martial, or family information, his evidence
suggested that the conditional independence of the instrument and outcome
(given $x_{i}$ and selection $s_{i}=1$) may indeed be violated for some
sub-groups (in particular, younger males with moderate levels of education).
Thus, rejection in the second test for males could indeed be related to this
feature.}

\begin{table}[H]
\begin{center}
\begin{tabular}{|c|c|c|}
\hline\hline
\multicolumn{3}{|c|}{Test 1 - 1995-1997} \\ \hline
& Males & Females \\ \hline
Statistic & $0.050 $ & $0.041 $ \\ 
10\% & $0.050 $ & $0.015 $ \\ 
20\% & $0.037 $ & $0.030 $ \\ 
30\% & $0.031 $ & $0.041 $ \\ 
40\% & $0.030 $ & $0.032 $ \\ 
50\% & $0.040 $ & $0.034 $ \\ 
60\% & $0.034 $ & $0.036 $ \\ 
70\% & $0.040 $ & $0.027 $ \\ 
80\% & $0.032 $ & $0.025 $ \\ 
90\% & $0.022 $ & $0.019 $ \\ \hline
90\%-CV & $0.050 $ & $0.049 $ \\ 
95\%-CV & $0.053 $ & $0.053 $ \\ 
P-Value & $0.10 $ & $0.32 $ \\ \hline
\# obs & $7623 $ & $7761 $ \\ \hline\hline
\end{tabular}%
\quad 
\begin{tabular}{|c|c|c|}
\hline\hline
\multicolumn{3}{|c|}{Test 1 - 1998-2000} \\ \hline
& Males & Females \\ \hline
Statistic & $0.045 $ & $0.049 $ \\ 
10\% & $0.033 $ & $0.021 $ \\ 
20\% & $0.039 $ & $0.042 $ \\ 
30\% & $0.044 $ & $0.049 $ \\ 
40\% & $0.041 $ & $0.031 $ \\ 
50\% & $0.045 $ & $0.032 $ \\ 
60\% & $0.039 $ & $0.047 $ \\ 
70\% & $0.035 $ & $0.042 $ \\ 
80\% & $0.033 $ & $0.026 $ \\ 
90\% & $0.032 $ & $0.032 $ \\ \hline
90\%-CV & $0.048 $ & $0.048 $ \\ 
95\%-CV & $0.049 $ & $0.051 $ \\ 
P-Value & $0.17 $ & $0.06 $ \\ \hline
\# obs & $6058 $ & $5931 $ \\ \hline\hline
\end{tabular}%
\end{center}
\caption{Results First Test. \textit{Note: Number of Bootstrap Replications
is 400.}}
\label{Table 1}
\end{table}

\begin{table}[H]
\begin{center}
\begin{tabular}{|c|c|c|c|c|}
\hline\hline
\multicolumn{5}{|c|}{Test 2} \\ \hline
\multicolumn{5}{|c|}{Males - 1995-1997} \\ \hline
& $\delta=.98$ & $\delta=.99$ & $\delta=1$ & $\delta=1$ \\ 
& $h_p =.02$ & $h_p=.02$ & $h_p=.02$ & $h_p=.01$ \\ \hline
Statistic & $9.013 $ & $9.828 $ & $8.488 $ & $6.185 $ \\ 
10\% & $5.961 $ & $7.118 $ & $5.899 $ & $4.439 $ \\ 
20\% & $5.961 $ & $8.358 $ & $7.377 $ & $5.716 $ \\ 
30\% & $7.097 $ & $9.397 $ & $8.033 $ & $7.758 $ \\ 
40\% & $8.614 $ & $9.828 $ & $8.488 $ & $7.554 $ \\ 
50\% & $9.013 $ & $9.165 $ & $7.704 $ & $7.101 $ \\ 
60\% & $8.267 $ & $8.800 $ & $7.132 $ & $7.064 $ \\ 
70\% & $8.170 $ & $7.082 $ & $5.185 $ & $6.245 $ \\ 
80\% & $7.310 $ & $5.939 $ & $4.422 $ & $5.309 $ \\ 
90\% & $6.166 $ & $4.267 $ & $3.203 $ & $3.043 $ \\ \hline
90\%-CV & $2.476 $ & $2.663 $ & $2.340 $ & $2.340 $ \\ 
95\%-CV & $2.682 $ & $2.936 $ & $2.589 $ & $2.589 $ \\ 
P-Value & $0.00 $ & $0.00 $ & $0.00 $ & $0.00 $ \\ \hline
\# obs & $583 $ & $491 $ & $375 $ & $191 $ \\ \hline\hline
\multicolumn{5}{|c|}{Females - 1998-2000} \\ \hline
& $\delta=.95$ & $\delta=.95$ & $\delta=.975$ & $\delta=.975$ \\ 
& $h_p =.075$ & $h_p=.05$ & $h_p=.05$ & $h_p=.025$ \\ \hline
Statistic & $1.249 $ & $1.431 $ & $2.346 $ & $1.894 $ \\ 
10\% & $0.492 $ & $0.323 $ & $0.176 $ & $1.071 $ \\ 
20\% & $0.312 $ & $0.024 $ & $1.032 $ & $1.829 $ \\ 
30\% & $0.664 $ & $0.054 $ & $0.247 $ & $0.172 $ \\ 
40\% & $0.696 $ & $0.108 $ & $1.209 $ & $0.719 $ \\ 
50\% & $1.249 $ & $0.736 $ & $1.557 $ & $1.304 $ \\ 
60\% & $1.125 $ & $1.144 $ & $2.061 $ & $1.894 $ \\ 
70\% & $0.477 $ & $1.431 $ & $2.346 $ & $1.799 $ \\ 
80\% & $0.739 $ & $0.310 $ & $0.993 $ & $0.919 $ \\ 
90\% & $0.913 $ & $0.591 $ & $0.522 $ & $0.361 $ \\ \hline
90\%-CV & $2.323 $ & $2.246 $ & $2.459 $ & $2.559 $ \\ 
95\%-CV & $2.528 $ & $2.453 $ & $2.706 $ & $2.765 $ \\ 
P-Value & $0.83 $ & $0.71 $ & $0.13 $ & $0.41 $ \\ \hline
\# obs & $669 $ & $343 $ & $100 $ & $42 $ \\ \hline\hline
\end{tabular}%
\end{center}
\caption{Results Second Test \textit{Note: Number of Bootstrap Replications
is 1,000}}
\label{Table 2}
\end{table}

\section{Conclusion}

\label{Conclusion}

This paper introduces two tests to detect sample selection
in conditional quantile functions, without imposing parametric assumptions
on either the outcome or the selection equation. The first test is an omitted predictor test, with the
estimated propensity score as omitted predictor. This test may be of particular interest to practitioners who rely on the three step estimator of \citet{AB2017}, but can check for the presence of sample selection without relying on functional form assumptions. Another feature of this test is that it may be carried out selecting the bandwidth parameters in a data-driven manner.  

As with any omnibus test,
rejection in the first test can be due to either selection or to the
omission of a predictor which is correlated with the estimated propensity
score. Since selection and misspecification have very different implications
for the estimation of nonparametric (conditional) quantile functions, we aim
at disentangling the two if we reject in the first step. That is, after
rejection in the first test we proceed to the second test using only observations with
(estimated) propensity score close to one. A rejection in this case
indicates the presence of misspecification, possibly in conjunction with
selection. 

Importantly, neither of our tests rely on a continuous exclusion restriction and may be executed over a (compact) subset of quantile ranks. Indeed, Monte Carlo evidence suggests that in particular the first test has good finite sample properties when the exclusion restriction is discrete, which renders it attractive for applied work.  In Appendix B, we also develop the first test for nonparametric conditional mean functions. Simulations confirm the favorable performance in this case, too.

In our empirical
illustration, we test for sample selection in log hourly wages of females
and males in the UK using data from the UK Family Expenditure Survey. We
find evidence for selection among females for the 1998-2000, but not for
males. In fact, what appears to be selection among males during the
1995-1997 period may actually be attributed to misspecification of the
conditional quantile function.

\singlespacing
\footnotesize
\section*{Appendix A}\label{Appendix A}

\noindent \textbf{Nonparametric Estimators}: \medskip

\noindent As detailed in the main text, to estimate the conditional quantile
function at some point $x_{i}=x$, we use an r-th order local polynomial
estimator based on the standard `check type' objective function: 
\begin{equation*}
l_{\tau }(v)=2v\left( \tau -1\left\{ v\leq 0\right\} \right) .\footnote{%
In the discrete case, we can set up a local constant estimator in the
direction of the discrete elements.}
\end{equation*}%
The local polynomial estimator is then given by: 
\begin{equation}
\widehat{b}_{h_{x}}\left( \tau ,x\right) =\arg \min_{b}\frac{1}{%
nh_{x}^{d_{x}}}\sum_{i=1}^{n}l_{\tau }\left( y_{i}-b_{0}-\sum_{0\leq |%
\mathbf{t}|\leq r}b_{\mathbf{t}}\left( x_{i}-x\right) ^{\mathbf{t}}\right)
s_{i}\mathbf{K}\left( \frac{x_{i}-x}{h_{x}}\right)   \label{LPQ}
\end{equation}%
is an estimator of $b_{h_{x}}^{\dag }\left( \tau ,x\right) $ with 
\begin{equation}
b^{\dag }\left( \tau ,x\right) =\arg \min_{b}\lim_{n\rightarrow \infty}\frac{1}{nh_{x}^{d_{x}}}%
\sum_{i=1}^{n}\mathrm{E}\left[ l_{\tau }\left( y_{i}-b_{0}-\sum_{0\leq |%
\mathbf{t}|\leq r}b_{\mathbf{t}}\left( x_{i}-x\right) ^{\mathbf{t}}\right)
s_{i}\mathbf{K}\left( \frac{x_{i}-x}{h_{x}}\right) \right] .\footnote{%
We borrow notation from \citet{Masry1996} letting $\mathbf{t}=(t_{1},\ldots
,t_{d_{x}})^{\prime }$, $|\mathbf{t}|=\sum_{j=1}^{d_{x}}t_{j}$, and $%
\sum_{0\leq |\mathbf{t}|\leq r}=\sum_{j=0}^{r}\sum_{t_{1}=0}^{j}\ldots
\sum_{t_{d_{x}}=0}^{j}$.}  \label{LPQLimit}
\end{equation}%
Here, $\mathbf{K}(\cdot )$ denotes a $d_{x}$ dimensional product kernel. We
use $\widehat{q}_{\tau }(x)=\widehat{b}_{0,h_{x}}\left( \tau ,x\right) $,
the first element of $\widehat{b}_{h_{x}}\left( \tau ,x\right) $, and set $%
q_{\tau }^{\dag }(x)=b_{0}^{\dag }\left( \tau ,x\right) $. If the quantile
functions are estimated using only observations with propensity score close
to one, a corresponding estimator can be defined as: 
\begin{equation}
\widehat{\widetilde{b}}_{h_{x}}\left( \tau ,x\right) =\arg \min_{b}\frac{1}{%
nh_{x}^{d_{x}}h_{p}}\sum_{i=1}^{n}l_{\tau }\left( y_{i}-b_{0}-\sum_{0\leq |%
\mathbf{t}|\leq r}b_{\mathbf{t}}\left( x_{i}-x\right) ^{\mathbf{t}}\right)
s_{i}\mathbf{K}\left( \frac{x_{i}-x}{h_{x}}\right) K\left( \frac{\widehat{p}%
_{i}-\delta _{n}}{h_{p}}\right) .  \label{LPQP1}
\end{equation}%
Finally, the kernel estimator of $F_{p|x,u_{\tau },s=1}\left(
p|x_{i},0,s_{i}=1\right) $ used to construct the bootstrap statistic of the
first test is given by: 
\begin{equation}
\widehat{F}_{p|x,u_{\tau },s=1}\left( p|x_{i,}0,s_{i}=1\right) =\frac{\frac{1%
}{nh_{F}^{d_{x}+1}}\sum_{j=1}^{n}s_{j}1\left\{ \widehat{p}_{j}\leq p\right\}
K\left( \frac{\widehat{u}_{j,\tau }-0}{h_{F}}\right) \mathbf{K}\left( \frac{%
x_{j}-x_{i}}{h_{F}}\right) }{\frac{1}{nh_{F}^{d_{x}+1}}\sum_{j=1}^{n}s_{j}K%
\left( \frac{\widehat{u}_{j,\tau }-0}{h_{F}}\right) \mathbf{K}\left( \frac{%
x_{j}-x_{i}}{h_{F}}\right) }.  \label{EQCDF}
\end{equation}%
\medskip

\noindent \textbf{Auxiliary Lemmas}: 

\noindent In the following, let $E_{\mathcal{S}%
_{n}}[\cdot ]$ denote the expectation operator conditional on the actual
sample realizations. Moreover, since $1\{\underline{p}\leq \widehat{p}%
_{i}\leq \overline{p}\}=1\{\widehat{p}_{i}\leq \overline{p}\}-1\{\widehat{p}%
_{i}\leq \underline{p}\}$, we will ignore the part of the statistic which
involves $1\{\widehat{p}_{i}\leq \underline{p}\}$ in the sequel. 
\medskip

\noindent \textbf{Lemma 1}: Let Assumptions \textbf{A.1}-\textbf{A.5} and 
\textbf{A.Q} hold. Moreover, let $h_{x}$ denote a deterministic bandwidth
sequence that satisfies $h_{x}\rightarrow 0$ as $n\rightarrow \infty $. If
as $n\rightarrow \infty $, $(nh_{x}^{2d_{x}})/\log n\rightarrow \infty $, $%
nh_{x}^{2r}\rightarrow 0$, then uniformly over $\mathcal{T}$, $\mathcal{X}$, and $\mathcal{P}$:

\noindent \textbf{(i)} Under $H_{0,q}^{(1)}$: 
\begin{eqnarray*}
&&\sqrt{n}E_{\mathcal{S}_{n}}\Biggl[s_{i}\left( 1\{\widehat{u}_{\tau
}(x_{i})\leq 0\}-\tau \right) \Pi _{j=1}^{d_{x}}1\{\underline{x}_{j}<x_{j,i}<\overline{x}_{j}\}1\{%
\widehat{p}_{i}\leq \overline{p}\} \\
&&-s_{i}\left( 1\{u_{\tau }(x_{i})\leq
0\}-\tau \right) \Pi _{j=1}^{d_{x}}1\{\underline{x}_{j}<x_{j,i}<\overline{x}%
_{j}\}1\{p_{i}\leq \overline{p}\}\Biggr] \\
&=&\frac{-1}{\sqrt{n}}\sum_{j=1}^{n}F_{p|x,u_{\tau },s=1}(\overline{p}%
|x_{j},0,s_{j}=1)\left( s_{j}(1\{u_{\tau }(x_{j})\leq 0\}-\tau )\right) \Pi
_{l=1}^{d_{x}}1\{\underline{x}_{l}<x_{l,j}<\overline{x}_{l}\}+o_{p}(1).
\end{eqnarray*}%
\noindent \textbf{(ii)} Under $H_{A,q}^{(1)}$: 
\begin{eqnarray*}
&&\sqrt{n}E_{\mathcal{S}_{n}}\Biggl[s_{i}\left( 1\{\widehat{u}_{\tau
}(x_{i})\leq 0\}-\tau \right)\times \Pi _{j=1}^{d_{x}}1\{\underline{x}_{j}<x_{j,i}<\overline{x}_{j}\}1\{%
\widehat{p}_{i}\leq \overline{p}\}-s_{i}\left( 1\{u_{\tau }(x_{i})\leq
0\}-\tau \right)   \\
&&\Pi _{j=1}^{d_{x}}1\{\underline{x}_{j}<x_{j,i}<\overline{x}%
_{j}\}1\{p_{i}\leq \overline{p}\}\Biggr] =O_{p}\left( \frac{\ln(n)}{\sqrt{h_{x}^{d_{x}}}}\right). 
\end{eqnarray*}%
\medskip

\noindent \textbf{Lemma 2}: Let Assumptions \textbf{A.1}, \textbf{A.3}, 
\textbf{A.5}, \textbf{A.6}, \textbf{A.7}, \textbf{A.8}, \textbf{A.9}, and 
\textbf{A.Q} hold. If as $n\rightarrow \infty$, $(nh_{x}^{2d_{x}})/\log
n\rightarrow \infty$, $nh_{x}^{2r}\rightarrow 0,$ $H\rightarrow 0,$ $%
H/h_{p}\rightarrow \infty $, $nh_{p}H^{2-\eta } \rightarrow 0$, and $%
nh_{p}H^{\eta }\rightarrow \infty$,
then uniformly over $\mathcal{T}$:

\noindent \textbf{(i)}
\begin{eqnarray*}
&&\frac{n}{\sqrt{nh_{p}H^{\eta } 
}}E_{\mathcal{S}_{n}}\left[ s_{i}(1\{\widehat{u}_{\tau }(x_{i})\leq 0\}-\tau
) K\left( \frac{\widehat{p}%
_{i}-\delta }{h_{p}}\right) \right.  \\
&&-\left. s_{i}(1\{u_{\tau }(x_{i})\leq 0\}-\tau ) K\left( \frac{p_{i}-\delta }{h_{p}}\right) \right] =o_{p}(1)
\end{eqnarray*}%
\noindent \textbf{(ii)} 
\begin{eqnarray*}
&&\frac{1}{\sqrt{nh_{p}H^{\eta } }}\sum_{i=1}^{n}\left( s_{i}(1\{\widehat{u}_{\tau }(x_{i})\leq
0\}-\tau ) K\left( \frac{\widehat{%
p}_{i}-\delta }{h_{p}}\right) \right.  \\
&&-\left. E_{\mathcal{S}_{n}}\left[ s_{i}(1\{\widehat{u}_{\tau }(x_{i})\leq
0\}-\tau ) K\left( \frac{\widehat{%
p}_{i}-\delta }{h_{p}}\right) \right] \right) \\
&&-\frac{1}{\sqrt{nh_{p}H^{\eta } }}\sum_{i=1}^{n}\left( s_{i}\left( 1\{u_{\tau }(x_{i})\leq
0\}-\tau \right) K\left( \frac{%
p_{i}-\delta }{h_{p}}\right) \right.  \\
&&-\left. E_{\mathcal{S}_{n}}\left[ s_{i}\left( 1\{u_{\tau }(x_{i})\leq
0\}-\tau \right)  K\left( \frac{%
p_{i}-\delta }{h_{p}}\right) \right] \right)  \\
=o_{p}(1).
\end{eqnarray*}%

\medskip

\noindent \textbf{Lemma 3}:
Let Assumptions \textbf{A.1}, \textbf{A.3}, 
\textbf{A.5}, \textbf{A.6}, \textbf{A.7}, \textbf{A.8}, \textbf{A.9}, and 
\textbf{A.Q} hold. If as $n\rightarrow \infty$, $(nh_{x}^{2d_{x}})/\log
n\rightarrow \infty$, $nh_{x}^{2r}\rightarrow 0,$ $H\rightarrow 0,$ $%
H/h_{p}\rightarrow \infty $, $nh_{p}H^{2-\eta } \rightarrow 0$, and $%
nh_{p} H^{\eta }\rightarrow \infty$,
then:

\noindent \textbf{(i)} Under $H_{0,q}^{(2)}$ and $H_{A,q}^{(2)}$, pointwise in $\tau\in\mathcal{T}$:
\begin{equation*}
\frac{\sum_{i=1}^{n}s_{i}(1\{u_{\tau }(x_{i})\leq 0\}-G_{u_{\tau}}(\tau
,1))K\left( \frac{p_{i}-\delta }{h_{p}}\right)}{\sqrt{\int
K^{2}(v)dv\sum_{i=1}^{n}s_{i}(1\{u_{\tau }(x_{i})\leq 0\}-G_{u_{\tau}}(\tau
,1))^{2} K\left( \frac{p_{i}-\delta }{h_{p}}\right)}} \stackrel{d}{\rightarrow}N(0,1).
\end{equation*}%

\noindent \textbf{(ii)} Under $H_{0,q}^{(2)}$, uniformly in $\tau\in\mathcal{T}$:
\begin{equation*}
\frac{1}{\sqrt{nh_{p}H^{\eta } }}\sum_{i=1}^{n}s_{i}(G_{u_{\tau}}\left( \tau ,p_{i}\right) -\tau
) K\left( \frac{p_{i}-\delta }{%
h_{p}}\right) =o_{p}(1)
\end{equation*}%

\medskip

\noindent \textbf{Proofs of Theorem 1 and 2}: 
\medskip

\noindent \textbf{Proof of Theorem 1}: \medskip

\textbf{(i)} Start by noting that we can decompose $Z_{1,n}\left( \tau ,\underline{x},\overline{x},\overline{p}\right)$ into the sum of the following three terms:
\begin{equation*}
I_{n}=\frac{1}{\sqrt{n}}\sum_{i=1}^{n}s_{i}(1\{u_{\tau }(x_{i})\leq 0\}-\tau
)\Pi _{j=1}^{d_{x}}1\{\underline{x}_{j}<x_{j,i}<\overline{x}%
_{j}\}1\{p_{i}\leq \overline{p}\}, 
\end{equation*}
\begin{eqnarray*}
II_{n}&=&+\sqrt{n}E_{\mathcal{S}_{n}}\Biggl[s_{i}\left( 1\{\widehat{u}_{\tau
}(x_{i})\leq 0\}-\tau \right) \Pi _{j=1}^{d_{x}}1\{\underline{x}_{j}<x_{j,i}<\overline{x}_{j}\}1\{%
\widehat{p}_{i}\leq \overline{p}\} \\
&&-s_{i}\left( 1\{u_{\tau }(x_{i})\leq
0\}-\tau \right) \Pi _{j=1}^{d_{x}}1\{\underline{x}_{j}<x_{j,i}<\overline{x}%
_{j}\}1\{p_{i}\leq \overline{p}\}\Biggr], 
\end{eqnarray*}
and
\begin{eqnarray*}
III_{n}&=&\frac{1}{\sqrt{n}}\sum_{i=1}^{n}\Biggl\{s_{i}\left( 1\{\widehat{u}_{\tau
}(x_{i})\leq 0\}-\tau \right) \Pi _{j=1}^{d_{x}}1\{\underline{x}_{j}<x_{j,i}<%
\overline{x}_{j}\}1\{\widehat{p}_{i}\leq \overline{p}\} \\
&&-E_{\mathcal{S}_{n}}%
\Biggl[s_{i}\left( 1\{\widehat{u}_{\tau }(x_{i})\leq 0\}-\tau \right)  \Pi _{j=1}^{d_{x}}1\{\underline{x}_{j}<x_{j,i}<\overline{x}_{j}\}1\{%
\widehat{p}_{i}\leq \overline{p}\}\Biggr]\Biggr\} \\
&&-\frac{1}{\sqrt{n}}\sum_{i=1}^{n}\Biggl\{s_{i}\left( 1\{u_{\tau
}(x_{i})\leq 0\}-\tau \right) \Pi _{j=1}^{d_{x}}1\{\underline{x}_{j}<x_{j,i}<%
\overline{x}_{j}\}1\{p_{i}\leq \overline{p}\}\\
&&-E_{\mathcal{S}_{n}}\Biggl[%
s_{i}\left( 1\{u_{\tau }(x_{i})\leq 0\}-\tau \right) \Pi _{j=1}^{d_{x}}1\{\underline{x}_{j}<x_{j,i}<\overline{x}%
^{j}\}1\{p_{i}\leq \overline{p}\}\Biggr]\Biggr\} .
\end{eqnarray*}%
From Lemma 1(i),%
\begin{equation*}
II_{n}=\frac{-1}{\sqrt{n}}\sum_{j=1}^{n}F_{p|x,u_{\tau },s=1}(\overline{p}%
|x_{j},0,s_{j}=1)\left( s_{j}(1\{u_{\tau }(x_{j})\leq 0\}-\tau )\right) \Pi
_{l=1}^{d_{x}}1\{\underline{x}_{l}<x_{l,j}<\overline{x}_{l}\}+o_{p}(1),
\end{equation*}%
where the $o_{p}(1)$ term holds uniformly over $\mathcal{T}$, $\mathcal{X}$, and $\mathcal{P}$. As for $III_{n}$, we first, we apply Lemma A.1 of %
\citet{EJCL2014} to the function classes $\mathcal{F}_{1}\equiv
\{f_{1}(s,\tau ,x)=s(1\{u_{\tau }(x)\leq 0\}-\tau )\Pi _{j=1}^{d_{x}}1\{%
\underline{x}_{j}<x_{j}<\overline{x}_{j}\}:\{\tau ,\underline{x},\overline{x}%
\}\in \mathcal{T}\times \mathcal{X}\}$ and $\mathcal{F}_{2}\equiv
\{f_{2}(p(z))=1\{p(z)\leq \overline{p}\}:\{p\}\in \mathcal{P}\}$ to conclude
that the product class $\mathcal{F}_{1}\times \mathcal{F}_{2}$ is Donsker.
Then, by Lemma A.3 in \citet{EJCL2014}, it follows that $III_{n}=o_{p}(1)$
uniformly over $\mathcal{T}$, $\mathcal{X}$, and $\mathcal{P}$. Thus,%
\begin{eqnarray*}
Z_{1,n}\left( \tau ,\underline{x},\overline{x},\overline{p}\right)  &=&\frac{%
1}{\sqrt{n}}\sum_{i=1}^{n}(1\{u_{\tau }(x_{i})\leq 0\}-\tau )\Pi
_{j=1}^{d}1\{\underline{x}_{j}<x_{j,i}<\overline{x}_{j}\}s_{i}\Biggl(%
1\{p_{i}\leq \overline{p}\} \\
&&-F_{p|x,u_{\tau },s=1}(\overline{p}|x_{i},0,s_{i}=1)\Biggr)+o_{p}(1),
\end{eqnarray*}%
with the $o_{p}(1)$ term holding uniformly in $\tau $,$p$, and $\left( 
\underline{x},\overline{x}\right) $. Thus, the process $Z_{1,n}^{q}\left(
\tau ,\underline{x},\overline{x},\underline{p},\overline{p}\right) $
converges weakly in $l^{\infty }(\mathcal{T}\times \mathcal{X}\times 
\mathcal{P})$, the Banach space of real bounded functions on $\mathcal{T}%
\times \mathcal{X}\times \mathcal{P}$, with covariance kernel 
\begin{eqnarray*}
&&cov\left( Z_{1,n}^{q}\left( \tau ,\underline{x},\overline{x},\underline{p},%
\overline{p}\right) ,Z_{1,n}^{q}\left( \tau ^{\prime },\underline{x}^{\prime
},\overline{x}^{\prime },\underline{p}^{\prime },\overline{p}^{\prime
}\right) \right)  \\
&=&E\left[ (1\{u_{\tau }(x_{i})\leq 0\}-\tau )\Pi _{j=1}^{d}1\{\underline{x}%
_{j}<x_{j,i}<\overline{x}_{j}\}\left( s_{i}(1\{p_{i}\leq \overline{p}%
\}-1\{p_{i}\leq \underline{p}\})\right. \right.  \\
&&\left. \left. -(F_{p|x,u_{\tau },s=1}(\overline{p}%
|x_{i},0,s_{i}=1)-F_{p|x,u_{\tau },s=1}(\underline{p}|x_{i},0,s_{i}=1))\Pr
(s_{i}=1|x_{i})\right) \right.  \\
&&\left. (1\{u_{\tau ^{\prime }}(x_{i})\leq 0\}-\tau ^{\prime })\Pi
_{j=1}^{d}1\{\underline{x}_{j}^{\prime }<x_{j,i}<\overline{x}_{j}^{\prime
}\}\left( s_{i}(1\{p_{i}\leq \overline{p}^{\prime }\}-1\{p_{i}\leq 
\underline{p}^{\prime }\})\right. \right.  \\
&&\left. \left. -(F_{p|x,u_{\tau },s=1}(\overline{p}^{\prime
}|x_{i},0,s_{i}=1)-F_{p|x,u_{\tau },s=1}(\underline{p}^{\prime
}|x_{i},0,s_{i}=1))\Pr (s_{i}=1|x_{i})\right) \right] 
\end{eqnarray*}%
As an immediate consequence, we also obtain the weak convergence of any
continuous functional and so: 
\begin{equation*}
Z_{1,n}^{q}\Rightarrow Z_{1}^{q}.
\end{equation*}%
\medskip 

\noindent \textbf{(ii)} Given Lemma 1(ii):%
\begin{eqnarray}
&&Z_{1,n}\left( \tau ,\underline{x},\overline{x},\overline{p}\right)   \notag
\\
&=&\frac{1}{\sqrt{n}}\sum_{i=1}^{n}\left( (1\{u_{\tau }(x_{i})\leq 0\}-\tau
)\Pi _{j=1}^{d}1\{\underline{x}_{j}<x_{j,i}<\overline{x}_{j}\}s_{i}\Biggl(%
1\{p_{i}\leq \overline{p}\}-F_{p|x,u_{\tau },s=1}(\overline{p}%
|x_{i},0,s_{i}=1)\Biggr)\right.   \notag \\
&&\left. -\mathrm{E}\left[ (1\{u_{\tau }(x_{i})\leq 0\}-\tau )\Pi
_{j=1}^{d}1\{\underline{x}_{j}<x_{j,i}<\overline{x}_{j}\}s_{i}\left(
1\{p_{i}\leq \overline{p}\}-F_{p|x,u_{\tau },s=1}(\overline{p}%
|x_{i},0,s_{i}=1)\right) \right) \right]   \notag \\
&&+\sqrt{n}\mathrm{E}\left[( 1\{u_{\tau }(x_{i})\leq 0\}-\tau )\Pi
_{j=1}^{d}1\{\underline{x}_{j}<x_{j,i}<\overline{x}_{j}\}s_{i}\left(
1\{p_{i}\leq \overline{p}\}-F_{p|x,u_{\tau },s=1}(\overline{p}%
|x_{i},0,s_{i}=1)\right) \right]   \notag \\
&&+O_{p}\left( \frac{\ln n}{\sqrt{h_{x}^{d_{x}}}}\right)   \label{Z1A}
\end{eqnarray}%
with the $O_{p}$ term holding uniformly over $\mathcal{T}$, $\mathcal{X}$, and $\mathcal{P}$. The statement then follows as the first term on
the right hand side (RHS) of (\ref{Z1A}) weakly converges, and $\frac{\ln n}{\sqrt{h_{x}^{d_{x}}%
}}$ diverges at a rate slower than $\sqrt{n}$ given that $%
nh_{x}^{2d_{x}}\rightarrow \infty$. \qedsymbol

\medskip

\noindent \textbf{Proof of Theorem 2}:

\noindent \textbf{(i)} Given Assumption A.6,%
\begin{eqnarray}
&&Z_{2,n,\delta}^{q}\left( \tau \right)    \label{Z2q} \\
&=&\frac{ZN_{2,n,\delta}^{q}\left( \tau\right) }{%
\sqrt{ \int K^{2}(v)dv\frac{1}{\sqrt{nh_{p}H^{\eta }}}\sum_{i=1}^{n}s_{i}1\{\widehat{u}%
_{\tau }(x_{i})\leq 0\}-\tau )^{2} K\left( \frac{\widehat{p}_{i}-\delta }{h_{p}}\right) }} \notag
\end{eqnarray}%
where:
\begin{equation*}
ZN_{2,n,\delta}^{q}\left( \tau \right) \equiv \frac{1}{%
\sqrt{nh_{p}H^{\eta }}}%
\sum_{i=1}^{n}s_{i}(1\{\widehat{u}_{\tau }(x_{i})\leq 0\}-\tau ) K\left( \frac{\widehat{p}_{i}-\delta }{%
h_{p}}\right) .
\end{equation*}%
Now:%
\begin{eqnarray*}
&&ZN_{2,n,\delta}^{q}\left( \tau \right)  \\
&=&\frac{1}{\sqrt{nh_{p}H^{\eta } }}\sum_{i=1}^{n}s_{i}(1\{u_{\tau }(x_{i})\leq 0\}-\tau
) K\left( \frac{p_{i}-\delta }{%
h_{p}}\right)  \\
&&+\frac{n}{\sqrt{nh_{p}H^{\eta } }}E_{\mathcal{S}_{n}}\left[ s_{i}(1\{\widehat{u}_{\tau
}(x_{i})\leq 0\}-\tau ) K\left( 
\frac{\widehat{p}_{i}-\delta }{h_{p}}\right) \right.  \\
&&-\left. s_{i}(1\{u_{\tau }(x_{i})\leq 0\}-\tau ) K\left( \frac{p_{i}-\delta }{h_{p}}\right) \right]  \\
&&-\frac{1}{\sqrt{nh_{p}H^{\eta } }}\sum_{i=1}^{n}\left( s_{i}\left( 1\{u_{\tau }(x_{i})\leq
0\}-\tau \right) K\left( \frac{%
p_{i}-\delta }{h_{p}}\right) \right.  \\
&&-\left. E_{\mathcal{S}_{n}}\left[ s_{i}\left( 1\{u_{\tau }(x_{i})\leq
0\}-\tau \right) K\left( \frac{%
p_{i}-\delta }{h_{p}}\right) \right] \right)  \\
&&+\frac{1}{\sqrt{nh_{p}H^{\eta } }}\sum_{i=1}^{n}\left( s_{i}(1\{\widehat{u}_{\tau }(x_{i})\leq
0\}-\tau ) K\left( \frac{\widehat{%
p}_{i}-\delta }{h_{p}}\right) \right.  \\
&&-\left. E_{\mathcal{S}_{n}}\left[ s_{i}(1\{\widehat{u}_{\tau }(x_{i})\leq
0\}-\tau )K\left( \frac{\widehat{%
p}_{i}-\delta }{h_{p}}\right) \right] \right)  \\
&=&I_{n}+II_{n}+III_{n}
\end{eqnarray*}%
Given Lemma 2(i)-(ii), $II_{n}$ and $III_{n}$ are $o_{p}(1)$ uniformly in $%
\tau \in \mathcal{T}$. Thus, it suffices to derive the limiting distribution
of $I_{n}$. Recalling that $G_{u_{\tau}}\left( \tau ,p_{i}\right) =\Pr
(y_{i}\leq q_{\tau }(x_{i})|p_{i})$, $I_{n}$ reads as: 
\begin{eqnarray}
I_{n} &=&\frac{1}{\sqrt{nh_{p}H^{\eta } }}\sum_{i=1}^{n}s_{i}(1\{u_{\tau }(x_{i})\leq
0\}-G_{u_{\tau}}\left( \tau ,p_{i}\right) ) K\left( \frac{p_{i}-\delta }{h_{p}}\right)   \notag \\
&&+\frac{1}{\sqrt{nh_{p}H^{\eta } }}\sum_{i=1}^{n}s_{i}(G_{u_{\tau}}\left( \tau ,p_{i}\right) -\tau
)K\left( \frac{p_{i}-\delta }{%
h_{p}}\right)   \label{TH21} \\
&=&I_{1,n}+I_{2,n}  \notag
\end{eqnarray}%
The first term drives the limiting the distribution, while the second
term can be thought of as bias since $G_{u_{\tau}}\left( \tau ,p\right) \rightarrow
\tau $ only as $p\rightarrow 1$. More specifically, by Lemma 3(i), $I_{1,n}$ satisfies a CLT for triangular arrays pointwise in $\tau\in\mathcal{T}$, while by Lemma 3(ii), $I_{2,n}=o_{p}\left( I_{1,n}\right)$ uniformly in $\tau \in\mathcal{T}$. We now need to study the
denominator in (\ref{Z2q}): 
\begin{eqnarray*}
&&\left( \int K^{2}(v)dv\right) \frac{1}{nh_{p}H^{\eta }}\sum_{i=1}^{n}s_{i}\left( 1\{%
\widehat{u}_{\tau }(x_{i})\leq 0\}-\tau \right) ^{2}K\left( \frac{\widehat{p}_{i}-\delta }{h_{p}}\right)  \\
&=&\left( \int_{-1}^{1}K(v)^{2}dv\right) \tau (1-\tau )g_{p}(1)+o_{p}(1)=O(1)
\end{eqnarray*}%
uniformly over $\mathcal{T}$. The covariance kernel of the statistic is
therefore given by: 
\begin{eqnarray*}
&&\text{cov}\left( Z_{2,n,\delta}^{q}\left( \tau \right)
,Z_{2,n,\delta}^{q}\left( \tau ^{\prime }\right) \right) 
\\
&=&\lim_{n\rightarrow \infty }E\left[ \frac{\sum_{i=1}^{n}s_{i}(1\{u_{\tau
}(x_{i})\leq 0\}-\tau )K\left( \frac{%
p_{i}-\delta }{h_{p}}\right) }{\left( \int K^{2}(v)dv\right)
\sum_{i=1}^{n}s_{i}\left( 1\{u_{\tau }(x_{i})\leq 0\}-\tau \right)
^{2}K\left( \frac{p_{i}-\delta }{h_{p}}%
\right) }\right.  \\
&&\left. \frac{\sum_{i=1}^{n}s_{i}(1\{u_{\tau ^{\prime }}(x_{i})\leq
0\}-\tau ^{\prime })K\left( \frac{%
p_{i}-\delta }{h_{p}}\right) }{\left( \int K^{2}(v)dv\right)
\sum_{i=1}^{n}s_{i}\left( 1\{u_{\tau ^{\prime }}(x_{i})\leq 0\}-\tau
^{\prime }\right) ^{2}K\left( \frac{%
p_{i}-\delta }{h_{p}}\right) }\right] .
\end{eqnarray*}%
Finally, by Lemma A.1 and B.3 of \citet{EJCL2014}, we can also conclude that
numerator and denominator of $Z_{2,n,\delta}^{q}\left( \tau \right) $ are Donsker, and hence by Theorem 2.10.6 of %
\citet{VW1996}, that $Z_{2,n,\delta}^{q}\left( \tau \right) $ is Donsker as well. Thus, it follows that 
\begin{eqnarray*}
&&Z_{2,n,\delta}^{q}\left( \tau \right)  \\
&=&\frac{ZN_{2,n,\delta}^{q}\left( \tau \right) }{%
\left( \left( \int K^{2}(v)dv\right) \frac{1}{nh_{p}H^{\eta }}\sum_{i=1}^{n}s_{i}(1\{%
\widehat{u}_{\tau }(x_{i})\leq 0\}-\tau )^{2}K\left( \frac{\widehat{p}_{i}-\delta }{h_{p}}\right) \right) ^{1/2}}
\end{eqnarray*}%
converges weakly in $l^{\infty }(\mathcal{T})$, and by continuous mapping,
so does the functional 
\begin{equation*}
\sup_{\tau \in \mathcal{T}}\left\vert Z_{2,n,\delta}^{q}\left( \tau \right) \right\vert 
\end{equation*}%
as postulated in the statement of part (i). \medskip 

\noindent \textbf{(ii)} Now, if there is an omitted relevant
regressor, given Asumption A.7,%
\begin{eqnarray*}
&&\frac{1}{nh_{p}H^{\eta }}%
\sum_{i=1}^{n}s_{i}(G_{u_{\tau}}\left( \tau ,p_{i}\right) -\tau )
K\left( \frac{\widehat{p}_{i}-\delta }{h_{p}}\right) \\
&=&\underbrace{\lim_{n\rightarrow \infty }\frac{1}{h_{p}H^{\eta }}E\left[ s_{i}(G_{u_{\tau}}\left( \tau
,p_{i}\right) -\tau )K\left( \frac{%
p_{i}-\delta }{h_{p}}\right) \right] }_{\neq 0}+o_{p}(1)
\end{eqnarray*}%
and%
\begin{eqnarray*}
&&\frac{1}{nh_{p}H^{\eta }}%
\sum_{i=1}^{n}s_{i}(1\{\widehat{u}_{\tau }(x_{i}) \leq 0\}-\tau
)^{2} K^2 \left( \frac{\widehat{p}_{i}-\delta }{h_{p}}\right) \\
&&\overset{p}{\rightarrow }\lim_{n\rightarrow \infty }\frac{1}{%
h_{p}H^{\eta }}E\left[
s_{i}\left(G_{u_{\tau}}(\tau ,p_{i})+\tau ^{2}-\tau G_{u_{\tau}}(\tau ,p_{i})\right)
^{2}K^2 \left( \frac{p_{i}-\delta }{h_{p}}%
\right) \right] >0.
\end{eqnarray*}%
Thus, the test has power against $\sqrt{nh_{p}H^{\eta }}$ alternatives. \qedsymbol

\section*{Appendix B}\label{Appendix B}

\noindent \textbf{First Test - Conditional Mean}: \medskip

\noindent In this section, we extend the first test to the case of the nonparametric conditional mean. Conditional mean functions are of primary interest in many applied studies, and in fact most of the sample selection literature has dealt with this case \citep[see e.g.][]{H1979,DNV2003,J15}. The testing set-up relies on the following assumption:
\medskip

\noindent\textbf{A.M} Assume that: 
\begin{equation}
E\left[ y_{i}\Bigl\vert x_{i},z_{i},s_{i}=1\right]=E\left[ y_{i}\Bigl\vert %
x_{i},p_{i}\right].  \label{EQ1b}
\end{equation}%
holds almost surely. 

\medskip

\noindent Assumption \textbf{A.M}, which is Assumption 2.1(i) from \citet[][p.35]{DNV2003}, is implied by standard threshold crossing selection models where, among other regularity conditions, $%
s_{i}=1\{p(z_{i})>v_{i}\}$ and the observable $z_{i}$ is statistcally independent of the unobservables $\varepsilon_{i}$ and $v_{i}$ given $x_{i}$. In what follows, let $m(x_{i})\equiv E[y_{i}|x_{i},s_{i}=1]$ be the conditional mean function given $x_{i}$ and $s_{i}=1$. The hypotheses for the first test in the conditional mean case then read as:
\begin{equation}
H_{0,m}^{(1)}: E\left[ (y_{i}-m(x_{i}))|x_{i}=x,p_{i}=p\right]=0  \label{H01m}
\end{equation}%
for all\ $x\in \mathcal{X},$\ and $p\in \mathcal{P}$, versus%
\begin{equation}
H_{A,m}^{(1)}: E\left[ (y_{i}-m(x_{i}))|x_{i}=x,p_{i}=p\right]\neq  0   \label{HA1m}
\end{equation}%
for some $x\in \mathcal{X}$, and $p\in \mathcal{P}$. The test statistic for the conditional mean is constructed on the basis of a statistic by \citet{DGM2001}  and given by:
\begin{equation*}
Z_{1,n}^{m}=\sup_{(\underline{x},\overline{x})\in 
\mathcal{X},(\underline{p},\overline{p})\in \mathcal{P}}|Z_{1,n}^{m}\left( \underline{x},%
\overline{x},\underline{p},\overline{p}\right) |,
\end{equation*}%
where 
\begin{equation*}
Z_{1,n}^{m}\left( \underline{x},%
\overline{x},\underline{p},\overline{p}\right) =\frac{1}{\sqrt{n%
}}\sum_{i=1}^{n}s_{i}(y_{i}-\widehat{m}(x_{i}) )\widehat{f}_{x}(x_{i})\dprod_{j=1}^{d_{x}}1\{\underline{x}_{j}<x_{j,i}<\overline{x}_{j}\}1\{\underline{p}\leq \widehat{p}%
_{i}\leq \overline{p}\}.
\end{equation*}%
with:
\[
\widehat{m}(x_{i})=\frac{1}{nh^{d_{x}}}\sum_{j=1}^{n}y_{j}s_{i}\mathbf{K}\left(\frac{x_{i}-x_{j}}{h}\right)/\widehat{f}_{x}(x_{i})\quad\text{and}\quad \widehat{f}_{x}(x_{i})=\frac{1}{nh^{d_{x}}}\sum_{j=1}^{n}\mathbf{K}\left(\frac{x_{i}-x_{j}}{h}\right)
\]
We require the following technical assumptions:\medskip

\noindent \textbf{S.1}\label{S1}  $(y_{i},x_{i}^{\prime },z_{i}^{\prime
},s_{i})\subset R_{y}\times R_{x}\times R_{z}\times \{0,1\}$ are identically
and independently distributed. Let $\mathcal{X}\equiv \mathcal{X}_{1}\times
\ldots \times \mathcal{X}_{d_{x}}$ denote a compact subset of the interior
of $R_{x}$. $z_{i}$ contains at least one variable which is not contained in 
$x_{i}$ and which is not $x_i$-measurable. The distributions of $x_{i}$ and $z_{i}$ have a probability density
function with respect to Lebesgue measure which is strictly positive and
continuously differentiable (with bounded derivatives) over the interior of
their respective supports. Assume that $\Pr (s_{i}=1|x,p)=\Pr (s_{i}=1|p)>0$ for
all $p\in\mathcal{P}$ and $x\in\mathcal{X}$. Moreover, assume that $E[|y_{i}|^{2+\delta}]<\infty$ for some $\delta>0$.
\medskip

\noindent \textbf{S.2}\label{S2}
Assume that $m(x_{i})$ and $f_{x}(x_{i})$ are $q$ times differentiable over $\mathcal{X}$ with uniformly bounded derivatives and $q>d_{x}/2$.\medskip

\noindent \textbf{S.3}\label{S3} The non-negative kernel function $K(\cdot)$ is a bounded, continuously differentiable function with uniformly bounded derivative and compact support on $[-1,1]$. It satisfies $\int K(v)dv=1$, $\int v^{l}K(v)dv=0$ for $l=1,\ldots,q-1$, and $\int |v^{q}|K(v)dv<\infty$.\medskip

\noindent \textbf{S.4}\label{S4}
The conditional distribution function $F_{p|x,s=1}(p_{i}\leq \cdot|\cdot,s_{i}=1)$ is continuously differentiable in $x$ and $p$ uniformly over $\mathcal{X}$ and $\mathcal{P}$, and for some $C>0$, satisfies:
\[
\vert \nabla_{p}F_{p|x,s=1}(p |x,s=1)-\nabla_{p}F_{p|x,s=1}(p^{\prime} |x^{\prime},s=1)\vert \leq C\Vert (x,p)-(x^{\prime},p^{\prime})\Vert
\] 
for every $(x,x^{\prime})\in\mathcal{X}$ and $(p,p^{\prime})\in\mathcal{P}$
\medskip

\noindent \textbf{S.5}\label{S5} There exists an estimator $\widehat{p}%
(z_{i})$ such that $\sup_{z\in\mathcal{Z}}|\widehat{p}(z)-p(z)|=o_{p}(n^{-%
\frac{1}{4}})$ with $\mathcal{Z}$ a compact subset of $R_{z}$, and that:
\[
\Pr\left(\exists i: z_{i}\in R_{z}\setminus \mathcal{Z}, p(z_{i})\in\mathcal{P}\right)=o(n^{-\frac{1}{4}}). 
\]\medskip

\noindent Note that primitive conditions for Assumption S.1 can for instance be found in \citet{EJCL2014}. 
\medskip

\noindent \textbf{Theorem S1}: Let Assumptions \textbf{S.1}-\textbf{S.5} and \textbf{A.M} hold. If, as $n\rightarrow \infty$, It holds that (i) $h_{x}\rightarrow 0$ (ii) $nh_{x}^{d_{x}}\rightarrow \infty$, (iii) $nh_{x}^{2q}\rightarrow 0$, then

\noindent (i) under $H_{0,m}^{(1)}$,%
\begin{equation*}
Z_{1,n}^{m}\Rightarrow Z_{1}^{m},
\end{equation*}
where $Z_{1}^{m}$ denotes the supremum of a zero mean Gaussian process with covariance kernel defined in the proof of Theorem S1.

\noindent (ii) and under $H^{(1)}_{A,m},$ there exists $\varepsilon >0,$ such that%
\begin{equation*}
\lim_{n\rightarrow \infty }\Pr \left( Z_{1,n}^{m}>\varepsilon \right) =1.
\end{equation*}
\medskip

\noindent	In the proof of Theorem S1 we establish that:
\begin{eqnarray*}
&&\frac{1}{\sqrt{n}}\sum_{i=1}^{n} s_{i}(y_{i}-\widehat{m}(x_{i})\widehat{f}_{x}(x_{i})\dprod_{j=1}^{d_{x}}1\{\underline{x}_{j}<x_{j,i}<\overline{x}_{j}\}1\{\underline{p}\leq \widehat{p}_{i}\leq \overline{p}\}\\
&=&\frac{1}{\sqrt{n}}\sum_{i=1}^{n} f_{x}(x_{i})(y_{i}-m(x_{i}))\dprod_{j=1}^{d_{x}}1\{\underline{x}_{j}<x_{j,i}<\overline{x}_{j}\}s_{i} \left( 1\{ p_{i}\leq \overline{p}\}-1\{p_{i}\leq \underline{p}\}\right. \\
&&\left. - \left( F_{p|x,s=1}(\overline{p}|x_{i},s_{i}=1) - F_{p|x,s=1}(\underline{p}|x_{i},s_{i}=1)\right)\Pr(s_{i}=1\vert x_{i})\right) +o_{p}(1).
\end{eqnarray*}
uniformly over $\mathcal{X}$ and $\mathcal{P}$. Thus, a natural bootstrap version of $Z_{1,n}^{m}(\underline{x},\overline{x},p)$ would be:
\begin{eqnarray*}
&&\widetilde{U}_{0}^{\ast}\left( \underline{x},%
\overline{x},\underline{p},\overline{p}\right)\\
&=&\frac{1}{\sqrt{n}}\sum_{i=1}^{n} v_{i}f_{x}(x_{i})(y_{i}-m(x_{i}))\dprod_{j=1}^{d_{x}}1\{\underline{x}_{j}<x_{j,i}<\overline{x}_{j}\}s_{i} \{ 1\{ p_{i}\leq \overline{p}\}- 1\{ p_{i}\leq \underline{p}\}\\
&& -  \left(F_{p|x,s=1}(\overline{p}|x_{i},s_{i}=1)-F_{p|x,s=1}(\underline{p}|x_{i},s_{i}=1)\right)\Pr(s_{i}=1\vert x_{i}) \},
\end{eqnarray*}
where $v_{i}$, $i=1,\ldots, n$ are i.i.d. random variables satisfying $E[v_{i}]=0$ and $E[v_{i}^{2}]=1$, which are independent of $(y_{i},x_{i}^{\prime})$. Of course, $\widetilde{U}_{0}^{\ast}\left( \underline{x},\overline{x},\underline{p},\overline{p}\right)$ is infeasible since $p_{i}$, $F_{p|x,s=1}(p|x_{i},s_{i}=1)$ as well as $\Pr(s_{i}=1\vert x_{i})$ need to be estimated. Instead, the feasible bootstrap version we use for computational reasons (cf. \citet{DGM2001}) is given by:
\begin{eqnarray*}
Z_{1,n}^{\ast m}\left( \underline{x},%
\overline{x},p\right)=\frac{1}{\sqrt{n}}\sum_{i=1}^{n} (y_{i}^{\ast}-\widehat{m}^{\ast}(x_{i}))s_{i}\widehat{f}_{x}(x_{i})\dprod_{j=1}^{d_{x}}1\{\underline{x}_{j}<x_{j,i}<\overline{x}_{j}\}s_{i} \{ 1\{\underline{p}\leq \widehat{p}_{i}\leq \overline{p}\}
\end{eqnarray*}
with $y_{i}^{\ast}=\widehat{m}(x_{i})+\widehat{\varepsilon}^{\ast}_{i}$ and $\widehat{\varepsilon}^{\ast}_{i}=v_{i}\widehat{\varepsilon}_{i}=v_{i}(y_{i}-\widehat{m}(x_{i}))$, and 
\[
\widehat{m}^{\ast}(x_{i})=\frac{1}{nh_{x}^{d_{x}}\widehat{f}_{x}(x_{i})} \sum_{j=1}^{n}y_{j}^{\ast} \mathbf{K}\left(\frac{x_{i}-x_{j}}{h}\right).
\]
The additional assumption we impose is:\medskip

\noindent \textbf{S.6}\label{S6} There exists an estimator $\widehat{m}(x)$ such that:  
\[
\sup_{x\in\mathcal{X}}\vert \widehat{m}(x)-m(x) \vert=o_{p}(n^{-\frac{1}{4}}),
\]
where $\mathcal{X}$ is a compact subset defined in A.1. Moreover, $f_{x}(x_{i})>c>0$ for all $x_{i}\in\mathcal{X}$.\medskip

\noindent Assumption S.6 is a high-level condition and rather standard in the non- and semiparametric literature. If the estimator is a kernel estimator with a kernel function as defined in S.3, this condition will be satisfied in our context when $d_{x}<4$.\medskip

\noindent \textbf{Theorem S1$^{\ast}$}: Let Assumptions \textbf{S.1}-\textbf{S.6} and \textbf{A.M} from the paper hold. If, as $n\rightarrow \infty$, it holds that (i) $h_{x}\rightarrow 0$ (ii) $nh_{x}^{d_{x}}\rightarrow \infty$, (iii) $nh_{x}^{2q}\rightarrow 0$, (iv) $R\rightarrow \infty$, then

\noindent (i) under $H_{0,q}^{(1)}$%
\begin{equation*}
\lim_{n,R\rightarrow \infty }\Pr \left( Z_{1,n}^{m}\geq c_{(1-\alpha
),n,R}^{\ast (1)}\right) =\alpha
\end{equation*}%
(ii) under $H_{A,q}^{(1)}$%
\begin{equation*}
\lim_{n,R\rightarrow \infty }\Pr \left( Z_{1,n}^{m}\geq c_{(1-\alpha
),n,R}^{\ast (1)}\right) =1.
\end{equation*}%

\noindent \textbf{Monte Carlo - Conditional Mean}: \medskip

\noindent In this section we examine the finite sample properties of our tests for the conditional mean function. The outcome equation of our simulation design (given selection $s_{i}=1$) is given by:
\[
y_{i}=x_{i}^2 + 0.5 x_{i} + \gamma_{1}\widetilde{z}_{i} +0.5\varepsilon_{i},
\]
where $x_{i}\sim U(0,1)$, and the marginal distribution of $\varepsilon_{i}$ is standard normal. Like in the quantile set-up, selection enters into this set-up via:
\[
s_{i}=1\{0.75(x_{i}-0.5)+0.75z_{i}>\gamma_{2} v_{i}\},
\]
with:
\[
\left(\begin{array}{c}\varepsilon_{i}\\ v_{i}\end{array}\right) \sim N\left(\left(\begin{array}{c}0\\ 0\end{array}\right),\left(\begin{array}{cc}1 & \rho \\ \rho & 1 \end{array}\right) \right).
\]
Thus, $\rho$ controls the degree of selection and we consider three scenarios, namely the case of `no selection' ($\rho=0$), `moderate selection' ($\rho=0.25$), and `strong selection' ($\rho=0.5$), while $\gamma_{1}$ sets the level of misspecification and $\gamma_{2}$ re-scales the error term $\varepsilon_{i}$.  As in the quantile case, the instrument $\widetilde{z}_{i}$ is simulated according to one of the following four designs:
\begin{eqnarray*}
\mathbf{\text{(i)}}: &&\widetilde{z}_{i} \sim N(0,1),\\
 \mathbf{\text{(ii)}}: && \widetilde{z}_{i}\sim Binom(0.5)-0.5,\\
 \mathbf{\text{(iii)}}: &&\widetilde{z}_{i}\sim 1.5-Poisson(1.5),\\
 \mathbf{\text{(iv)}}: && \widetilde{z}_{i}\sim Discrete Unif.(0,7).
\end{eqnarray*}
We consider two sample sizes $n=\{400 , 1000\}$, which, given a selection probability of approximately $.5$, imply an effective sample size for the estimation of the conditional mean function of around 200 to 500 observations, respectively. 

For the first test we consider six different cases: Cases I-IV use a differently distributed instrument according to designs (i)-(iv) above, and the oracle propensity score. Cases V and VI on the other hand use $\widetilde{z}_{i}\sim N(0,1)$ and $\widetilde{z}_{i}\sim Binom(0.5)-0.5$, respectively, an a nonparametrically estimated propensity score. More specifically, for Cases V and VI we first estimate the propensity score using a standard local constant estimator with second order Epanechnikov kernel, where the bandwidth is chosen using cross-validation as allowed by our conditions (cf. footnote 7).\footnote{To construct this estimator as well as the estimators for the conditional mean, we use routines from the \texttt{np package} of \citet{HR2008}.}  In order to restrict ourselves to a compact subset $\mathcal{X}$, we trim the outer 2.5\% observations of the selected sample. The bandwidth is chosen according to $h_{x}=c\cdot \text{sd}(x_{i})n^{-1/3}$ with $c\in\{0.125,0.25,0.5\}$. Finally, we use the `warp speed' procedure of \citet{GPW2013} with 999 Monte Carlo replications and fix the nominal level of the test to $\alpha_{1}=0.05$ and $\alpha_{2}=0.1$. All results can be found in Table \ref{Table S1}.

As for the quantile case, we can observe that we have throughout a good control of the size and power, which increases with the degree of sample selection. More specifically, the size converges to the nominal level across the different scaling parameters for $h_{x}$ and irrespective of the discreteness or continuity of the underlying instrument (cf. Cases I-IV). As expected, this also holds when the propensity score is estimated nonparametrically choosing the bandwidth via cross-validation (Cases V-VI). On the other hand, turning to power, we see that power generally increases with the sample size. Interestingly, while there is a `power loss' of around 15 precentage points when the instrument is binary (Case II and VI), there is almost no loss when the instrument follows a Poisson or discrete uniform distribution.

\begin{table}[H]
\footnotesize
\begin{center}
\begin{tabular}{|c|c|c|c|c|c|c|c|c|c|c|}
\hline\hline
\multicolumn{11}{|c|}{Conditional Mean - First Test} \\ \hline\hline
 & & \multicolumn{3}{|c|}{$\rho=0$} &  \multicolumn{3}{|c|}{$\rho=0.25$} &  \multicolumn{3}{|c|}{$\rho=0.5$} \\ \hline
Case I & & $c=0.125$ & $c=0.25$ & $c=0.5$ & $c=0.125$ & $c=0.25$ & $c=0.5$ & $c=0.125$ & $c=0.25$ & $c=0.5$ \\\hline
\multirow{2}{*}{$\alpha=0.05$}&	$n=400$	&$	0.073	$&$	0.061	$&$	0.056	$&$	0.204	$&$	0.179	$&$	0.238	$&$	0.455	$&$	0.535	$&$	0.568	$\\	
&	$n=1000$	&$	0.054	$&$	0.046	$&$	0.053	$&$	0.355	$&$	0.350	$&$	0.414	$&$	0.917	$&$	0.931	$&$	0.905	$\\	\hline
\multirow{2}{*}{$\alpha=0.10$}&	$n=400$	&$	0.142	$&$	0.126	$&$	0.117	$&$	0.302	$&$	0.298	$&$	0.355	$&$	0.648	$&$	0.662	$&$	0.709	$\\	
&	$n=1000$	&$	0.118	$&$	0.094	$&$	0.098	$&$	0.459	$&$	0.480	$&$	0.537	$&$	0.957	$&$	0.966	$&$	0.960	$\\	\hline
																					
Case II & & $c=0.125$ & $c=0.25$ & $c=0.5$ & $c=0.125$ & $c=0.25$ & $c=0.5$ & $c=0.125$ & $c=0.25$ & $c=0.5$ \\\hline
\multirow{2}{*}{$\alpha=0.05$}&	$n=400$	&$	0.067	$&$	0.072	$&$	0.084	$&$	0.115	$&$	0.143	$&$	0.164	$&$	0.369	$&$	0.435	$&$	0.384	$\\	
&	$n=1000$	&$	0.047	$&$	0.040	$&$	0.052	$&$	0.224	$&$	0.215	$&$	0.299	$&$	0.770	$&$	0.702	$&$	0.730	$\\	\hline
\multirow{2}{*}{$\alpha=0.10$}&	$n=400$	&$	0.129	$&$	0.126	$&$	0.163	$&$	0.211	$&$	0.224	$&$	0.244	$&$	0.487	$&$	0.530	$&$	0.499	$\\	
&	$n=1000$	&$	0.102	$&$	0.112	$&$	0.108	$&$	0.337	$&$	0.346	$&$	0.390	$&$	0.826	$&$	0.792	$&$	0.815	$\\	\hline
																					
Case III & & $c=0.125$ & $c=0.25$ & $c=0.5$ & $c=0.125$ & $c=0.25$ & $c=0.5$ & $c=0.125$ & $c=0.25$ & $c=0.5$ \\\hline
\multirow{2}{*}{$\alpha=0.05$}&	$n=400$	&$	0.051	$&$	0.043	$&$	0.055	$&$	0.153	$&$	0.168	$&$	0.211	$&$	0.545	$&$	0.585	$&$	0.560	$\\	
&	$n=1000$	&$	0.044	$&$	0.032	$&$	0.038	$&$	0.393	$&$	0.377	$&$	0.358	$&$	0.908	$&$	0.903	$&$	0.911	$\\	\hline
\multirow{2}{*}{$\alpha=0.10$}&	$n=400$	&$	0.103	$&$	0.090	$&$	0.129	$&$	0.265	$&$	0.279	$&$	0.305	$&$	0.647	$&$	0.675	$&$	0.699	$\\	
&	$n=1000$	&$	0.110	$&$	0.081	$&$	0.092	$&$	0.501	$&$	0.518	$&$	0.484	$&$	0.954	$&$	0.946	$&$	0.953	$\\	\hline
																					
Case IV & & $c=0.125$ & $c=0.25$ & $c=0.5$ & $c=0.125$ & $c=0.25$ & $c=0.5$ & $c=0.125$ & $c=0.25$ & $c=0.5$ \\\hline
\multirow{2}{*}{$\alpha=0.05$}&	$n=400$	&$	0.077	$&$	0.059	$&$	0.056	$&$	0.218	$&$	0.218	$&$	0.239	$&$	0.606	$&$	0.587	$&$	0.586	$\\	
&	$n=1000$	&$	0.061	$&$	0.074	$&$	0.062	$&$	0.396	$&$	0.465	$&$	0.422	$&$	0.964	$&$	0.967	$&$	0.948	$\\	\hline
\multirow{2}{*}{$\alpha=0.10$}&	$n=400$	&$	0.134	$&$	0.118	$&$	0.117	$&$	0.316	$&$	0.339	$&$	0.349	$&$	0.714	$&$	0.674	$&$	0.709	$\\	
&	$n=1000$	&$	0.111	$&$	0.122	$&$	0.146	$&$	0.531	$&$	0.577	$&$	0.583	$&$	0.984	$&$	0.989	$&$	0.980	$\\	\hline
																					
Case V & & $c=0.125$ & $c=0.25$ & $c=0.5$ & $c=0.125$ & $c=0.25$ & $c=0.5$ & $c=0.125$ & $c=0.25$ & $c=0.5$ \\\hline
\multirow{2}{*}{$\alpha=0.05$}&	$n=400$	&$	0.073	$&$	0.068	$&$	0.071	$&$	0.172	$&$	0.177	$&$	0.178	$&$	0.536	$&$	0.526	$&$	0.640	$\\	
&	$n=1000$	&$	0.056	$&$	0.039	$&$	0.040	$&$	0.349	$&$	0.343	$&$	0.405	$&$	0.916	$&$	0.911	$&$	0.918	$\\	\hline
\multirow{2}{*}{$\alpha=0.10$}&	$n=400$	&$	0.128	$&$	0.132	$&$	0.152	$&$	0.280	$&$	0.295	$&$	0.287	$&$	0.660	$&$	0.686	$&$	0.730	$\\	
&	$n=1000$	&$	0.120	$&$	0.102	$&$	0.093	$&$	0.462	$&$	0.438	$&$	0.512	$&$	0.954	$&$	0.955	$&$	0.969	$\\	\hline
																					
Case VI & & $c=0.125$ & $c=0.25$ & $c=0.5$ & $c=0.125$ & $c=0.25$ & $c=0.5$ & $c=0.125$ & $c=0.25$ & $c=0.5$ \\\hline
\multirow{2}{*}{$\alpha=0.05$}&	$n=400$	&$	0.066	$&$	0.053	$&$	0.064	$&$	0.160	$&$	0.134	$&$	0.142	$&$	0.375	$&$	0.350	$&$	0.375	$\\	
&	$n=1000$	&$	0.049	$&$	0.063	$&$	0.046	$&$	0.212	$&$	0.211	$&$	0.229	$&$	0.718	$&$	0.695	$&$	0.712	$\\	\hline
\multirow{2}{*}{$\alpha=0.10$}&	$n=400$	&$	0.111	$&$	0.124	$&$	0.132	$&$	0.220	$&$	0.197	$&$	0.225	$&$	0.474	$&$	0.453	$&$	0.488	$\\	
&	$n=1000$	&$	0.082	$&$	0.100	$&$	0.096	$&$	0.329	$&$	0.303	$&$	0.338	$&$	0.811	$&$	0.808	$&$	0.808	$\\	\hline
\hline
\end{tabular}%
\end{center}
\normalsize
\caption{First Test }
\label{Table S1}
\end{table}

\bibliographystyle{chicago}
\bibliography{QuantileSelection}

\end{document}